\documentclass[twocolumn]{aastex61}

\usepackage{graphicx}
\usepackage{times}
\usepackage{amsmath}
\usepackage{color}
\usepackage{graphics}
\usepackage{subfigure}
\usepackage{txfonts}
\usepackage{natbib}
\usepackage{wasysym}
\usepackage{afterpage}

\newcommand{\kps}{km\,s$^{-1}$}

\definecolor{carnelian}{rgb}{0.7, 0.11, 0.11}
\definecolor{ao(english)}{rgb}{0.0, 0.5, 0.0}

\received{\today}
\revised{}
\accepted{}
\submitjournal{ApJ}

\shorttitle{Expanding and Contracting Coronal Loops}
\shortauthors{Dud\'{i}k et al.}
\watermark{draft}

\begin{document}

\title{Expanding and Contracting Coronal Loops as Evidence of Vortex Flows \\ Induced by Solar Eruptions}

\correspondingauthor{J. Dud\'{i}k}
\email{jaroslav.dudik@asu.cas.cz}

\author[0000-0003-1308-7427]{J. Dud\'{i}k}
\affil{Astronomical Institute of the Czech Academy of Sciences, Fri\v{c}ova 298, 251 65 Ond\v{r}ejov, Czech Republic}
\affiliation{LESIA, Observatoire de Paris, Psl Research University, CNRS, Sorbonne Universités, UPMC Univ. Paris 06, Univ. Paris Diderot, Sorbonne Paris Cité, 5 place Jules Janssen, F-92195 Meudon, France}

\author{F. P. Zuccarello}
\affiliation{Center for Mathematical Plasma Astrophysics, Department of Mathematics, KU Leuven, Celestijnenlaan 200B, B-3001 Leuven, Belgium}
\affiliation{UCL-Mullard Space Science Laboratory, Holmbury St. Mary, Dorking, Surrey, RH5 6NT, UK}
\affiliation{LESIA, Observatoire de Paris, Psl Research University, CNRS, Sorbonne Universités, UPMC Univ. Paris 06, Univ. Paris Diderot, Sorbonne Paris Cité, 5 place Jules Janssen, F-92195 Meudon, France}

\author{G. Aulanier}
\affiliation{LESIA, Observatoire de Paris, Psl Research University, CNRS, Sorbonne Universités, UPMC Univ. Paris 06, Univ. Paris Diderot, Sorbonne Paris Cité, 5 place Jules Janssen, F-92195 Meudon, France}

\author{B. Schmieder}
\affiliation{LESIA, Observatoire de Paris, Psl Research University, CNRS, Sorbonne Universités, UPMC Univ. Paris 06, Univ. Paris Diderot, Sorbonne Paris Cité, 5 place Jules Janssen, F-92195 Meudon, France}

\author{P. D\'{e}moulin}
\affiliation{LESIA, Observatoire de Paris, Psl Research University, CNRS, Sorbonne Universités, UPMC Univ. Paris 06, Univ. Paris Diderot, Sorbonne Paris Cité, 5 place Jules Janssen, F-92195 Meudon, France}


\begin{abstract}
Eruptive solar flares were predicted to generate large-scale vortex flows at both sides of the erupting magnetic flux rope. This process is analogous to a well-known hydrodynamic process creating vortex rings. The vortices lead to advection of closed coronal loops located at peripheries of the flaring active region. Outward flows are expected in the upper part and returning flows in the lower part of the vortex. Here, we examine two eruptive solar flares, an X1.1-class flare SOL2012-03-05T03:20 and a C3.5-class SOL2013-06-19T07:29. In both flares, we find that the coronal loops observed by the Atmospheric Imaging Assembly in its 171\,\AA, 193\,\AA, or 211\,\AA~passbands show coexistence of expanding and contracting motions, in accordance with the model prediction. In the X-class flare, multiple expanding/contracting loops coexist for more than 35 minutes, while in the C-class flare, an expanding loop in 193\,\AA~appears to be close-by and co-temporal with an apparently imploding loop arcade seen in 171\,\AA. Later, the 193\,\AA~loop also switches to contraction. These observations are naturally explained by vortex flows present in a model of eruptive solar flares.
\end{abstract}

\keywords{Sun: flares --- Sun: corona --- Sun: UV radiation --- methods: observational --- methods: data analysis}

%
\section{Introduction}
\label{Sect:1}


Eruptive solar flares are one of the most sudden and violent manifestation of the solar activity. They exhibit a multitude of observed dynamic phenomena arising from release of energy stored within the pre-flare magnetic field \citep[e.g.,][]{Fletcher11,Schmieder15}. The destabilization of the magnetic field via the torus instability results in an eruption \citep[e.g.,][]{Aulanier12,Zuccarello15,Zuccarello16}, with the eruption driving other dynamic phenomena, such as slipping motion of flare loops and expansion/contraction behavior of the neighboring coronal loops \citep[e.g.,][]{Janvier13,Dudik14a,Dudik16}. In this paper, we are concerned with the expansion/contraction behavior of closed coronal loops at the periphery of active regions with respect to the flare and/or eruption site.

Observational evidence suggests that while the contracting loop motions in flares are not common, they may happen at any phase of solar flares, including the early, impulsive, and gradual one \citep[e.g.,][]{Liu09a,Liu09b,Liu10,Liu12a,Sun12,Gosain12,Simoes13a,Yan13,Shen14,Kushwaha15,Thalmann16,Petrie16}. \citet{Gosain12} and \citet{Simoes13a} have shown that the onset of the contraction depends on the location of the loops with respect to the flare: loops located progressively further away from the flare contract later. \citet{Russell15} showed that these loops can also display oscillations while contracting.

The relation of these loop contractions to an eruption (as opposed to a given phase of a flare) is clearer. Loop contractions occur only after, or are simultaneous with the onset of the eruption \citep[e.g.,][]{Liu12a,Shen14}, with strong coronal loop contractions occuring only after the onset of the fast eruption \citep{Dudik16,Wang16}. The latter authors reported on the behavior of a coronal loop arcade during a filament eruption and a C-class flare. They have shown that the onset of the filament eruption coincides with the start of the 171\,\AA~arcade expansion, and is followed by a rapid arcade contraction after the filament has erupted.

This behavior has been interpreted in terms of the implosion conjecture proposed originally by \cite{Hudson00}. The conjecture holds that, for low-$\beta$ plasmas where gravity is unimportant, liberation of magnetic energy in a flare or an eruption must somehow ``originate in a magnetic implosion'' \citep[][Section 5, p. L76 therein]{Hudson00}, i.e, the decrease of the $\int B^2/ 2 \mu \mathrm{d}V $ over a volume $V$ of the solar corona. In other words, a portion of the solar corona must contract (implode) to power a flare or an eruption. In the case of an eruptive flare the decrease of volume must be stronger than for a confined flare of same energy in order to compensate the eruption-related increase in $V$. The understanding of implosion later evolved to contracting motions being a consequence of the reduction of magnetic pressure in the flare or erupting region \citep{Janse07,Russell15,Wang16}. The contracting motions of the coronal loops at peripheries of active regions are then interpreted as a change of ``the position in which nearby coronal loops are in equilibrium'' \citep[][Section 5, p. A5
therein]{Russell15}, where the equilibrium occurs between outward magnetic pressure and the inward magnetic tension of curved coronal loops.

In a three-dimensional magnetohydrodynamic model of an erupting magnetic flux rope \citep[][ hereafter, ZAD17]{Zuccarello17}, we have shown that contracting coronal loops naturally exist in the proximity of the flare and eruption site. A detailed analysis of the evolution of the Lorentz forces as well as flows in the model has shown that flow vorticity is generated around the erupting flux rope as soon as the eruption begins. This is a consequence of the passage of an Alfv\'en wave. The magnetic arcades near the legs of the erupting flux rope are advected by these vortex flows and display expansion or contraction depending on their location with respect to the vortex. An important consequence of the vortices advecting the coronal loops is that both expanding and contracting loops should be observed co-temporally. 

The aim of this paper is to provide an observational evidence of coexistence of expanding and contracting motions of the peripheral coronal loops predicted by the eruption model, and to show that their presence and intensity do not depend on the energy of the flare. To this purpose we study the evolution of two eruptive flares, one X-class, and one C-class. The model is described in Section \ref{Sect:2}, while Sects. \ref{Sect:3} and \ref{Sect:4} describe the coronal loop dynamics in the X- and C-class events, respectively. The results are summarized and discussed in Section \ref{Sect:5}.

%
\begin{figure*}[ht]
	
\begin{center}
\subfigure{ 
\includegraphics[width=0.49\textwidth]{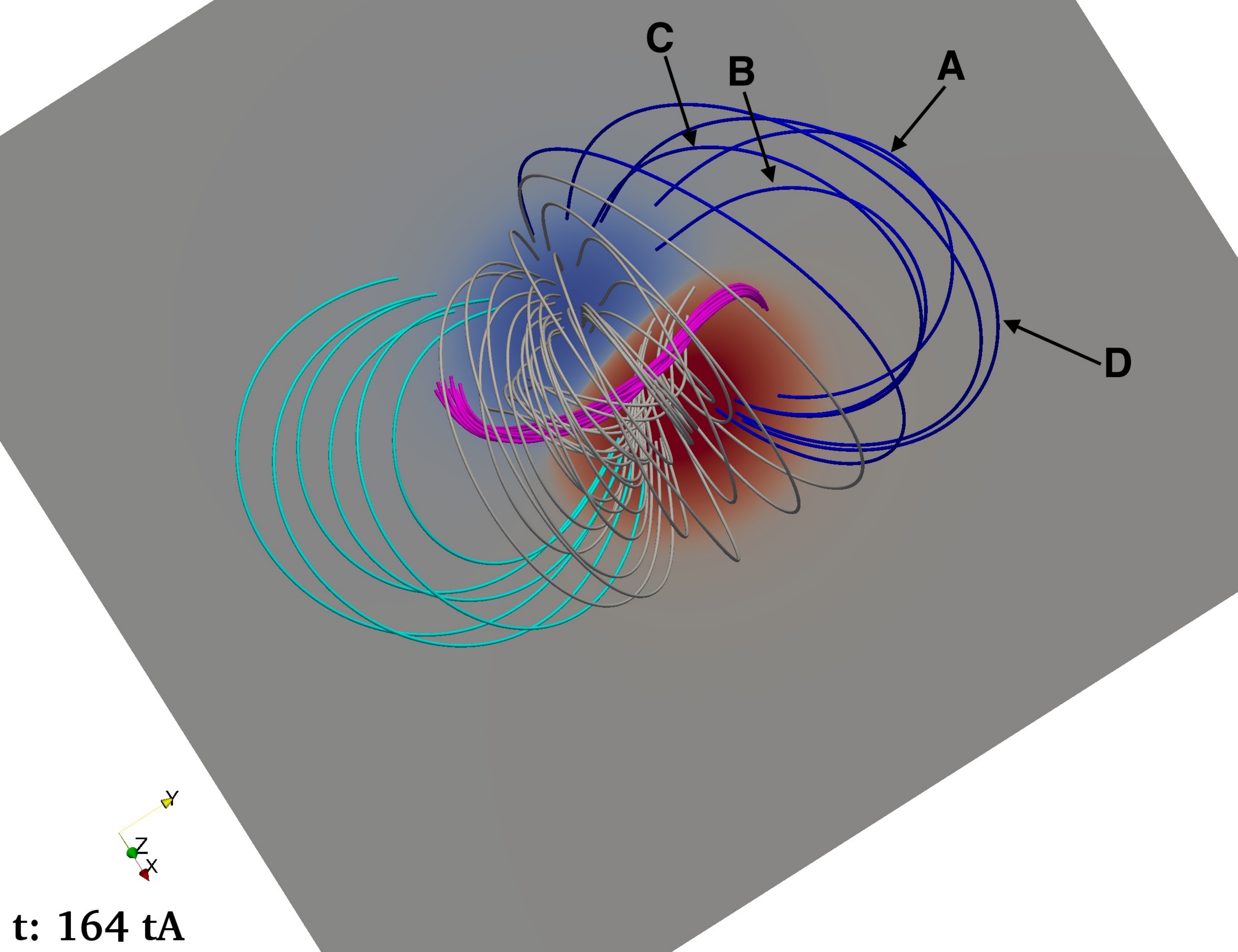}\label{Fig:Model1}}
\subfigure{ 
\includegraphics[width=0.49\textwidth]{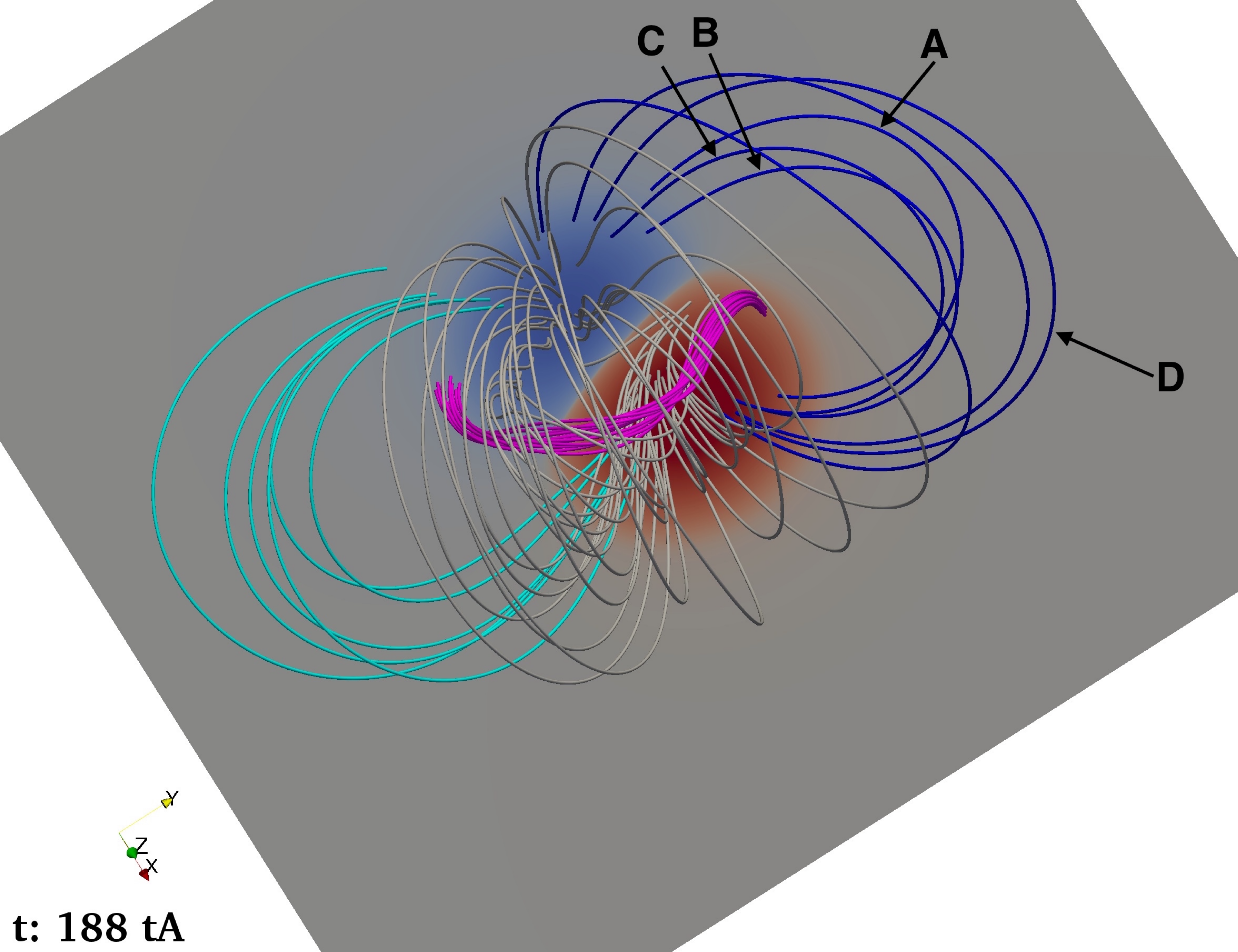}\label{Fig:Model2}}
\\
\subfigure{
\includegraphics[width=0.49\textwidth]{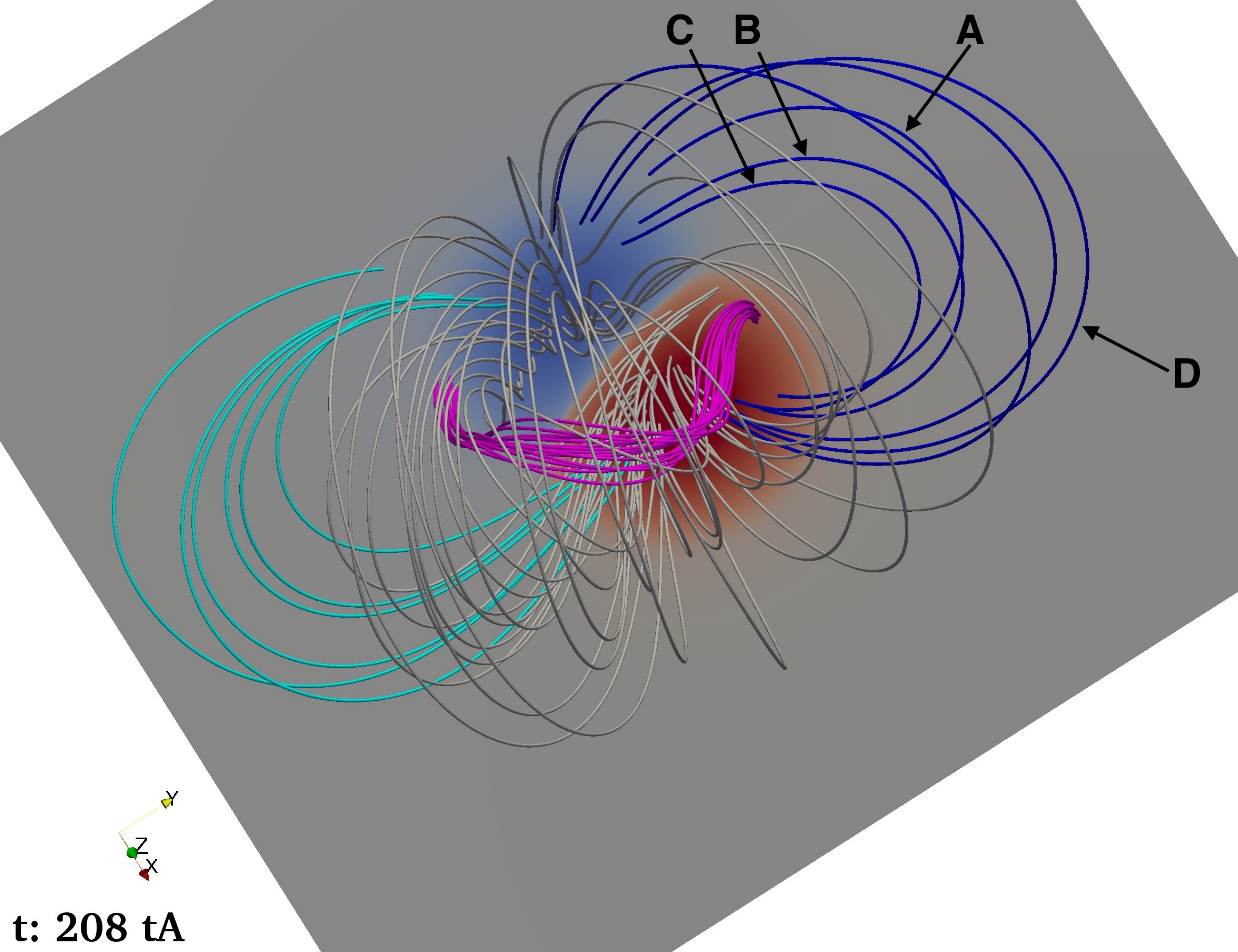}\label{Fig:Model4}}
\subfigure{
\includegraphics[width=0.49\textwidth]{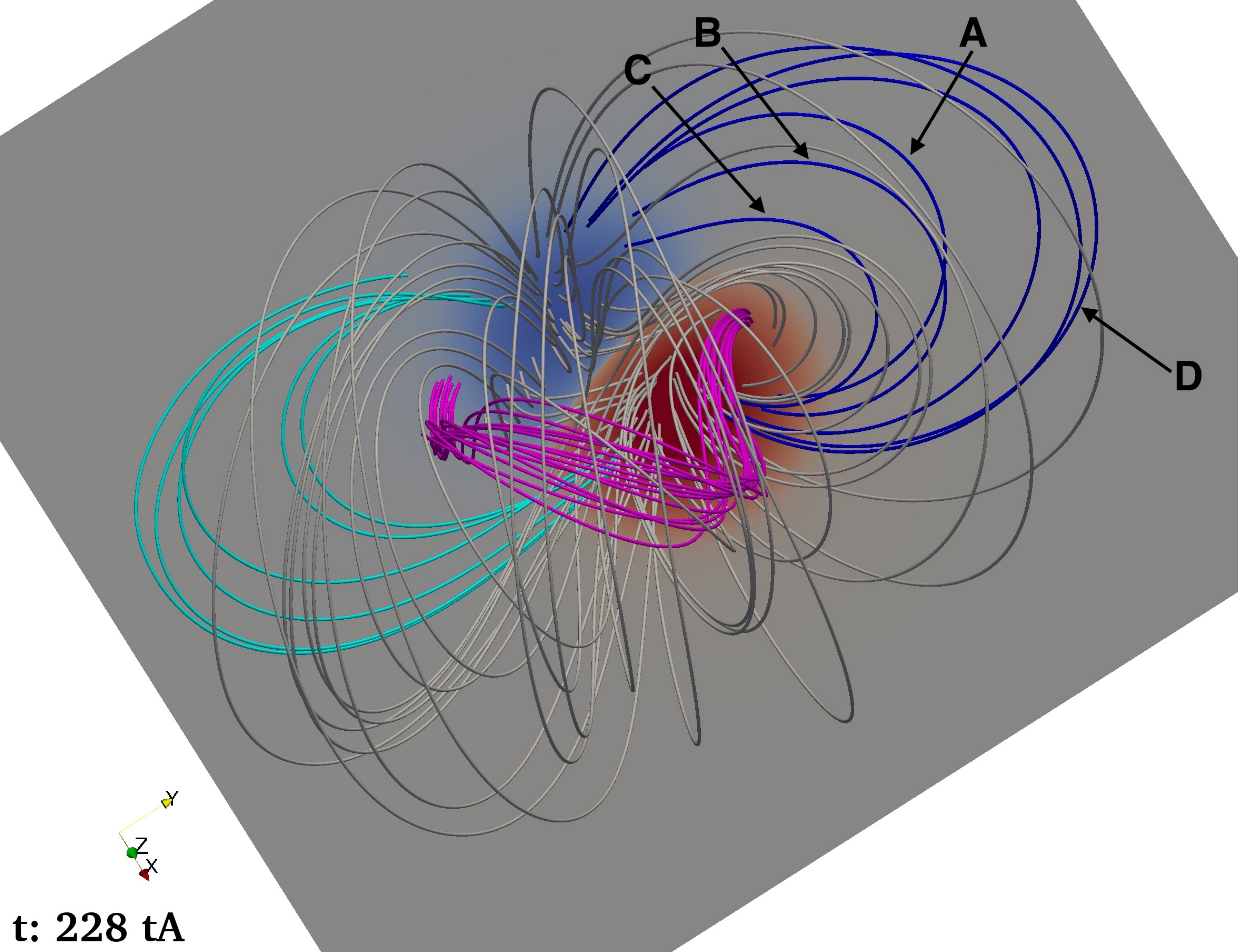}\label{Fig:Model5}}
\caption{Selected snapshots of the evolution of the system. Red/blue indicate positive/negative magnetic field (component normal to the boundary $z=0$). The magenta field lines highlight the central part of the erupting magnetic flux rope. The gray field lines highlight the portion of the overlying field that is rooted around the center of the strongest field (the center of the polarities), while the blue/cyan field lines are rooted in the polarities' periphery. The labels in the top left panel are used to identify selected field lines which dynamics is described in Section~\ref{Sect:2.1}. In each row field lines with the same photospheric anchorage are shown.
(An animation of this figure from t = 164 to 244 $t_\mathrm{A}$ is available in the online version of the article.)
\label{Fig:Model}}
\end{center}
\end{figure*}
%

%
%
\section{Vortex Flows in the Model}
\label{Sect:2}

The dynamics of peripheral loops during the occurrence of eruptive flares has been modeled and described in ZAD17. Here, we briefly summarize the most salient points.

\subsection{The MHD Model}
\label{Sect:2.1}

To model the dynamics of peripheral magnetic loops during an eruptive flare we solved the three-dimensional MHD equations by using the visco-resistive OHM-MPI code \citep{Aulanier05,Aulanier10,Zuccarello15}. As detailed in \cite{Zuccarello15} the MHD equations are solved in their dimensionless form and on a non uniform mesh that expands from the origin at $x=y=z=0$.

The initial conditions for the simulation are as follows. The magnetic field is generated by two unbalanced monopoles placed at different heights below the photospheric boundary. The initial density is $\rho(t=0)=B^2(t=0)$,  so that the Alfv\'{e}n speed is uniform within the numerical domain. Finally, at $t=0$ there are no flows present. We apply `open' boundary conditions for the side and top boundaries and line-tied boundary conditions for the bottom boundary placed at $z=0$ \citep[see][]{Aulanier05}.

During the first stage of the simulation, asymmetric rotational boundary flows centered around the center of the two polarities are imposed at the photospheric boundary $z=0$. The strongest component of these flows occur close to the polarity inversion line and decreases towards the center of the polarity. These motions result in a sheared arcade magnetic field configuration.

During the second stage of the simulation a different class of photopheric flows is applied. The mathematical formulation of these flows is described in \citet[][Run D2 therein]{Zuccarello15}. A key property of these flows is that they have a component converging towards the polarity inversion line. A magnetic flux rope is formed as a consequence of flux cancellation driven by these boundary flows. The flux rope being formed undergoes a slow rise until a point when it experiences a full eruption. To study the stability of the magnetic flux rope and to determine the moment of the eruption, the imposed boundary motions are slowed to zero, by using a hyperbolic tangent time profile of width $\Delta t =6 t_A$, at different times during the evolution of the system. It was found that the flux rope becomes torus unstable and experiences a full eruption at around $t =165t_\mathrm{A}$ \citep{Zuccarello15}.  

During the eruption, an increased coronal resistivity has been applied to delay the onset of a numerical instability at the current sheet. The numerical instability would halt the simulation. Finally, the simulation lasts until $t=244~t_\mathrm{A}$, when the numerical instability at the collapsing current sheet is no longer prevented by the enhanced resistivity.

\subsection{Vortex Motion in the Model}
\label{Sect:2.2}


The dynamics of the erupting system as well as the evolution of the Lorentz forces in the computational domain from the moment of the onset of the eruption onward has been discussed in detail in ZAD17. There, we have shown that vortex flows develop as soon as the flux rope enters the torus unstable regime. These vortex flows are the result of the propagation of an  Alfv\'{e}n wave generated as a consequence of the initial upward motion of the plasma-carrying magnetic structure. The compressible component of the flow leaves the numerical domain within an Alfv\'{e}n crossing time, while the solenoidal component, i.e., the vorticity carrying component, remains in the domain and manifests itself as two slowly (compared to the Alfv\'{e}n crossing time) propagating and expanding vortex arcs at each sides of the legs of the flux rope. This dynamics is equivalent to the generation of vortex rings by a vortex cannon. However, the magnetic field introduces an anisotropy that does not exist in the pure hydrodynamic case. While in the hydrodynamic case a toroidal vorticity ring is generated around the moving fluid element, in the zero plasma-$\beta$ MHD case the magnetic tension of the overlying field inhibits the vortex flows along planes perpendicular to the flux rope axis. Therefore, the vortex flows only develop on planes almost parallel to the axis of the flux rope. There, the vortex flow is orthogonal to the inclined peripheral loops and cannot be inhibited by their magnetic tension that only acts along the axis of loop curvature (see Section~3 of ZAD17 for more details). This is consistent with the observational findings that contracting loops are predominately observed at the peripheries of erupting active regions. Highly inclined loops are prone to have higher density since apparent hydrostatic density scale-height changes with inclination \citep[see Figure 3.12 of][]{Aschwanden05}, and such loops are expected to be caught in the returning part of the vortex.

\subsection{Evolution of the magnetic loops}
\label{Sect:2.3}

The evolution of the system as seen from planes nearly perpendicular and parallel to the flux rope axis has been shown and discussed in ZAD17 (Figure~1 and Section~2 therein), and we refer the reader to that paper for the discussion of the forces and physical processes that drive this dynamics. Here we briefly discuss the evolution when the system is seen from a top view, as is often the case for solar active regions seen on solar disk. 

The evolution of the system between $t = 164$ and 244\,$t_\mathrm{A}$, as seen from above, is presented in Figure~\ref{Fig:Model} and accompanying animation 1. While the analysis of the Figure and the accompanying animation seems to show that the cyan/blue field lines globally undergo an initial expansion followed by a global contraction starting from about 232\,$t_\mathrm{A}$, 
the timing and importance of the different kind of motions is different for different field lines. In fact, expanding and contracting motions coexist within a given arcade.

In particular, the field line labeled as `A' shows a minor expansion until $t$\,=\,204\,$t_\mathrm{A}$ before contracting; the field line labeled as `B' shows a more pronounced expansion and starts to contract only after $t$\,=\,208\,$t_\mathrm{A}$. The field line labeled as `C' displays a similar behavior as the field line `B', but starts to contract quite soon after the onset of the torus instability (t$\simeq$ 188\,$t_\mathrm{A}$) and undergoes a significant contraction while the other field lines of the same arcade are still expanding. Finally, the field line labeled as `D' undergoes an expansion until about 228\,$t_\mathrm{A}$ when finally starts to contract towards the site of the eruption. The cyan field lines show a similar behavior with coexistence of expanding/contracting motions.

Both the blue and cyan field lines never become part of the flux rope during its eruption, i.e., they do not reconnect at the current sheet formed below the erupting flux rope. This is not true for some of the gray field lines shown in Figure~\ref{Fig:Model}. (In ZAD17, the reconnecting field lines have been distinguished by green color.) This means that the blue and cyan field lines studied here belong to a connectivity domain that is separated from the connectivity domain of the flux rope at all times. Therefore, the dynamics of the blue and cyan field lines is not related to the magnetic pressure driven expansion of the flux rope that would, at most, only affect the flux rope's field lines. Instead, it is due to the vorticity that has been generated by the passage of the initial Alfv\'en wave front, as already discussed in ZAD17. As a final remark, we note that an eventual filament \citep[modeled as the dipped portion of the field lines, see][]{Zuccarello16} would be located below the central part of the flux rope shown as magenta field lines, therefore even further away from the blue/cyan field lines.

We note that the footpoints of some of the modeled loops undergo slow slipping motions. These motions are likely connected to the presence of broad quasi separatrix layers \citep[QSLs][]{Demoulin96a,Demoulin96b} that were formed before the eruption, as can be seen in Figure~3 of \citet{Janvier13}. The increased coronal resistivity, applied during the eruption stage in order to delay the onset of a numerical instability at the current sheet (see Section \ref{Sect:2.1}), could also contribute to the slow slippage of the loop footpoints. However, it is not a dominant effect, as evidenced by presence of other loops that display almost no slipping motion.

%
\begin{figure*}[ht]
	\centering
	\includegraphics[width=8.8cm,clip,viewport= 0 40 495 345]{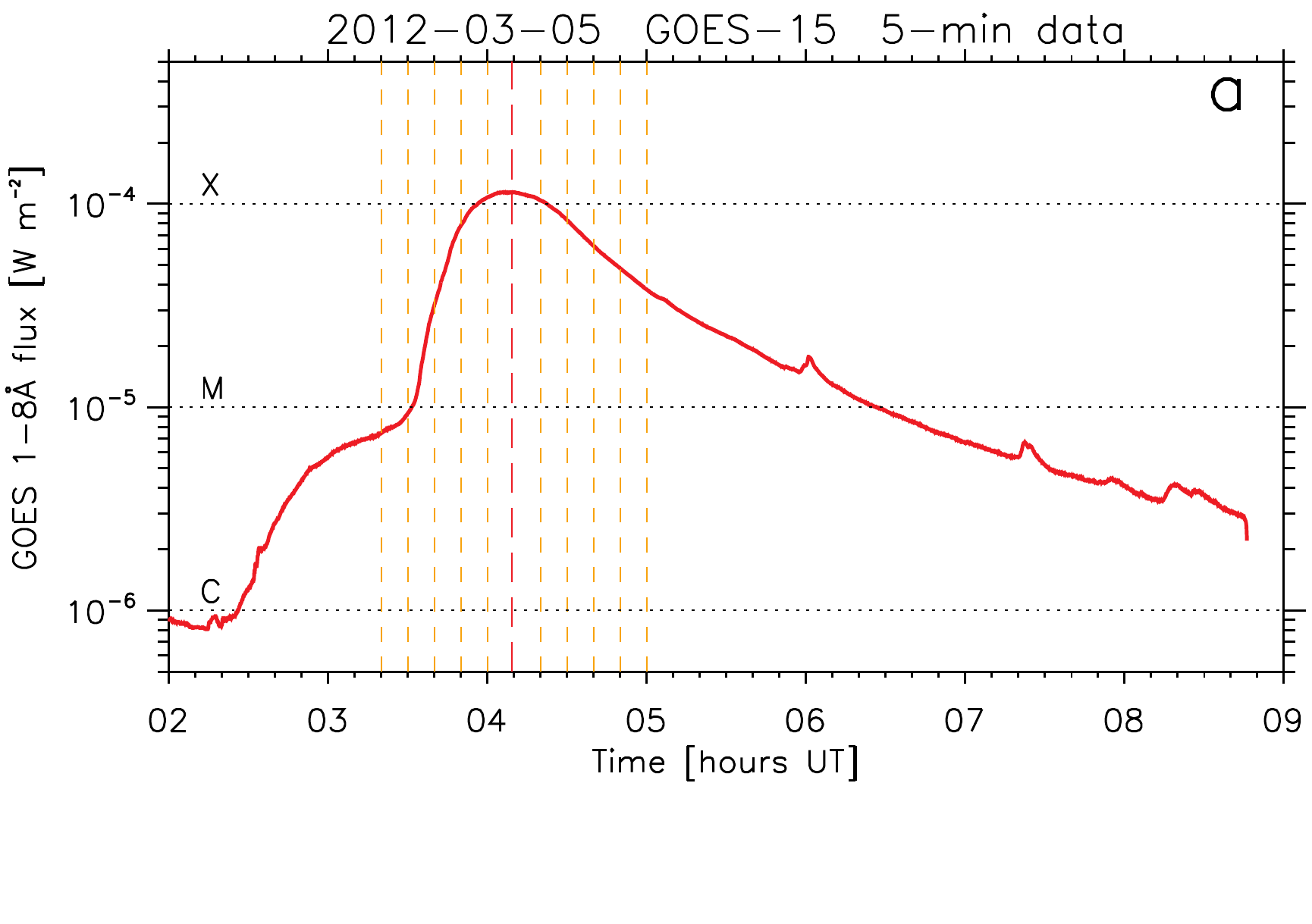}
	\includegraphics[width=8.8cm,clip,viewport= 0 40 495 345]{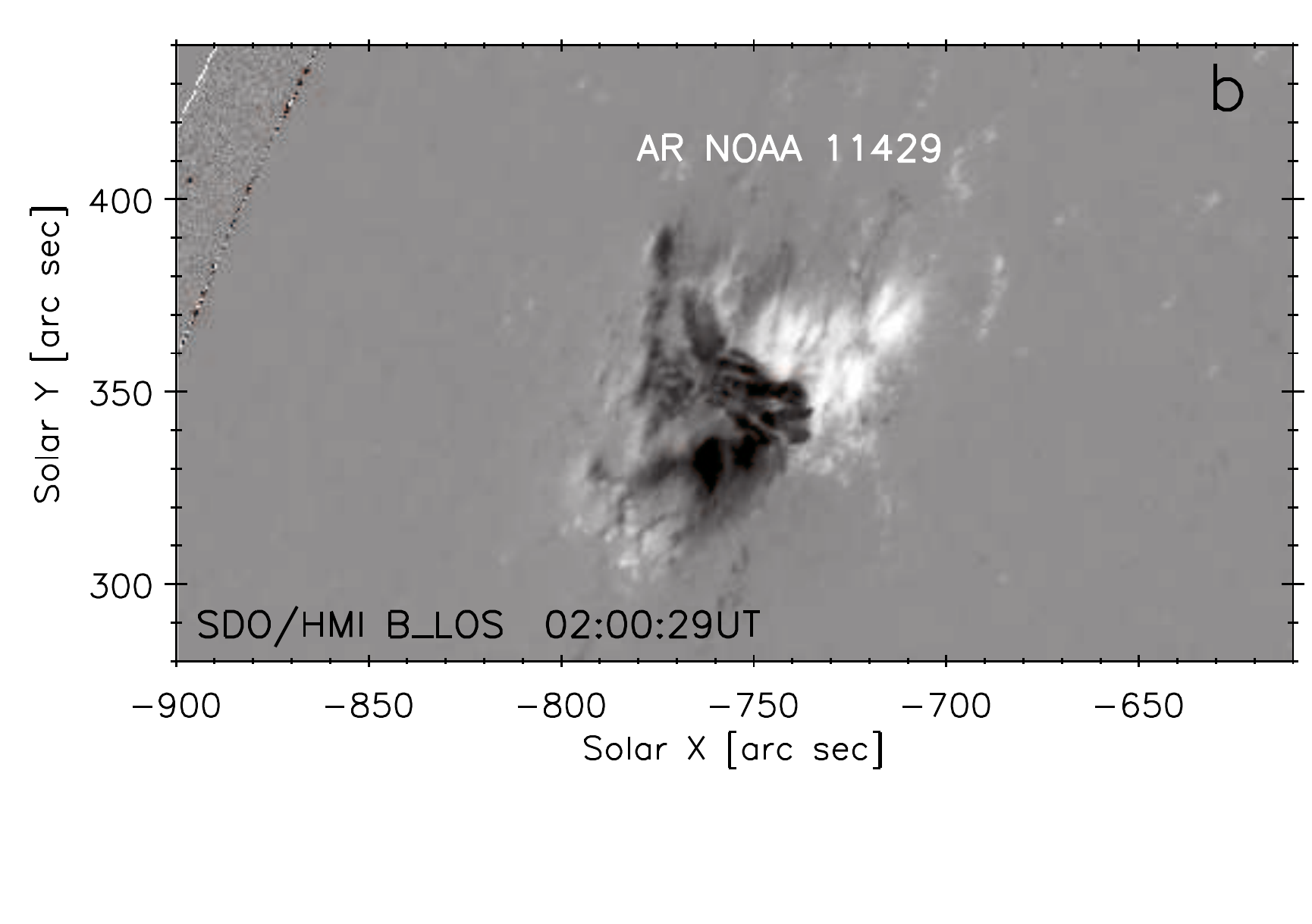}
	\includegraphics[width=8.8cm,clip,viewport= 0  0 495 345]{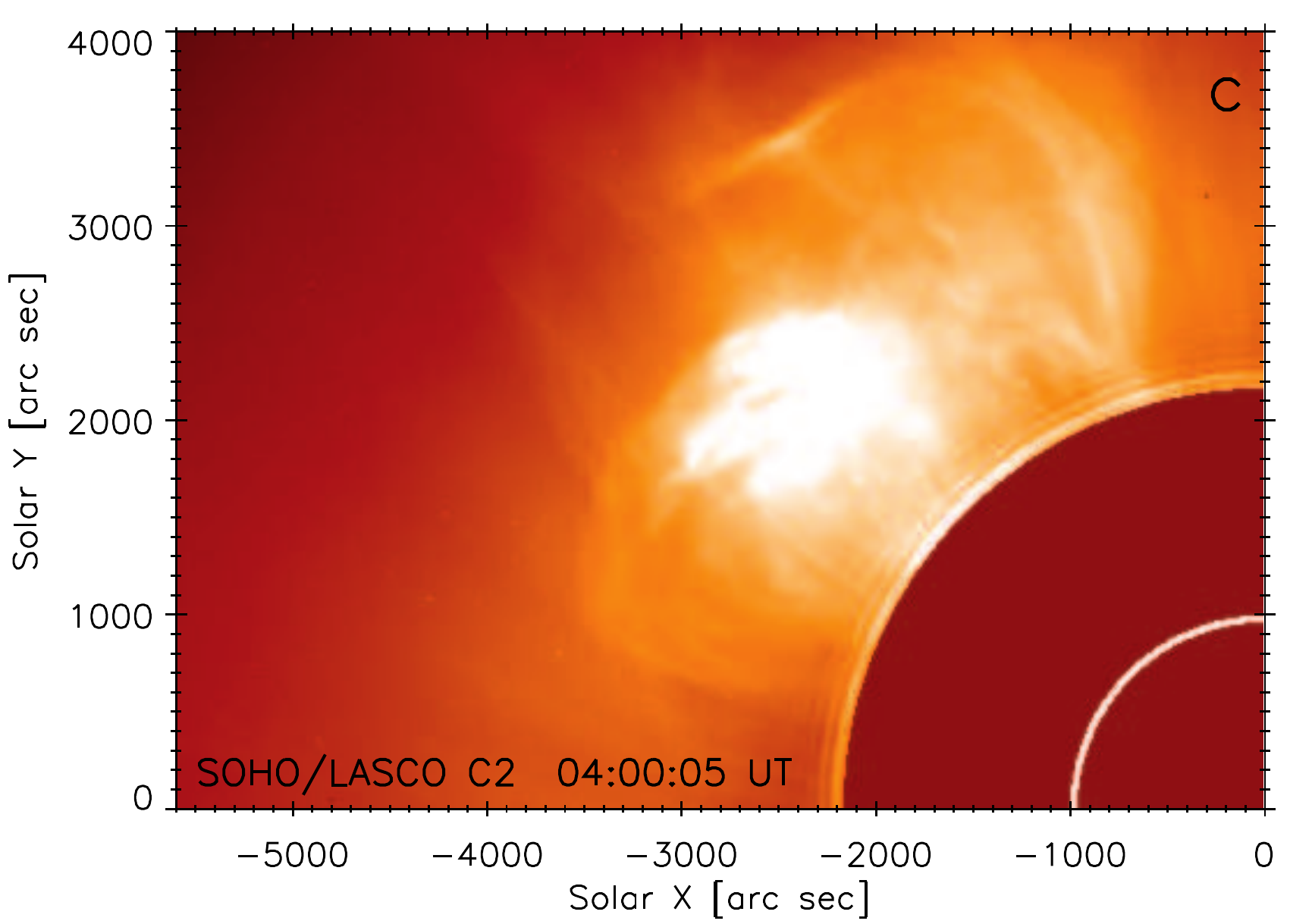}
	\includegraphics[width=8.8cm,clip,viewport= 0  0 495 345]{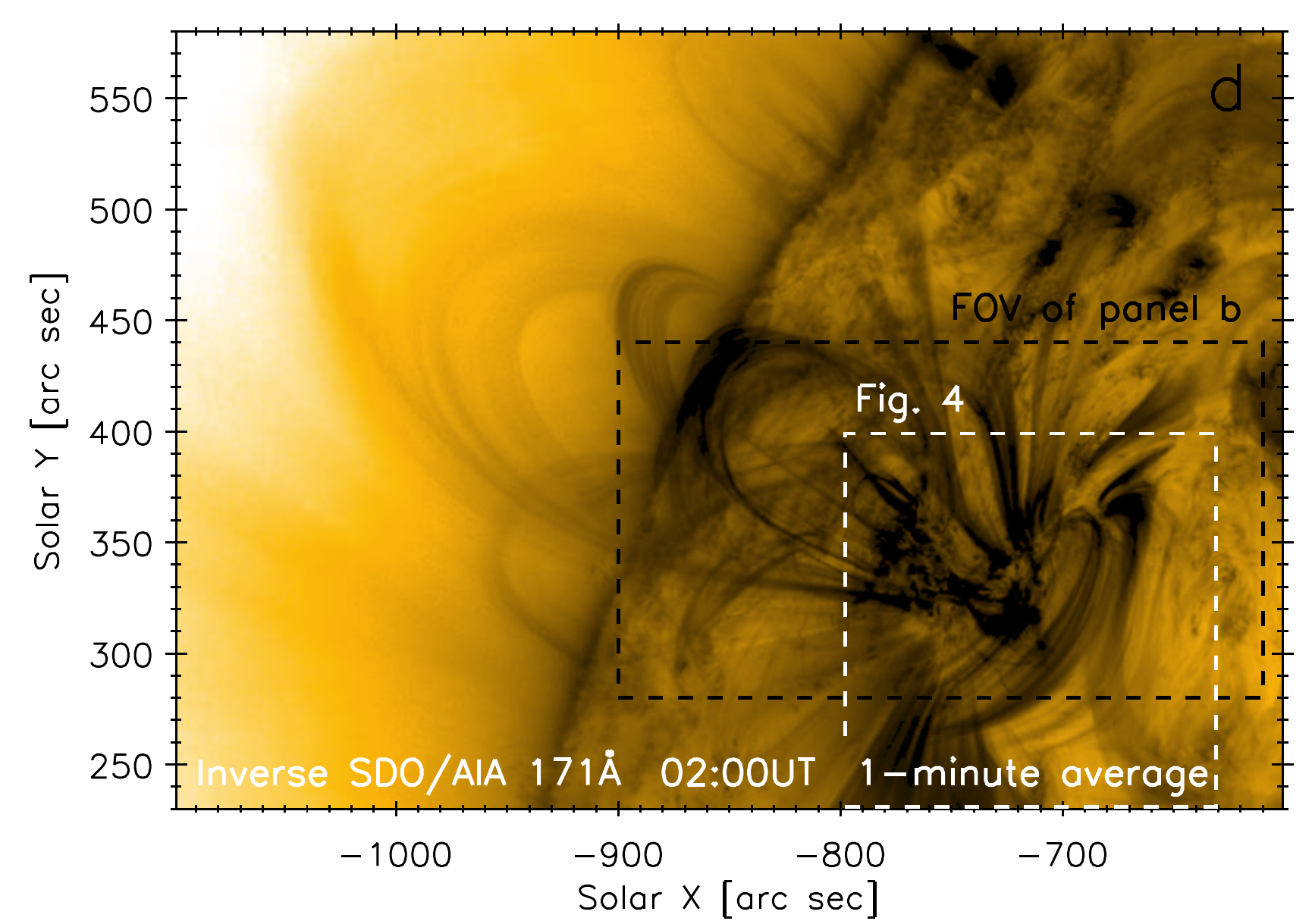}
\caption{Context observations of the active region and the CME produced during the X1.1 flare on 2012 March 05. (a): \textit{GOES} soft X-ray flux at 1--8\AA. Vertical red line shows the time of the flare maximum, while the vertical orange lines correspond to times shown in Figure \ref{Fig:X_Vortex_loops}. (b): \textit{SDO}/HMI longitudinal magnetogram of the AR 11429. (c): \textit{SOHO}/LASCO C2 observation of a CME during the flare. (d): The corona of the active region as observed by \textit{SDO}/AIA 171\,\AA. The colorscale is inverse to enhance weaker coronal structures. The black and white rectangles corresponds to the field of view shown in the panel (b) and Figure \ref{Fig:X_Vortex_loops}, respectively. \\
\label{Fig:X_Context}}
\end{figure*}
%
%
\begin{figure*}[ht]
	\centering
	\includegraphics[width=8.19cm,clip,viewport=10  0 495  85]{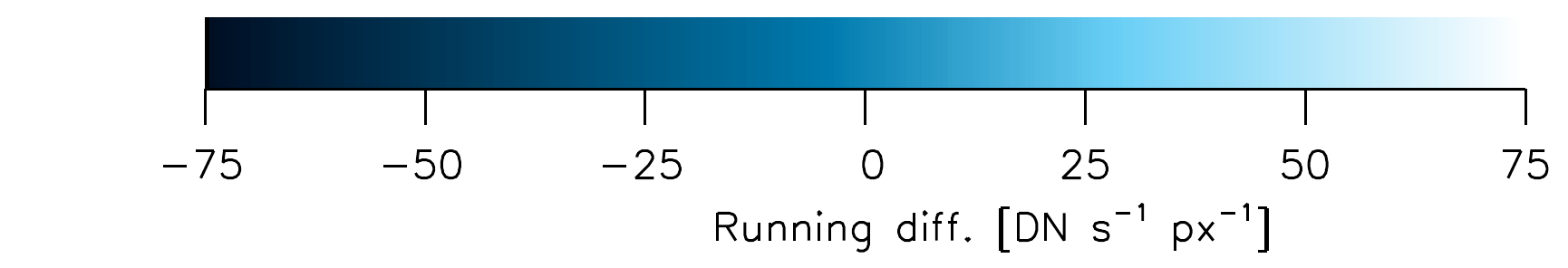}
	\includegraphics[width=7.11cm,clip,viewport=55  0 495  85]{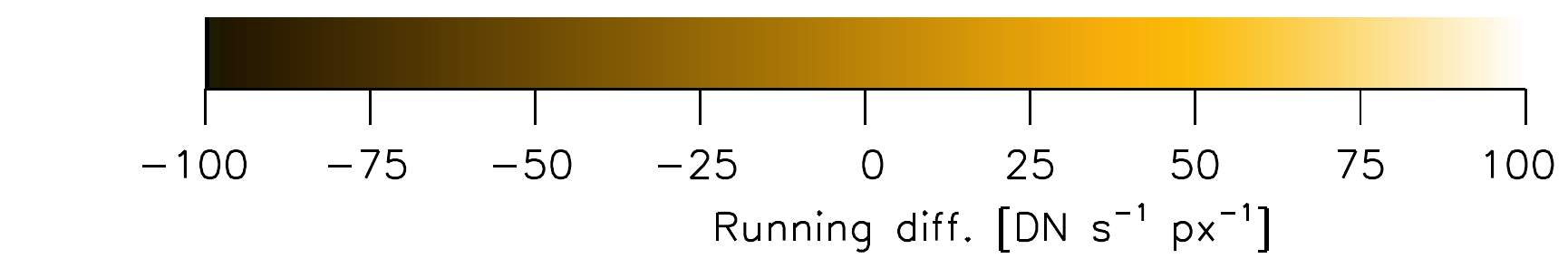}
	\includegraphics[width=8.19cm,clip,viewport= 0 40 495 345]{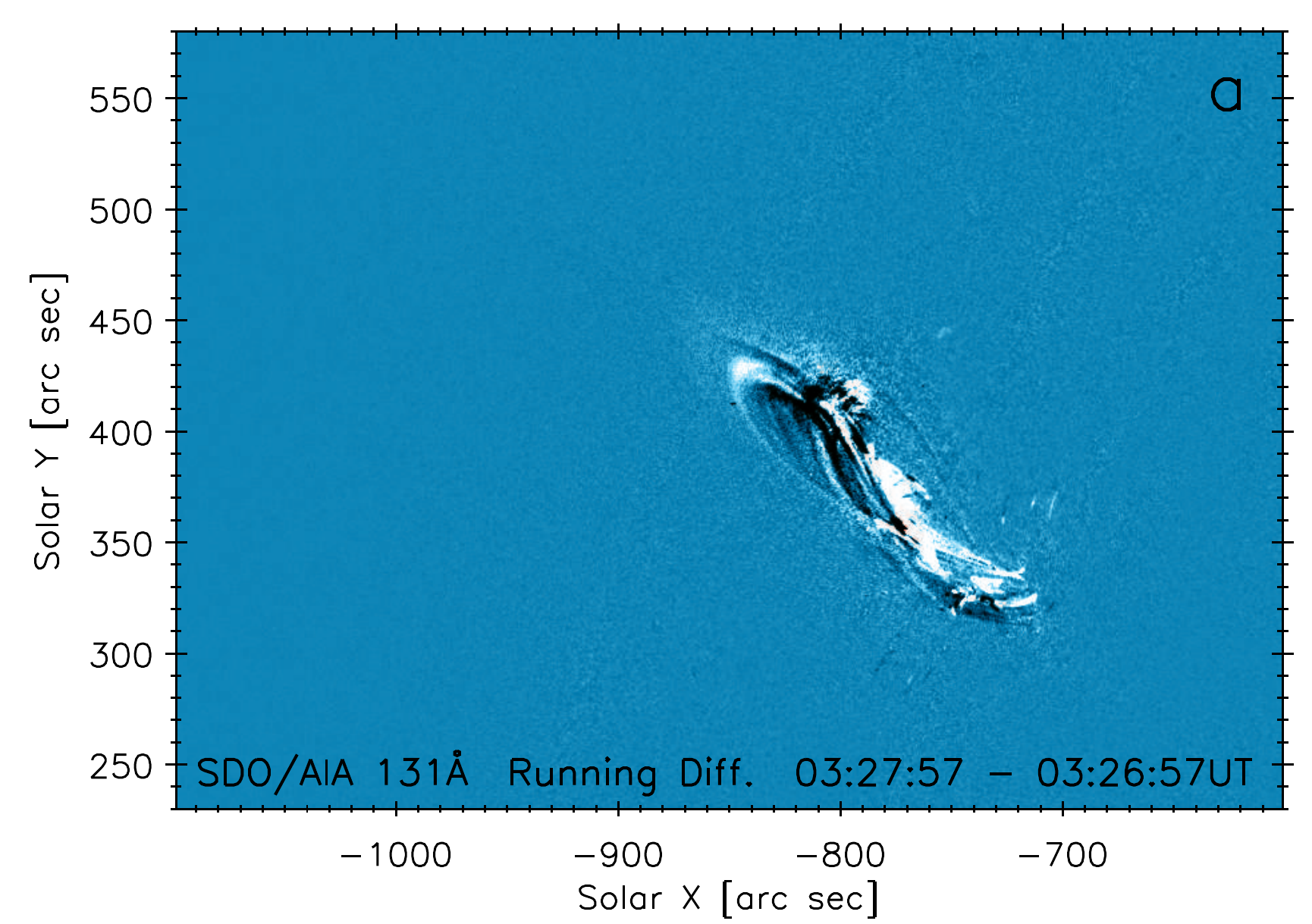}
	\includegraphics[width=7.11cm,clip,viewport=65 40 495 345]{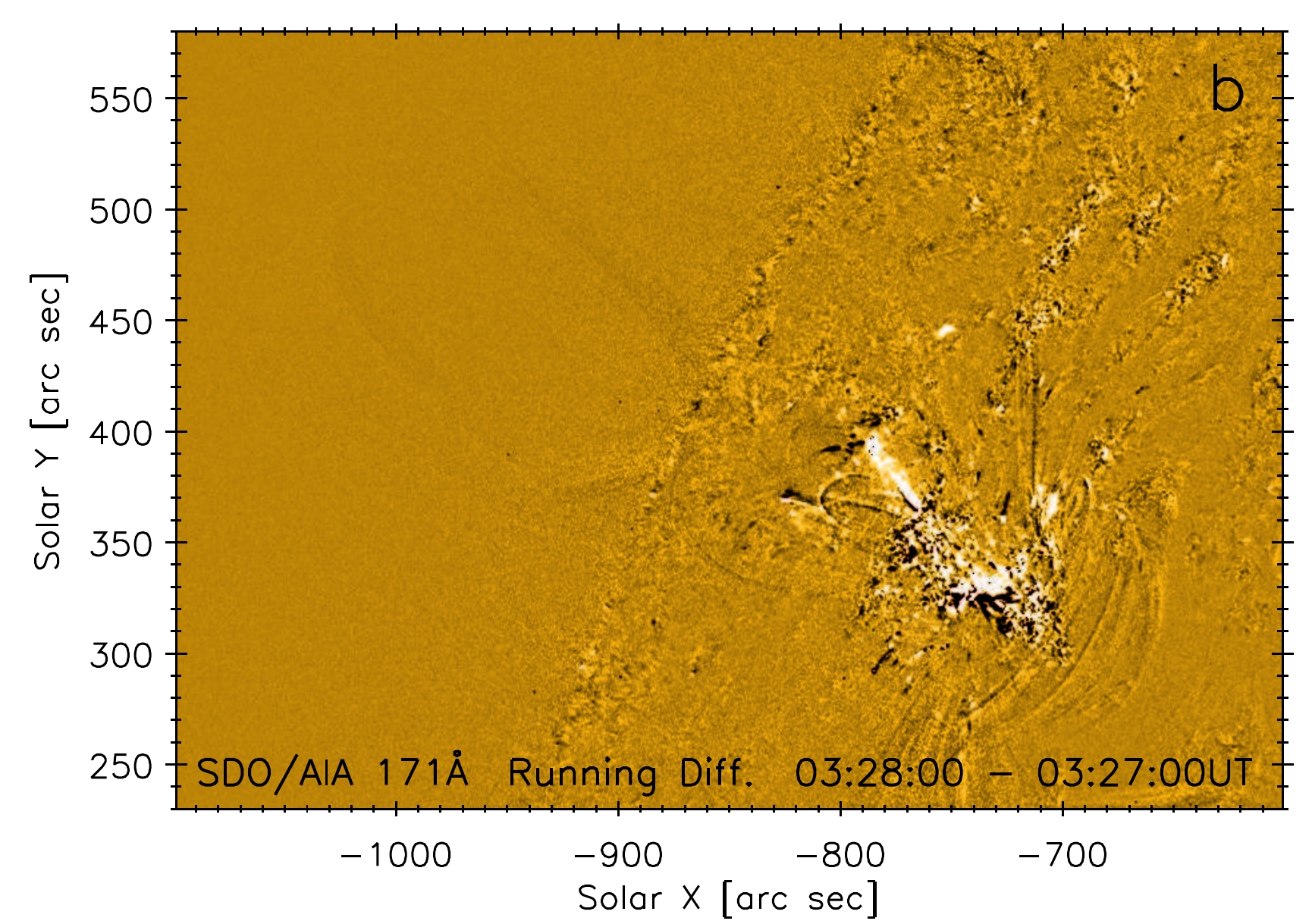}
	\includegraphics[width=8.19cm,clip,viewport= 0 40 495 345]{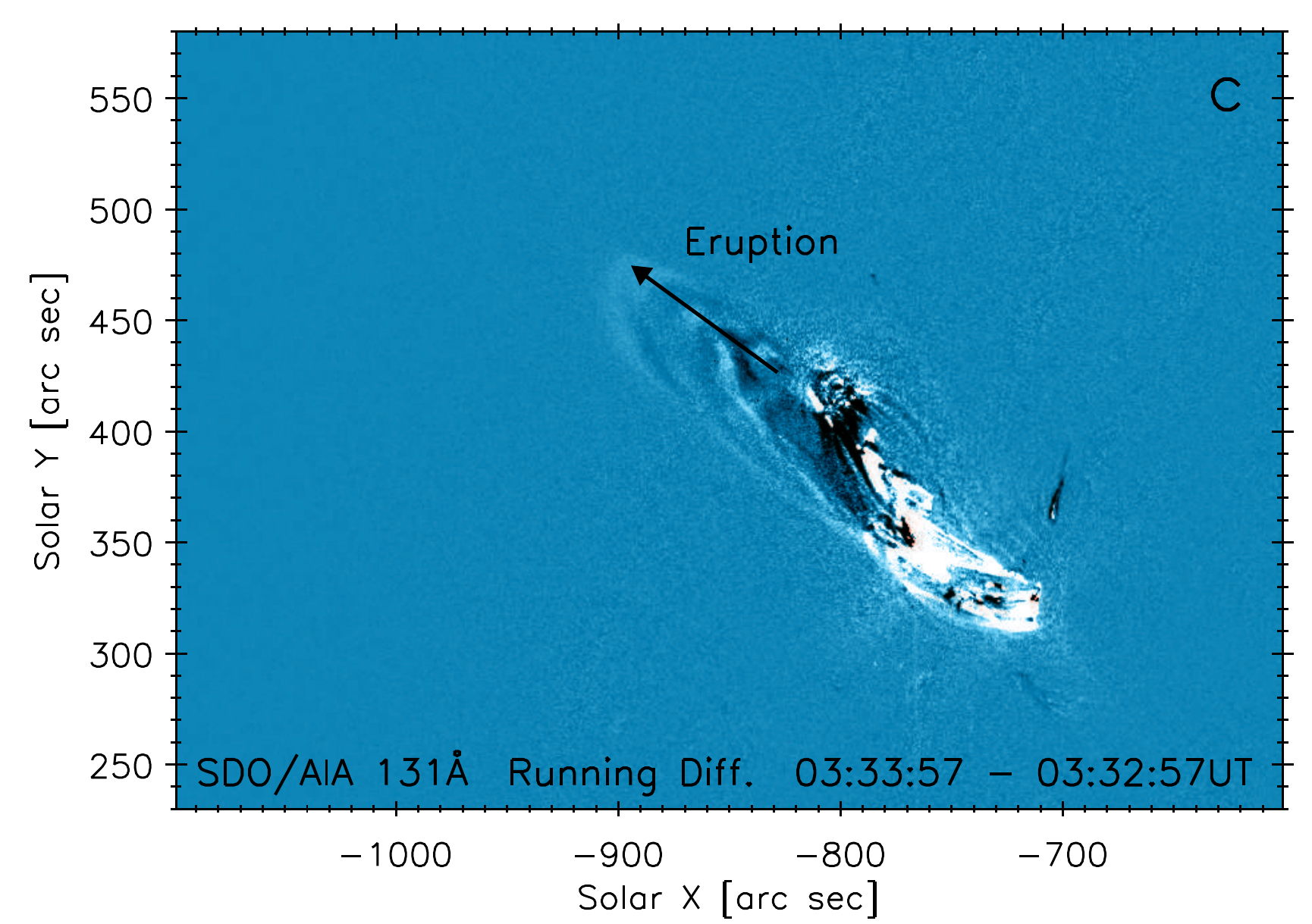}
	\includegraphics[width=7.11cm,clip,viewport=65 40 495 345]{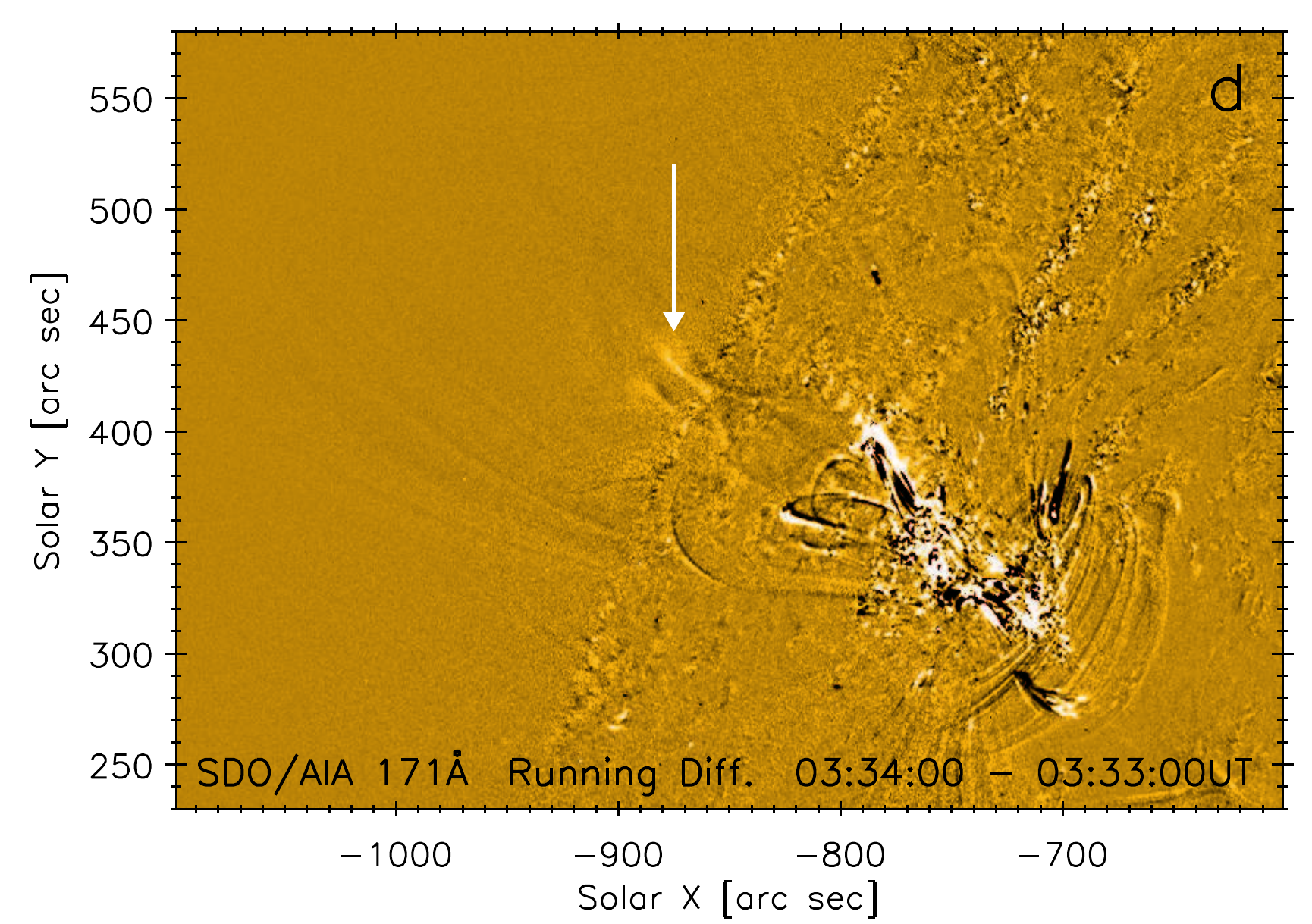}
	\includegraphics[width=8.19cm,clip,viewport= 0 40 495 345]{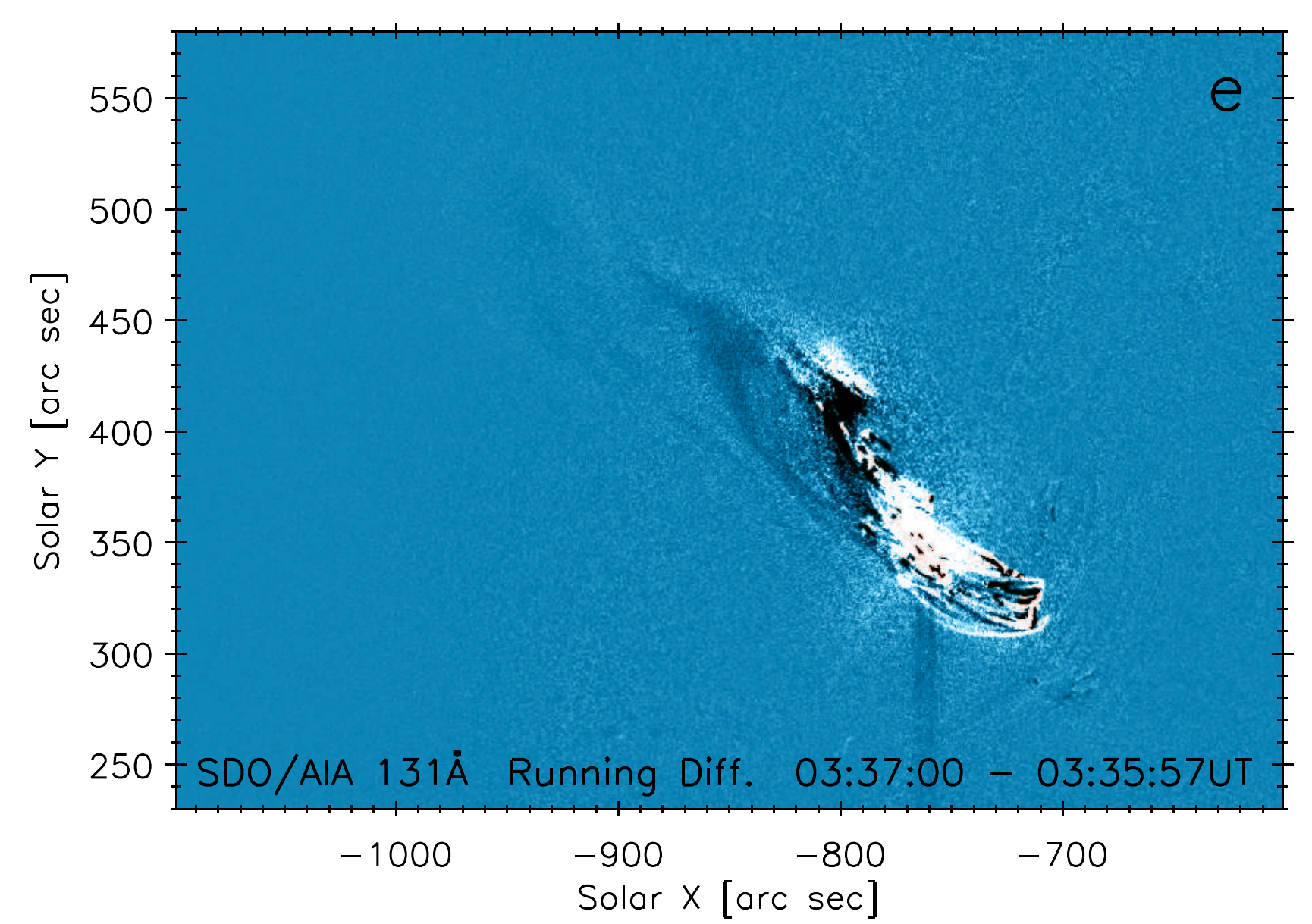}
	\includegraphics[width=7.11cm,clip,viewport=65 40 495 345]{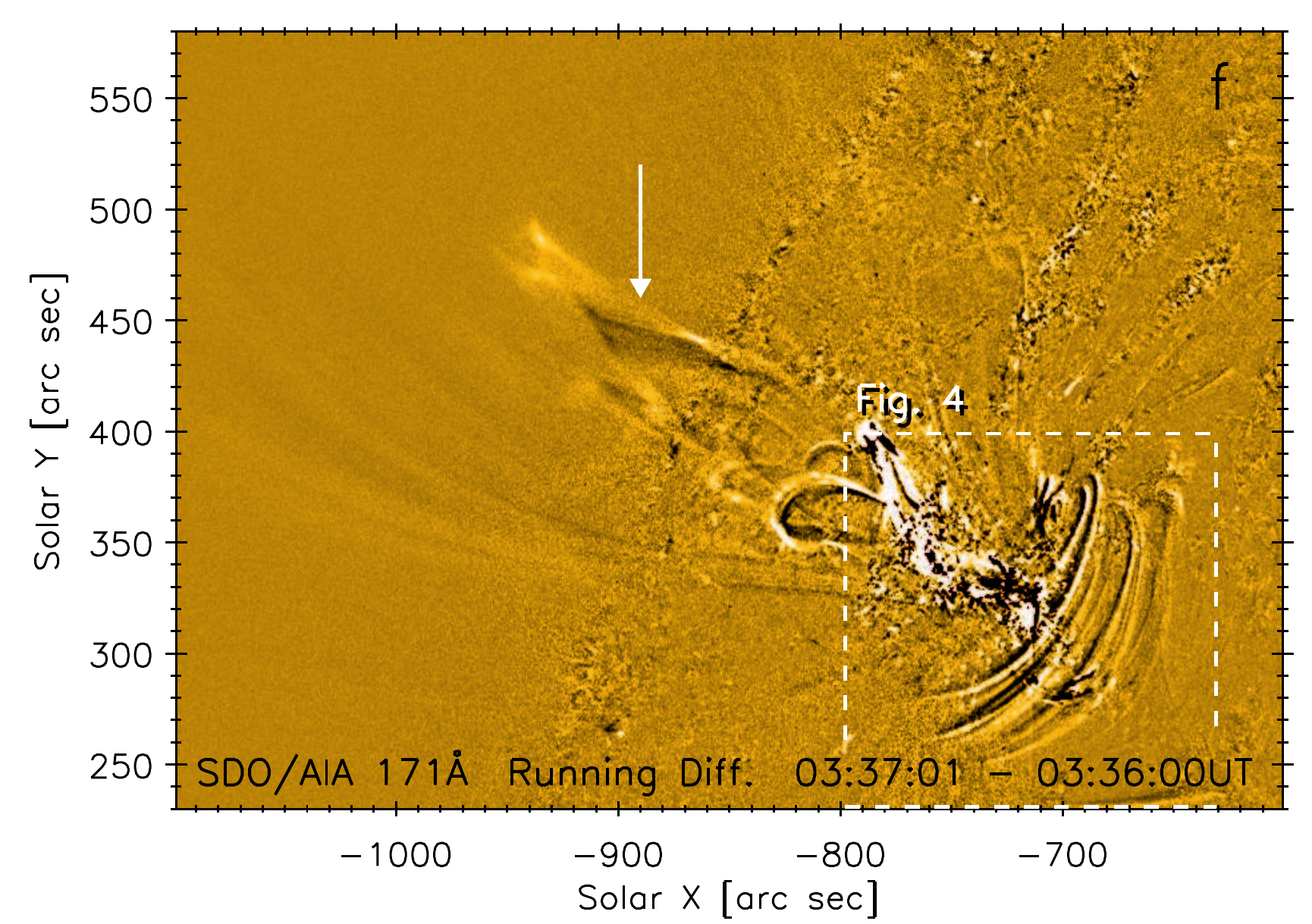}
	\includegraphics[width=8.19cm,clip,viewport= 0  0 495 345]{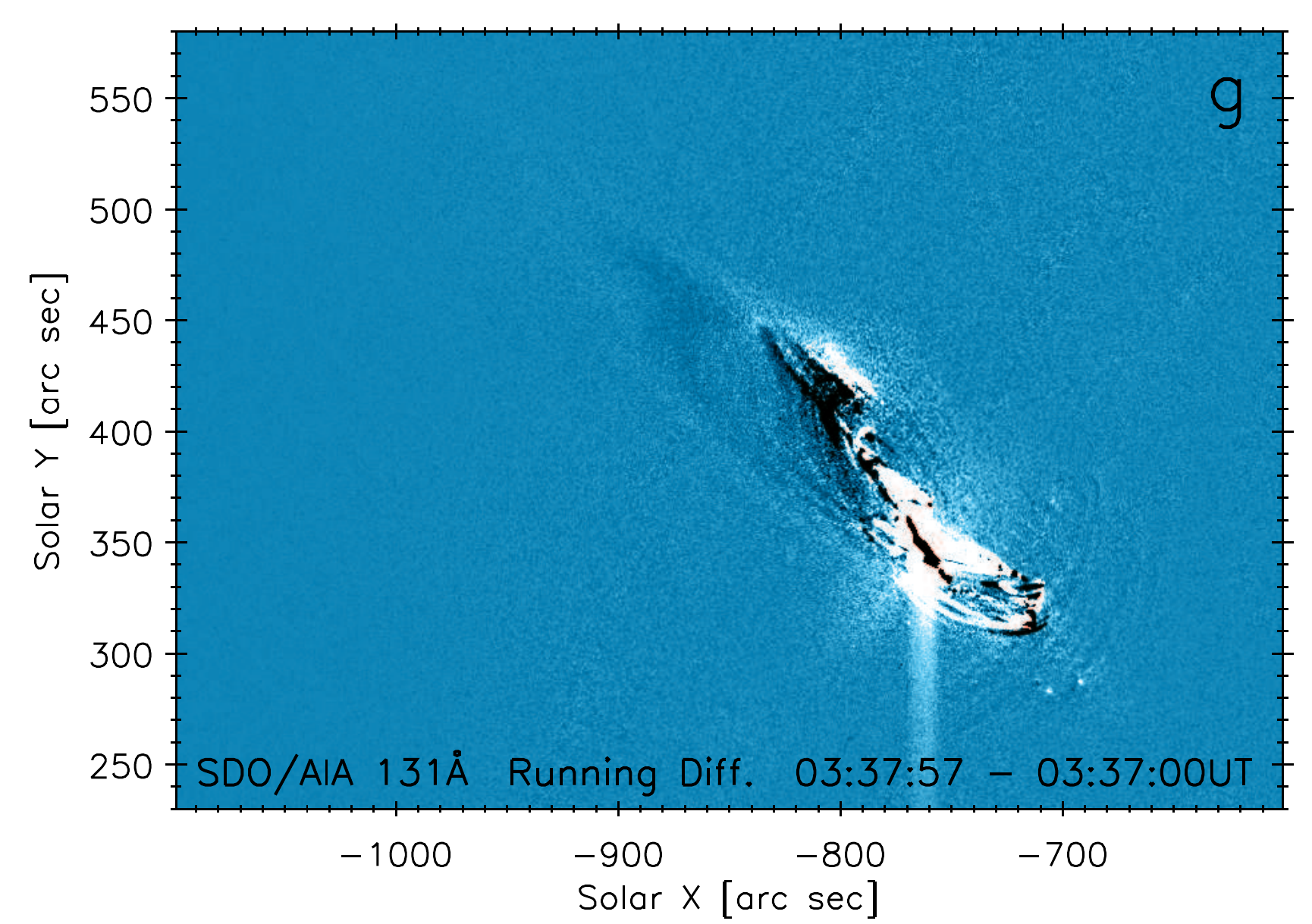}
	\includegraphics[width=7.11cm,clip,viewport=65  0 495 345]{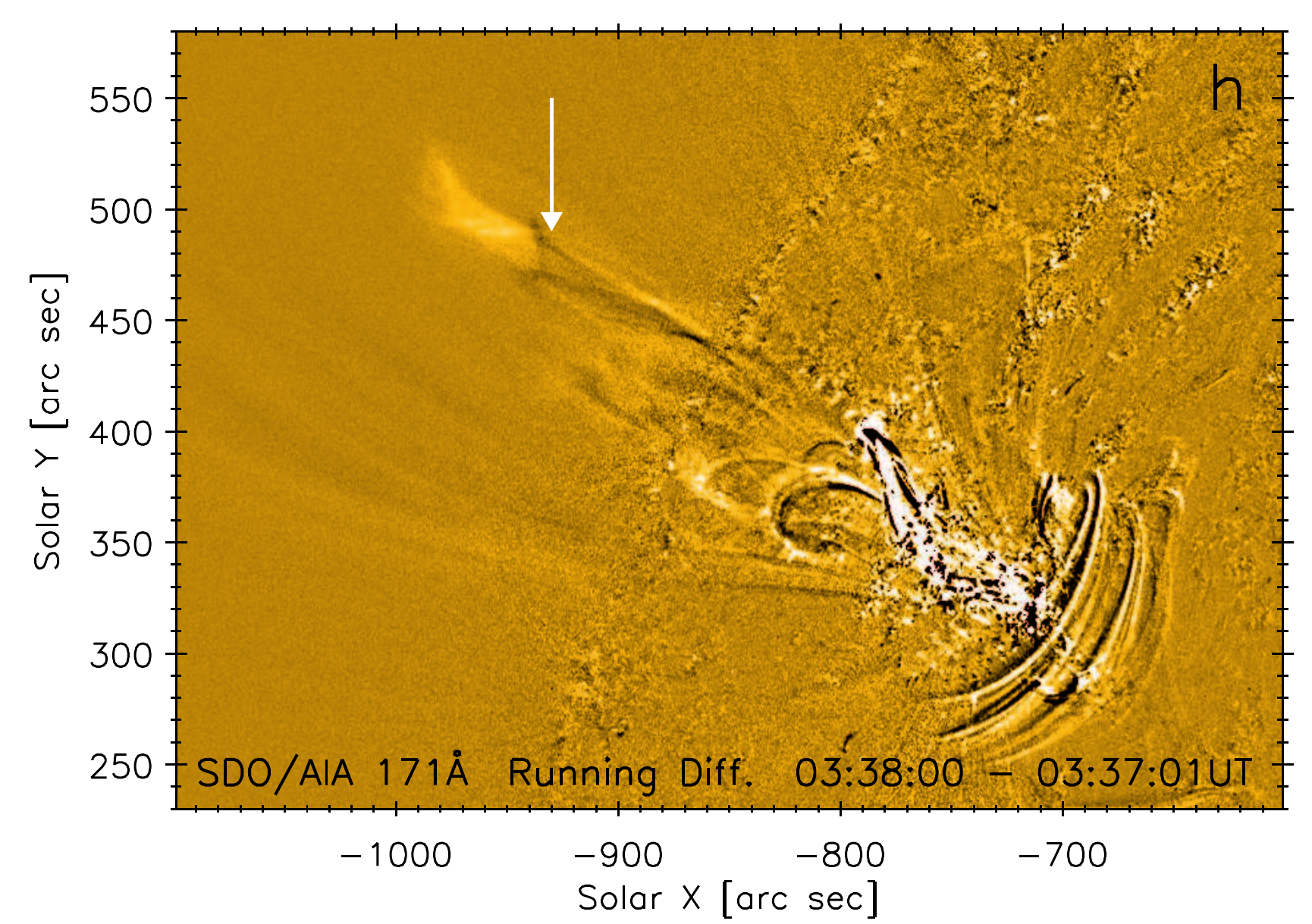}
\caption{AIA 131 and 171\,\AA~running difference images of the erupting structure. Black and white arrows denote erupting structures in 131 and 171\,\AA, respectively. The white box in panel (f) denotes the field of view of Figure \ref{Fig:X_Vortex_loops}. An animation of this figure is available in the online version of the article.
\label{Fig:X_Eruption}}
\end{figure*}
%

%
%
%
\section{The 2012 March 5 X-class flare}
\label{Sect:3}

To demonstrate the presence of vortices in eruptive solar flares, we first analyze the imaging observations of the 2012 March 5 X1.1-class solar flare and eruption (solar object locator SOL2012-03-05T02:30). This flare occurred in a complex, flare-producing active region NOAA 11429 (hereafter, AR 11429). On 2012 March 5 alone, AR 11429 produced 15 C-class, 3 M-class, and one X1.1-class flare.

The GOES 1--8\AA~X-ray lightcurve of the X1.1 flare is shown in Figure \ref{Fig:X_Context}a. It is readily seen that the flare is a long-duration event. It starts at about 02:30\,UT and its GOES X-ray flux peaks at about 04:09\,UT (red dashed line in Figure \ref{Fig:X_Context}a). The pre-flare configuration of the underlying line-of-sight component of the magnetic field and of the overlying solar corona at 02:00\,UT are also shown in Figure \ref{Fig:X_Context}. These observations are made by the Helioseismic Magnetic Imager \citep[HMI,][]{Scherrer12,Schou12} and Atmospheric Imaging Assembly \citep[AIA,][]{Lemen12,Boerner12} onboard \textit{Solar Dynamics Observatory} \citep[\textit{SDO},][]{Pesnell12}. An arcade of closed coronal loops is observed at the location of about Solar $[X,Y]$\,=\,$[-700\arcsec,300\arcsec]$, i.e., extending from the AR towards the direction of the disk centre. Additionally, a series of large-scale coronal loops, with some extending well above the solar limb, is also present at larger altitudes above the AR core. 

The flare is an eruptive one with a CME detected by \textit{SOHO}/LASCO-C2 \citep{Brueckner95}, shown in Figure~\ref{Fig:X_Context}c.

\subsection{Eruption during the X-class flare}
\label{Sect:3.1}

Since the vortex flows are predicted to be generated by a flare-related eruption, we first discuss observations of an erupting structure occurring during the X1.1-class flare. The eruption is well observed by \textit{SDO}/AIA instrument, which observes the full Sun with a cadence of 12\,s and spatial resolution of 1.5$\arcsec$ (0.6$\arcsec$ pixel size) in 10 EUV and UV spectral bands selected by the corresponding filters. The 6 EUV filters are typically centered on strong, well-known spectral lines belonging to various ionization stages of Fe. This allows for sampling of the solar corona at temperatures in the range 0.4--20\,MK.

Figure \ref{Fig:X_Eruption} and the accompanying Movie~2 show the evolution of the eruption based on the AIA observations in the 131\,\AA~and 171\,\AA~bands. To enhance the presence of the erupting structures in these two passbands, we show the running-difference images constructed as a difference between an image obtained at a given time and an image obtained 1 minute earlier in the same passband. We chose the 131 and 171\,\AA~bands because they represent the morphology of both the hot flare plasma together with the neighboring warm solar corona. We note that under flare conditions, the 131\,\AA~band is dominated by the \ion{Fe}{21} emission at 128.75\,\AA, occurring at temperatures of about 10\,MK \citep{ODwyer10,Petkaki12}. The 131\,\AA~band however also contains the \ion{Fe}{8} 130.94\,\AA~and 131.24\,\AA~emission lines originating at about 0.5\,MK. This emission is morphologically similar to the \ion{Fe}{9} 171.07\,\AA~line which dominates the 171\,\AA~band. This means that structures common to both 131\,\AA~and 171\,\AA~bands likely originate at coronal temperatures, while structures seen only in 131\,\AA~likely represent the \ion{Fe}{21} emission. Such inferences are however only approximate, as a detailed analysis requires differential emission measure modeling using observations in all 6 AIA EUV channels \citep[e.g.][]{Guennou12a,Guennou12b,Hannah13,Sun14,Dudik14a}.

First signatures of the eruption are detected at about 03:26 in 131\,\AA~band. A series of hot loops, accelerating in the NE direction outward from the AR is seen, lasting till about 03:39\,UT. Their location is shown by the black arrow in Figure \ref{Fig:X_Eruption}c. There is no corresponding signature in 171\,\AA, indicating that this erupting structure is hot and emits in \ion{Fe}{21}. In 171\,\AA, the eruption manifests itself as a writhed erupting loop (white arrows in Figure \ref{Fig:X_Eruption}d,f,h) following the 131\,\AA~eruption. The writhed loop has a complex shape, which consists of alternating bright and dark structures in the AIA 171\,\AA~running difference images, reminiscent of a weakly twisted envelope of an erupting flux rope, which is also a feature of the simulation (Section \ref{Sect:2}).

We note that hot magnetic flux ropes emitting in 131\,\AA~are commonly observed during eruptive flares \citep[e.g.,][]{Zhang12,Cheng14a,Dudik14a,Dudik16,Li15,Li16,Zhu16}. In our case, the erupting structure is probably also a hot magnetic flux rope, as indicated by the occurrence of the eruption first in 131\,\AA~band, followed by a writhed series of loops in 171\,\AA.

%
\begin{table}[ht]
\begin{center}
\scriptsize
\caption{Summary of projected velocities of individual coronal loops as measured in time-distance plots in Figures \ref{Fig:X_Stackplots_loops}. Positive velocities are set away from the AR. 
The times shown are in UT and rounded to the next minute.
\label{Table:1}}
\tabletypesize{\normalsize}
\begin{tabular}{llll}
\tableline\tableline
\multicolumn{4}{c}{Vortex loops (171\,\AA) -- Figure \ref{Fig:X_Stackplots_loops}c}		\\
Structure	& Velocity $v$ [\kps]	& $\pm$ $\sigma(v)$ [\kps]	& Time [UT]	\\
\tableline
U1		&  $+1.7$				& $\pm$ 1.0			& 03:09 -- 03:34\\
U2		& $+16.5$				& $\pm$ 3.4			& 03:34 -- 03:41\\
U3		& $+29.7$				& $\pm$ 5.8			& 03:34 -- 03:39\\
U4		&  $+1.6$				& $\pm$ 1.1			& 04:09 -- 04:32\\
U5		&  $-5.9$				& $\pm$ 1.3			& 04:32 -- 04:53\\
U6		&  $+3.4$				& $\pm$ 1.3			& 04:20 -- 04:39\\
U7		& $-20.6$				& $\pm$ 5.8			& 04:39 -- 04:43\\
U8		&  $-3.8$				& $\pm$ 2.1			& 04:41 -- 04:53\\
U9		&  $-3.5$				& $\pm$ 0.7			& 04:19 -- 04:54\\
\tableline\tableline
\multicolumn{4}{c}{Vortex loops (211\,\AA) -- Figure \ref{Fig:X_Stackplots_loops}d}		\\
Structure	& Velocity $v$ [\kps]	& $\pm$ $\sigma(v)$ [\kps]	& Time [UT]		\\
\tableline
V1		& $+32.8$				& $\pm$ 5.7			& 03:35 -- 03:39\\
V2		& $+27.1$				& $\pm$ 4.1			& 03:35 -- 03:41\\
V3		&  $-1.5$				& $\pm$ 0.6			& 04:04 -- 04:48\\
V4		&  $+1.5$				& $\pm$ 1.0			& 04:18 -- 04:32\\
V5		&  $-1.3$				& $\pm$ 1.9    			& 04:32 -- 04:46\\
V6		&  $+5.4$				& $\pm$ 1.3			& 04:46 -- 05:05\\
V7		&  $+9.5$				& $\pm$ 2.7			& 04:32 -- 04:41\\
V8		& $-18.6$				& $\pm$ 5.3			& 04:48 -- 04:53\\
V9		& $-39.1$				& $\pm$ 13.8			& 04:53 -- 04:55\\
\tableline\tableline
\end{tabular}
\end{center}
\end{table}
%
\begin{figure*}[ht]
	\centering
	\includegraphics[width=4.42cm,clip,viewport= 0 40 243 225]{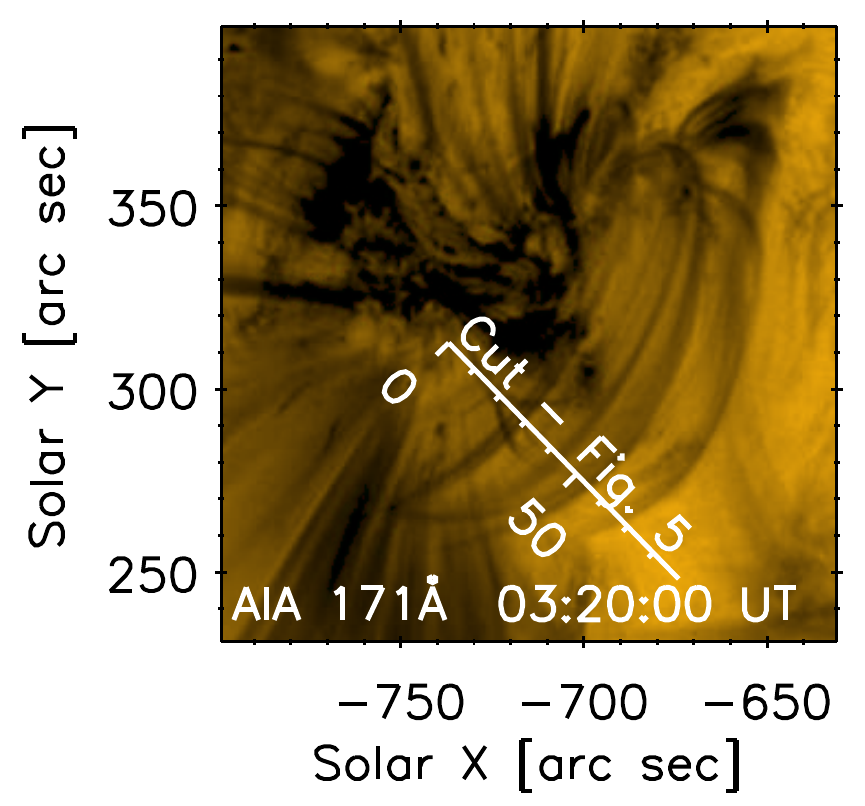}
	\includegraphics[width=3.29cm,clip,viewport=62 40 243 225]{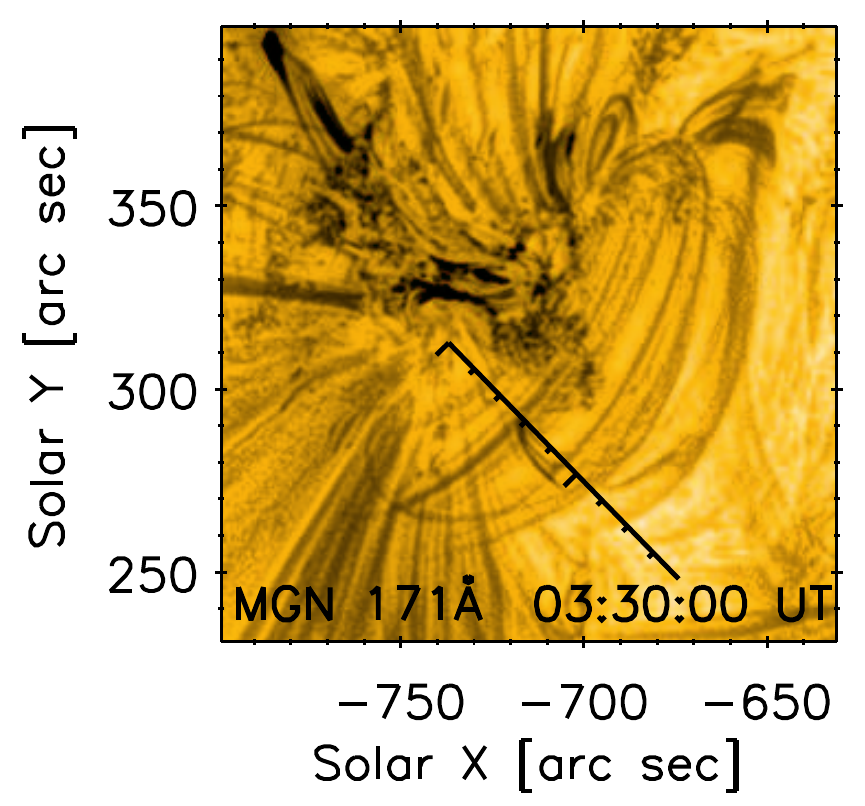}
	\includegraphics[width=3.29cm,clip,viewport=62 40 243 225]{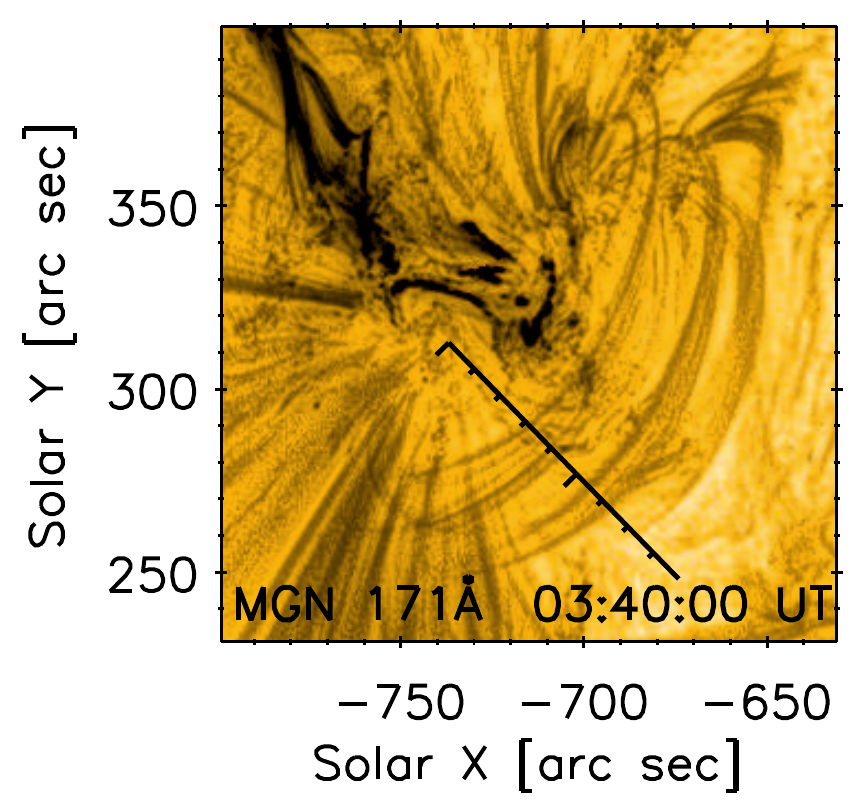}
	\includegraphics[width=3.29cm,clip,viewport=62 40 243 225]{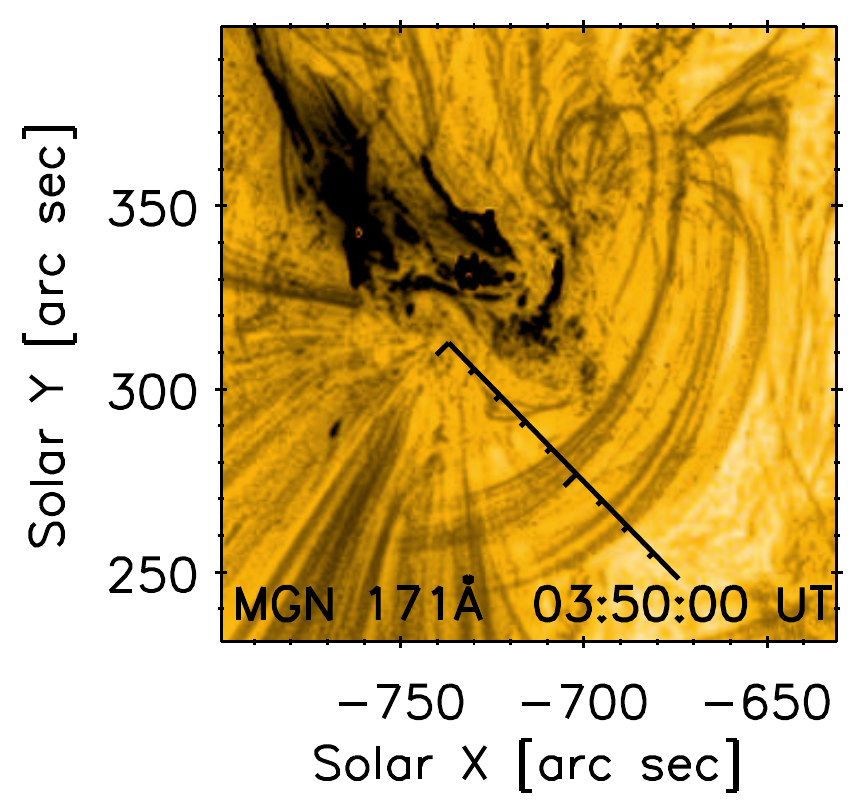}
	\includegraphics[width=3.29cm,clip,viewport=62 40 243 225]{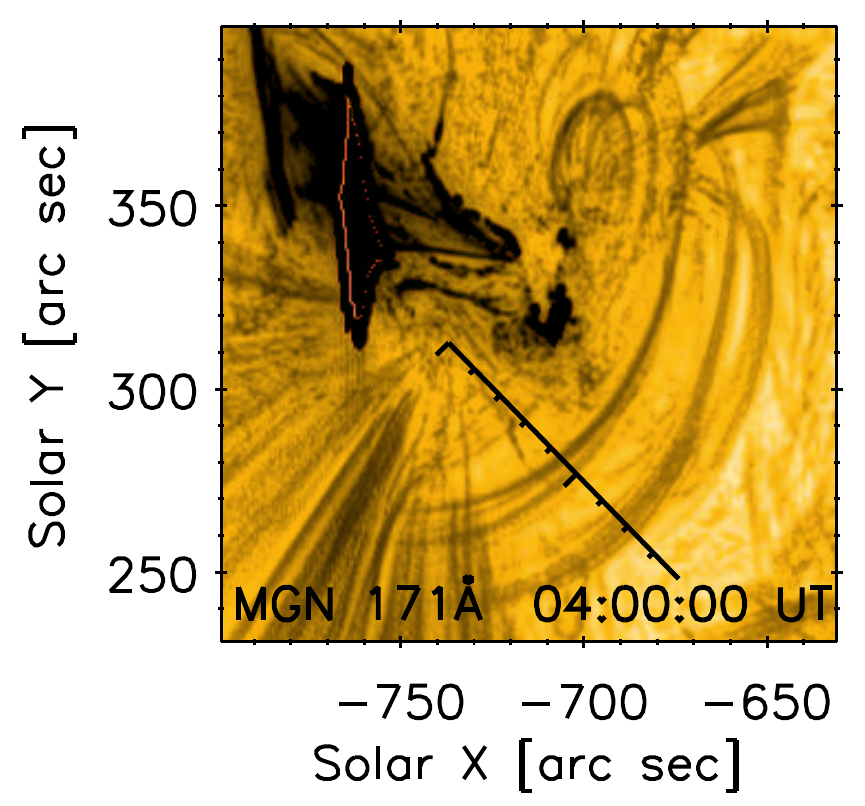}
	\includegraphics[width=4.42cm,clip,viewport= 0  0 243 225]{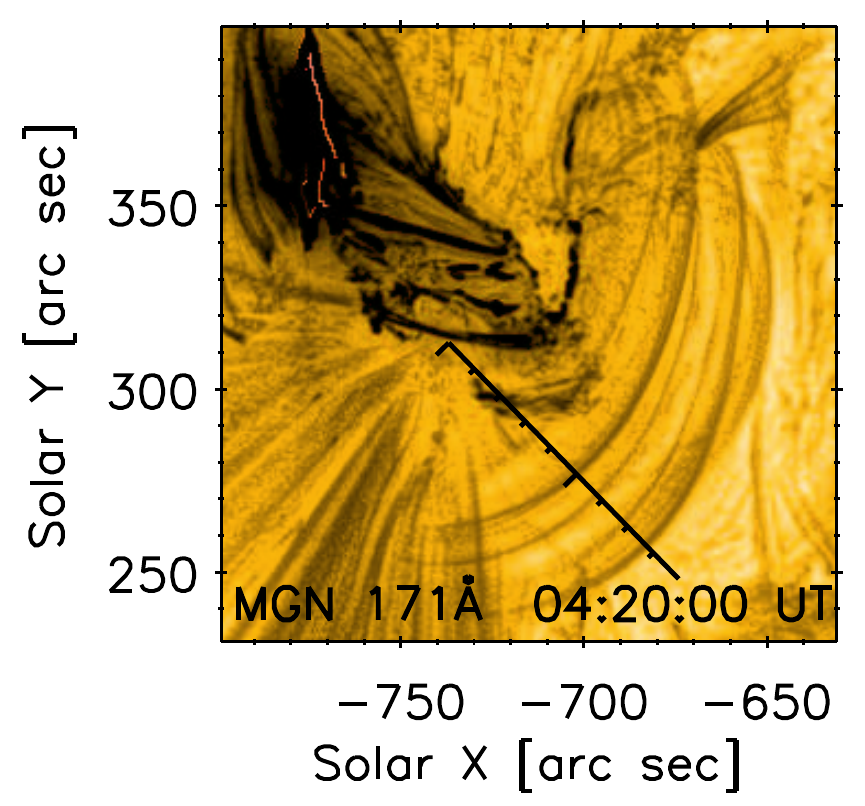}
	\includegraphics[width=3.29cm,clip,viewport=62  0 243 225]{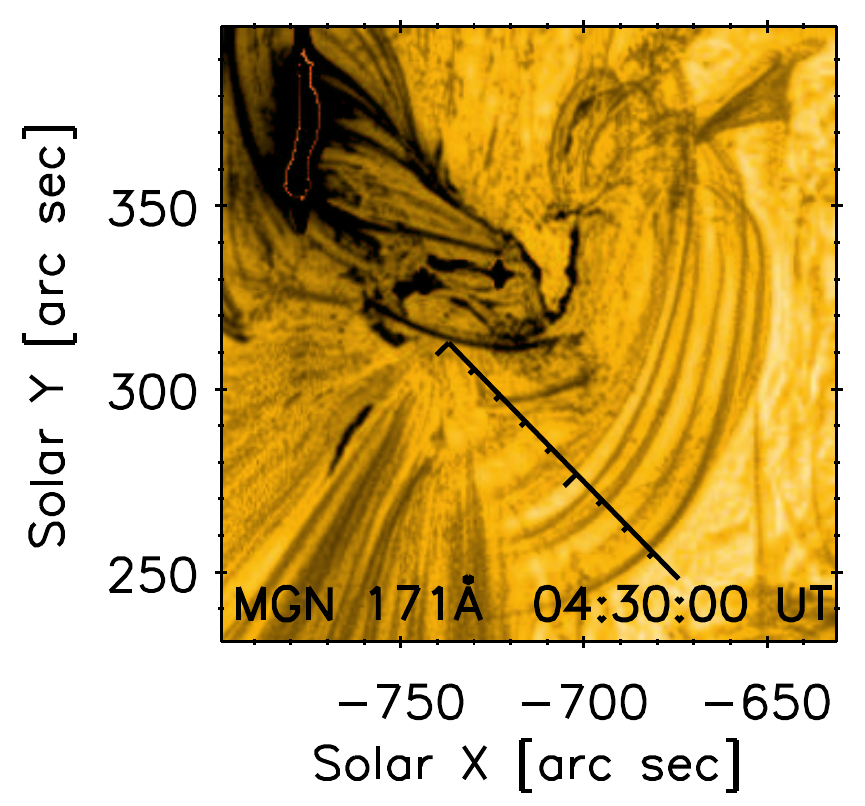}
	\includegraphics[width=3.29cm,clip,viewport=62  0 243 225]{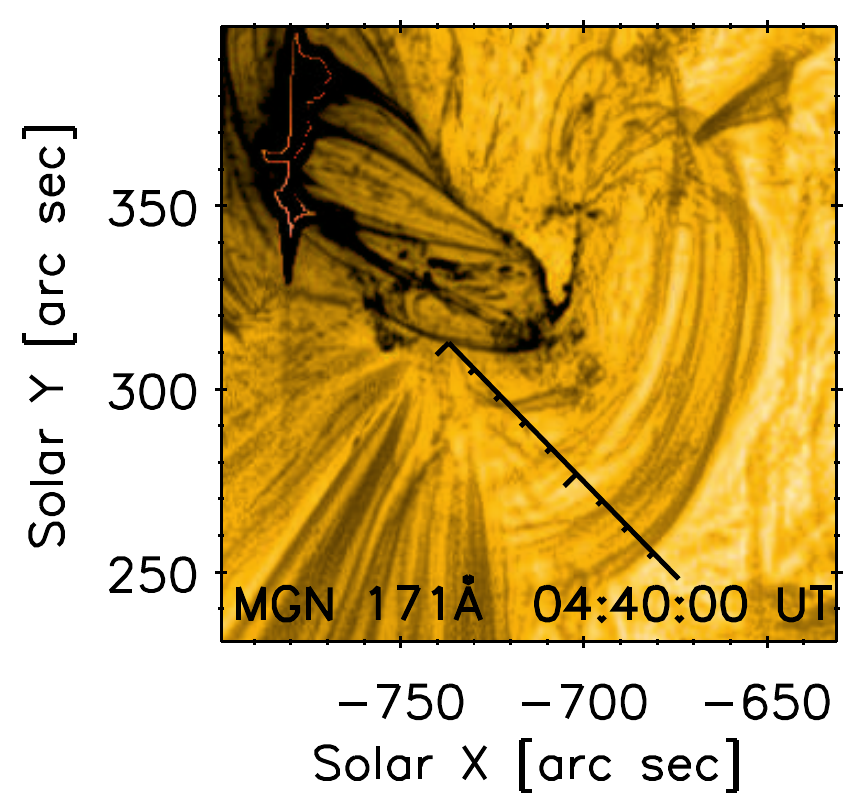}
	\includegraphics[width=3.29cm,clip,viewport=62  0 243 225]{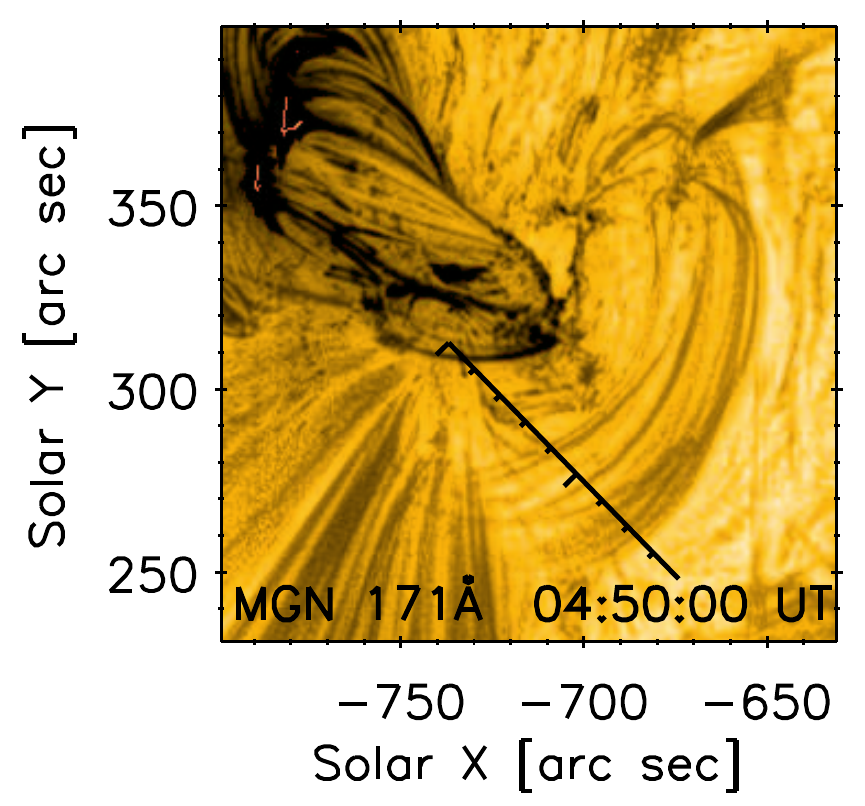}
	\includegraphics[width=3.29cm,clip,viewport=62  0 243 225]{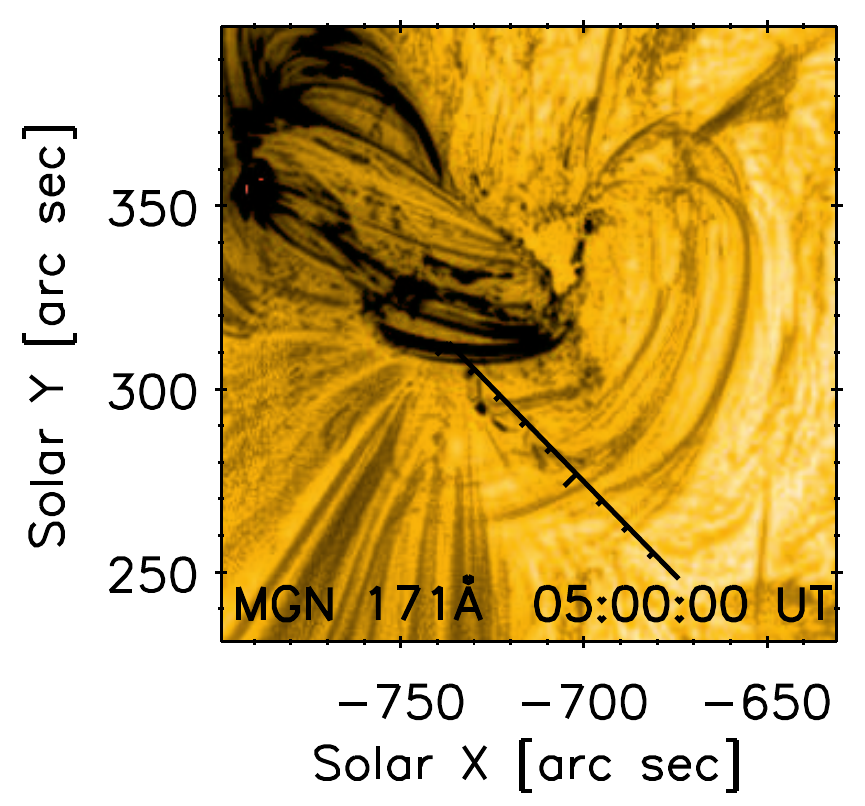}
	\includegraphics[width=4.42cm,clip,viewport= 0 40 243 225]{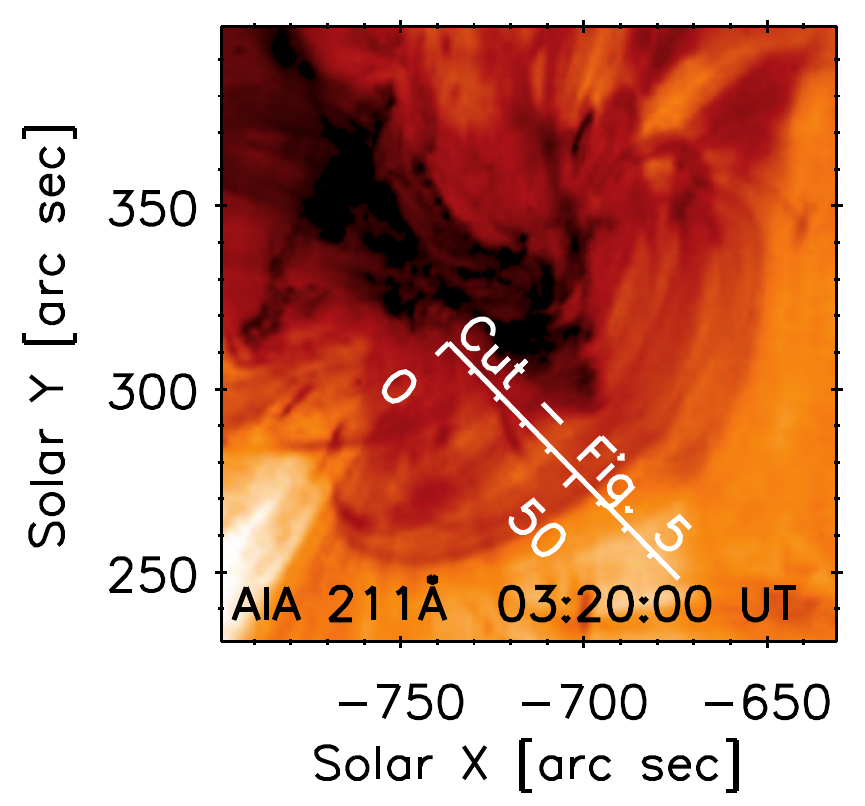}
	\includegraphics[width=3.29cm,clip,viewport=62 40 243 225]{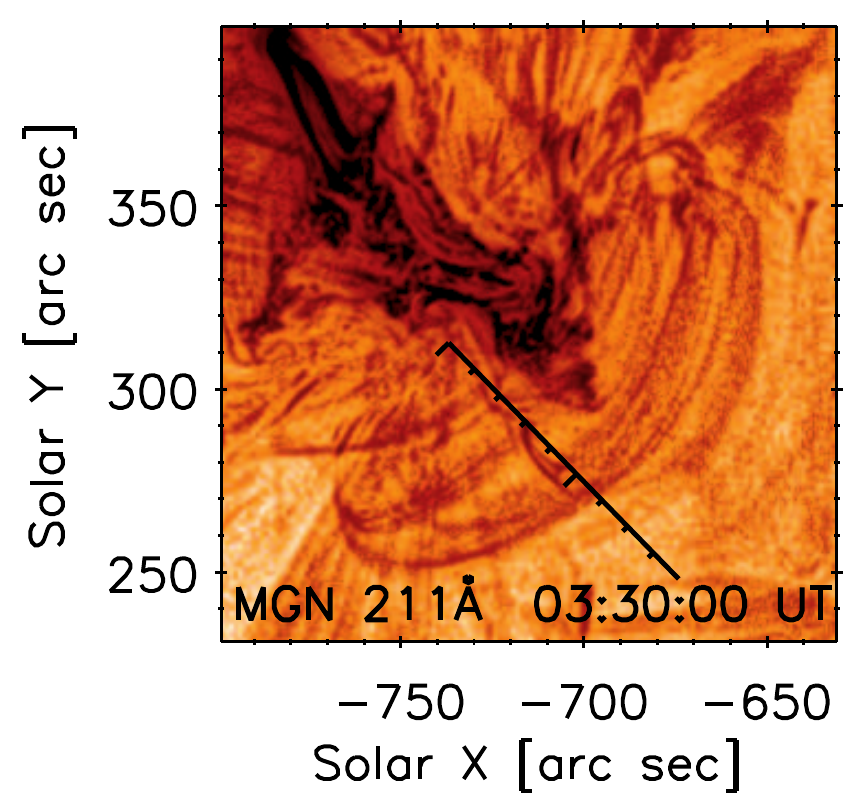}
	\includegraphics[width=3.29cm,clip,viewport=62 40 243 225]{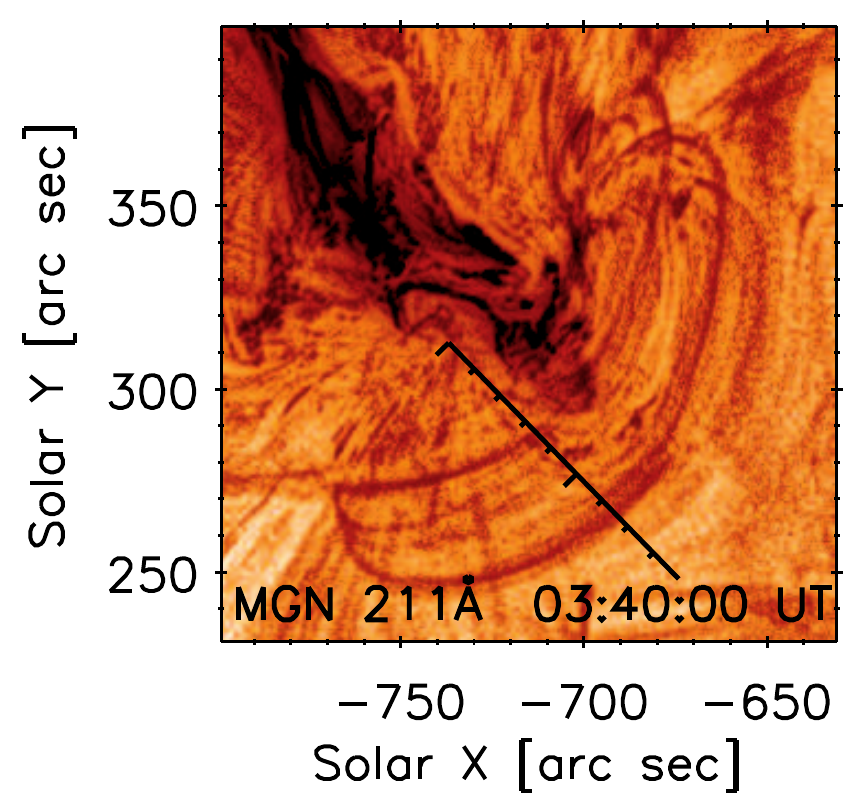}
	\includegraphics[width=3.29cm,clip,viewport=62 40 243 225]{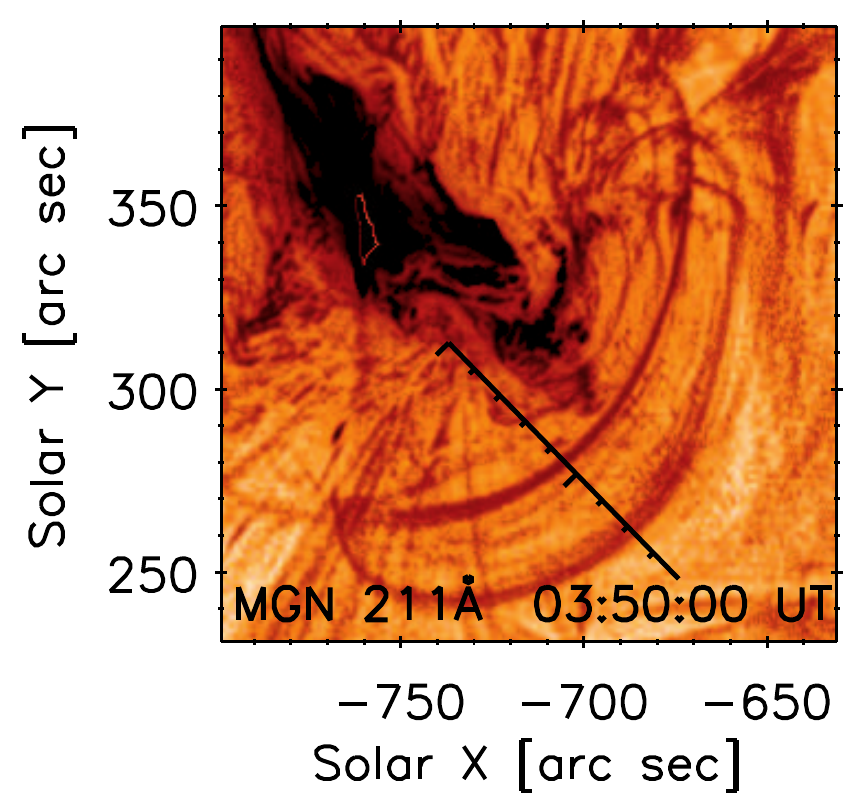}
	\includegraphics[width=3.29cm,clip,viewport=62 40 243 225]{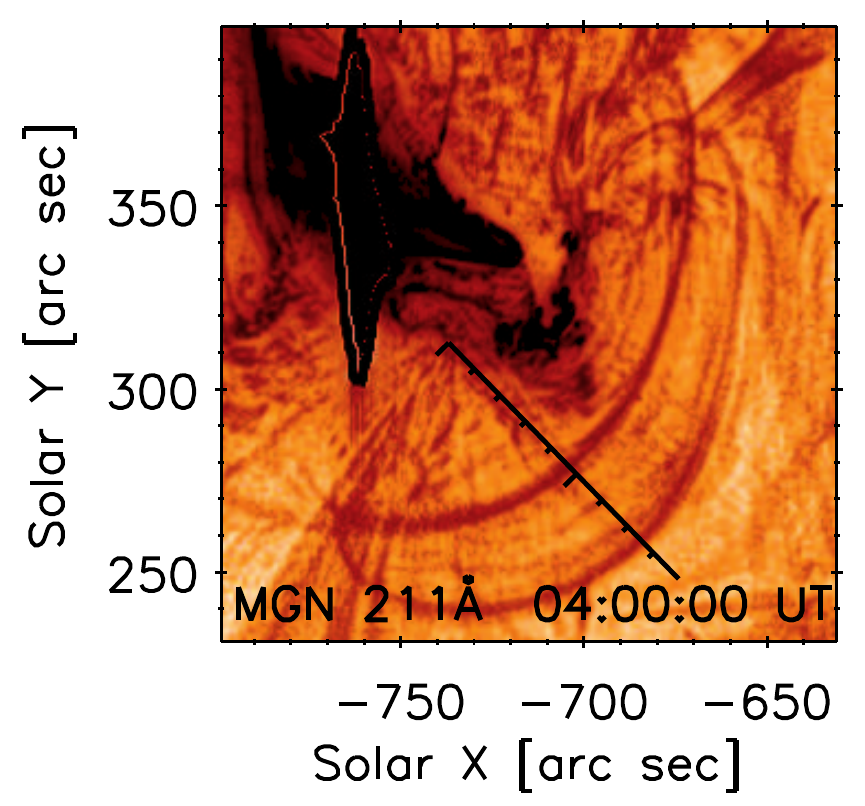}
	\includegraphics[width=4.42cm,clip,viewport= 0  0 243 225]{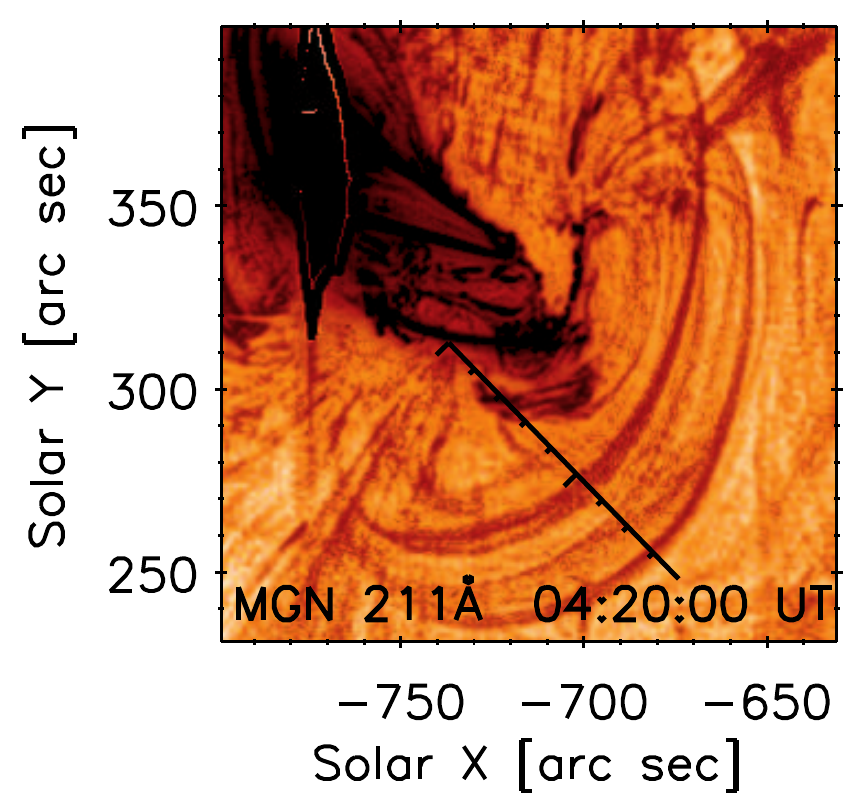}
	\includegraphics[width=3.29cm,clip,viewport=62  0 243 225]{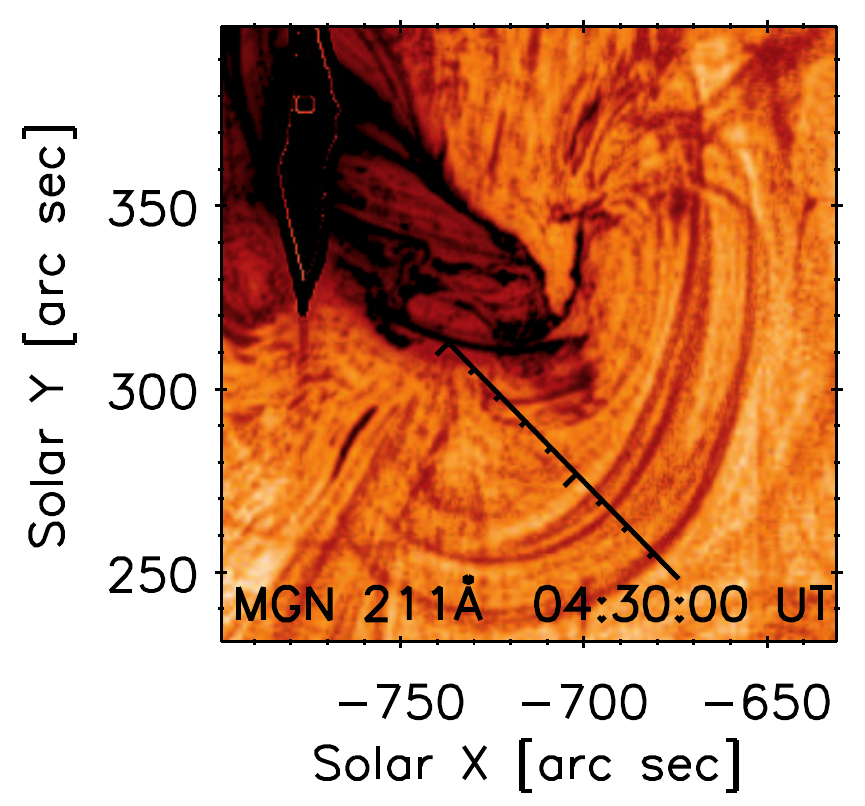}
	\includegraphics[width=3.29cm,clip,viewport=62  0 243 225]{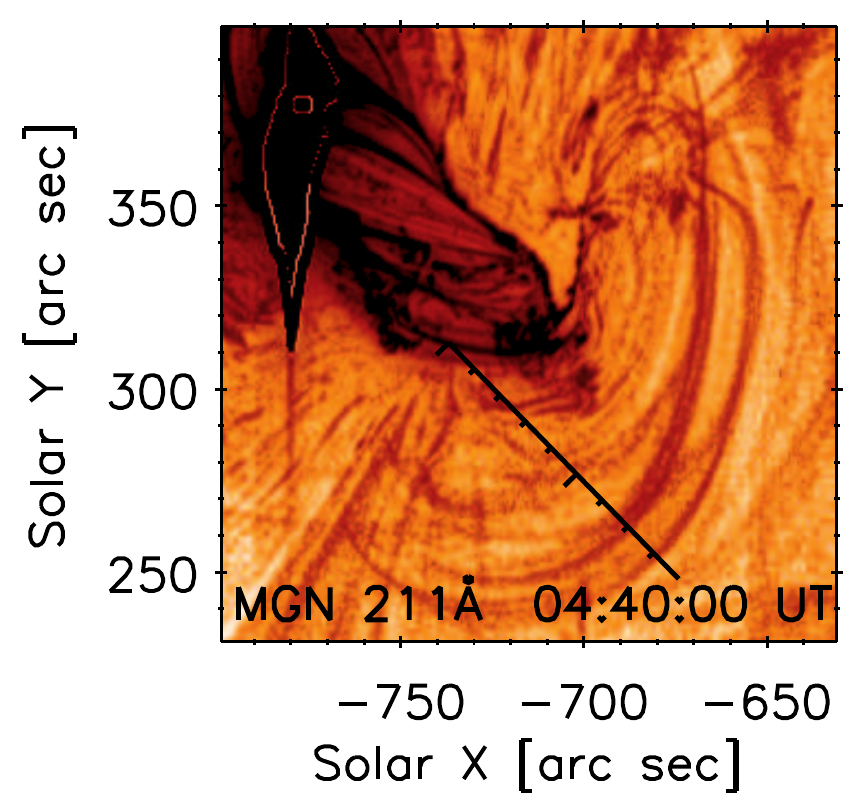}
	\includegraphics[width=3.29cm,clip,viewport=62  0 243 225]{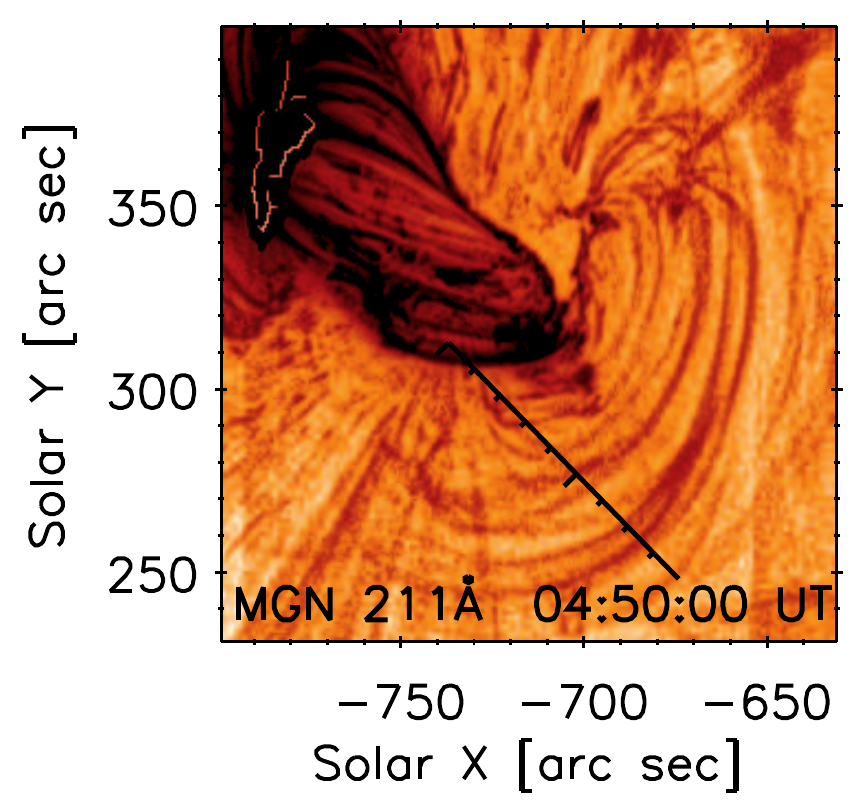}
	\includegraphics[width=3.29cm,clip,viewport=62  0 243 225]{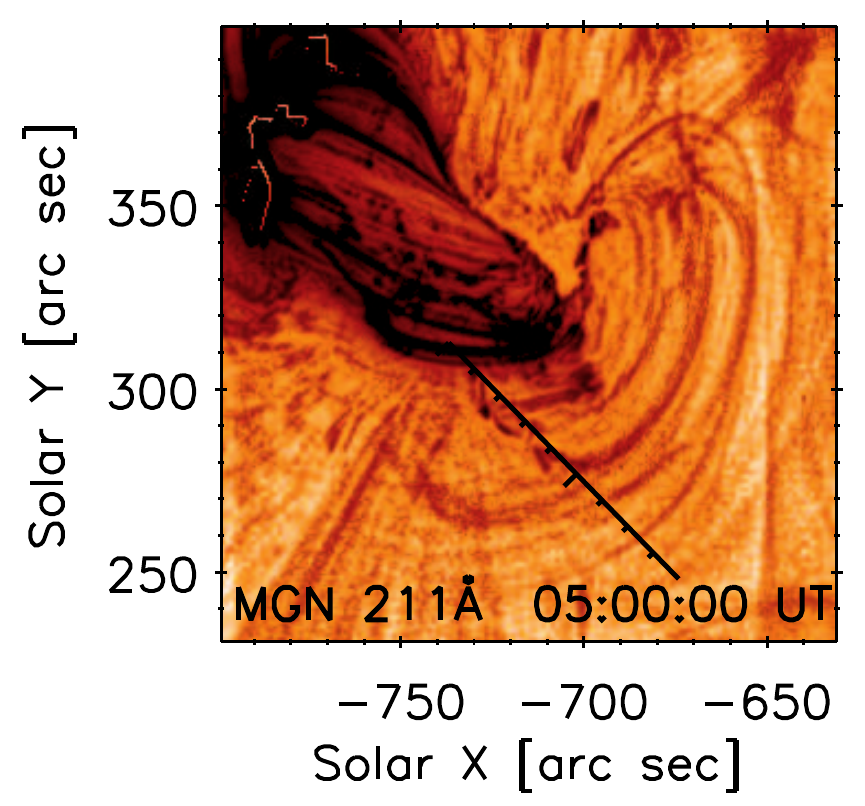}
\caption{Evolution of the peripheral coronal loops in AR 11429 showing vortex motions. Multi-Gaussian normalized figures are denoted by the label ``MGN'' and are shown in inverse scale (see Figure \ref{Fig:X_Stackplots_loops}). Position of the cut used to construct the time-distance plots shown in Figure \ref{Fig:X_Stackplots_loops} is shown in white in the first panel for each filter and in black on all other panels.  An animation of this figure is available in the online version of the article.}
\label{Fig:X_Vortex_loops}
\end{figure*}

%
\begin{figure*}[ht]
	\centering
	\includegraphics[width=8.60cm,clip]		 {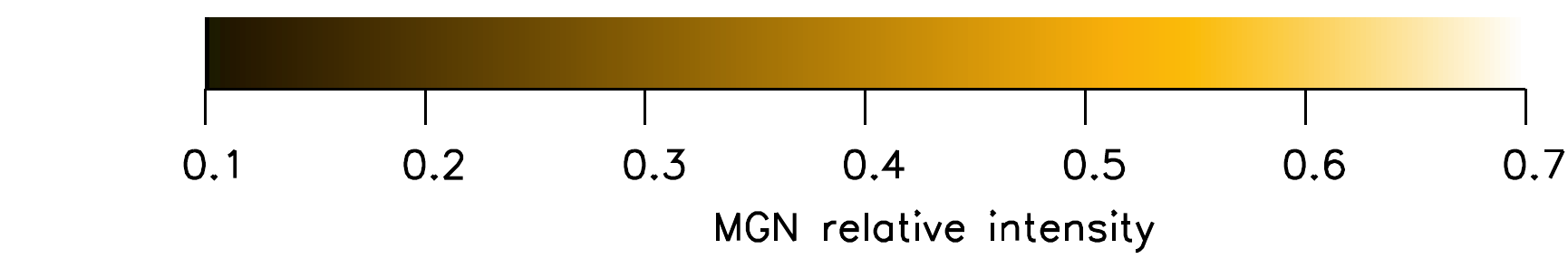}
	\includegraphics[width=8.60cm,clip]		 {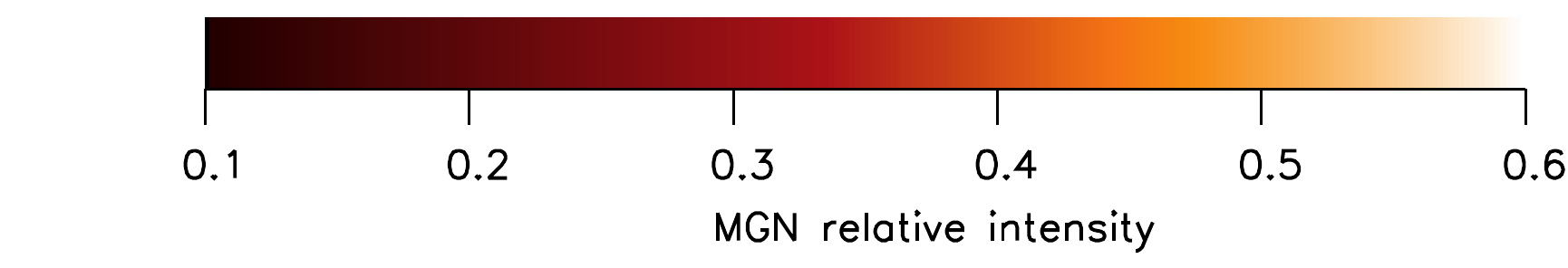}
	\includegraphics[width=8.60cm,clip,viewport=0 40 498 340] {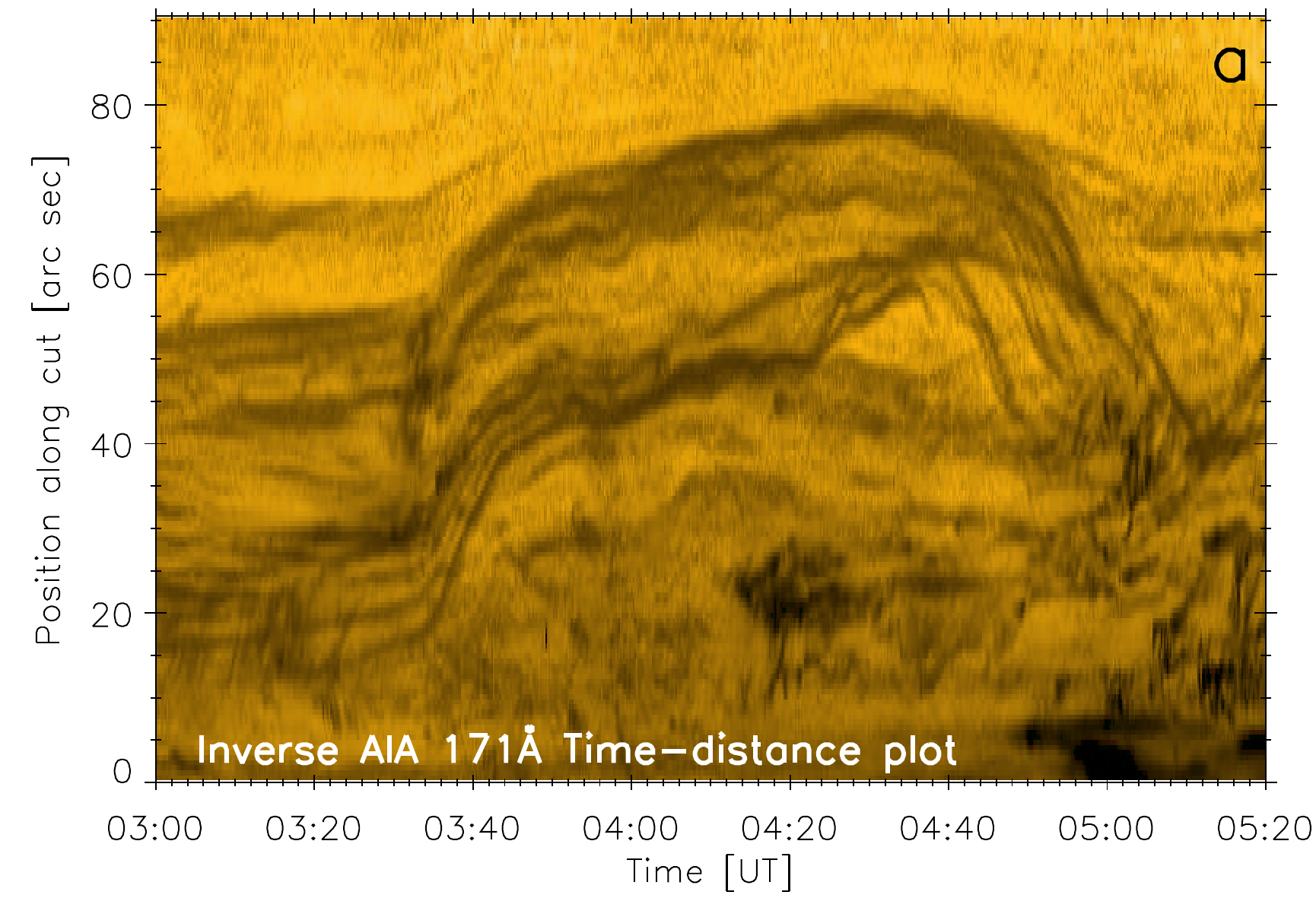}
	\includegraphics[width=8.60cm,clip,viewport=0 40 498 340] {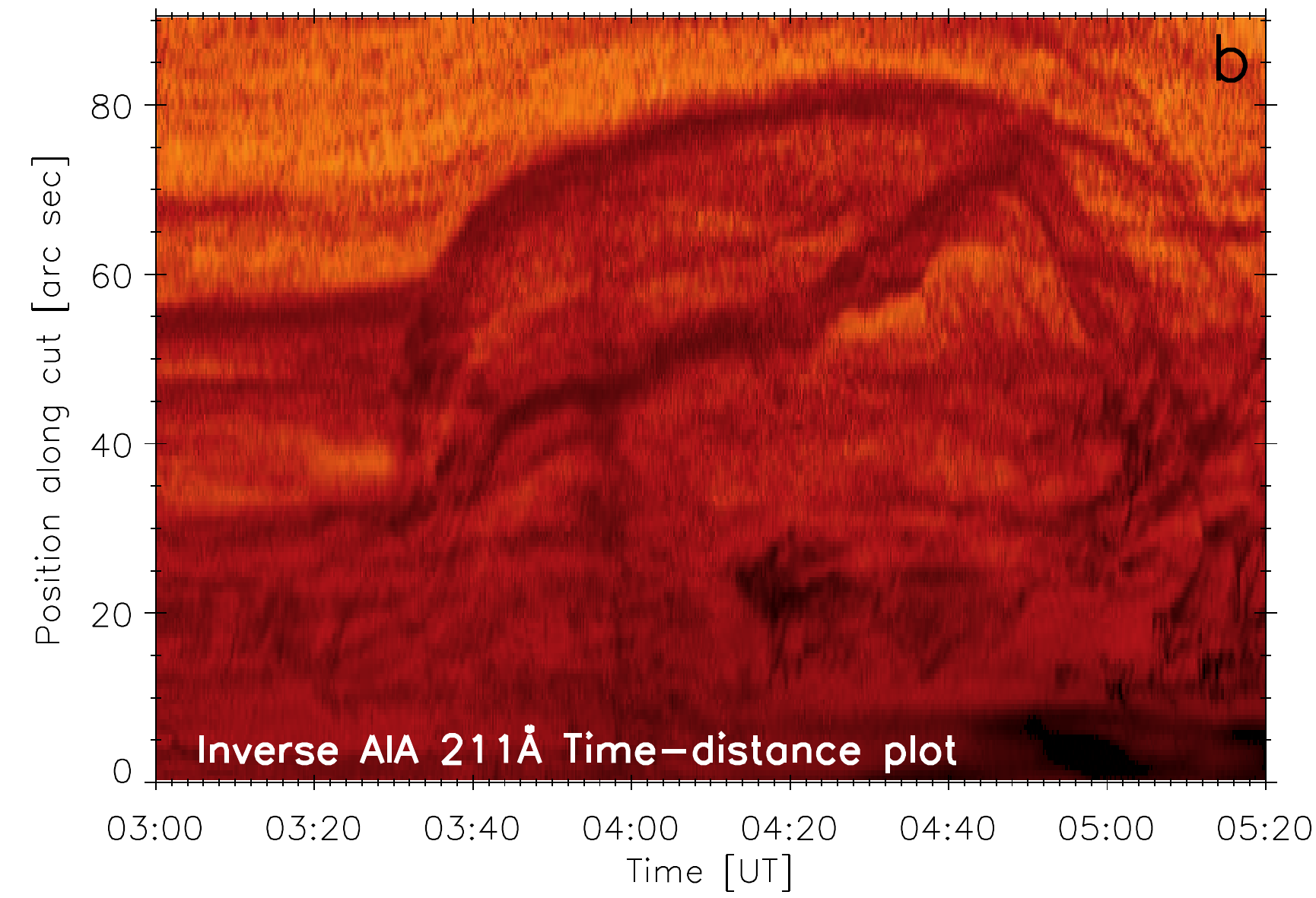}
	\includegraphics[width=8.60cm,clip,viewport=0  0 498 340] {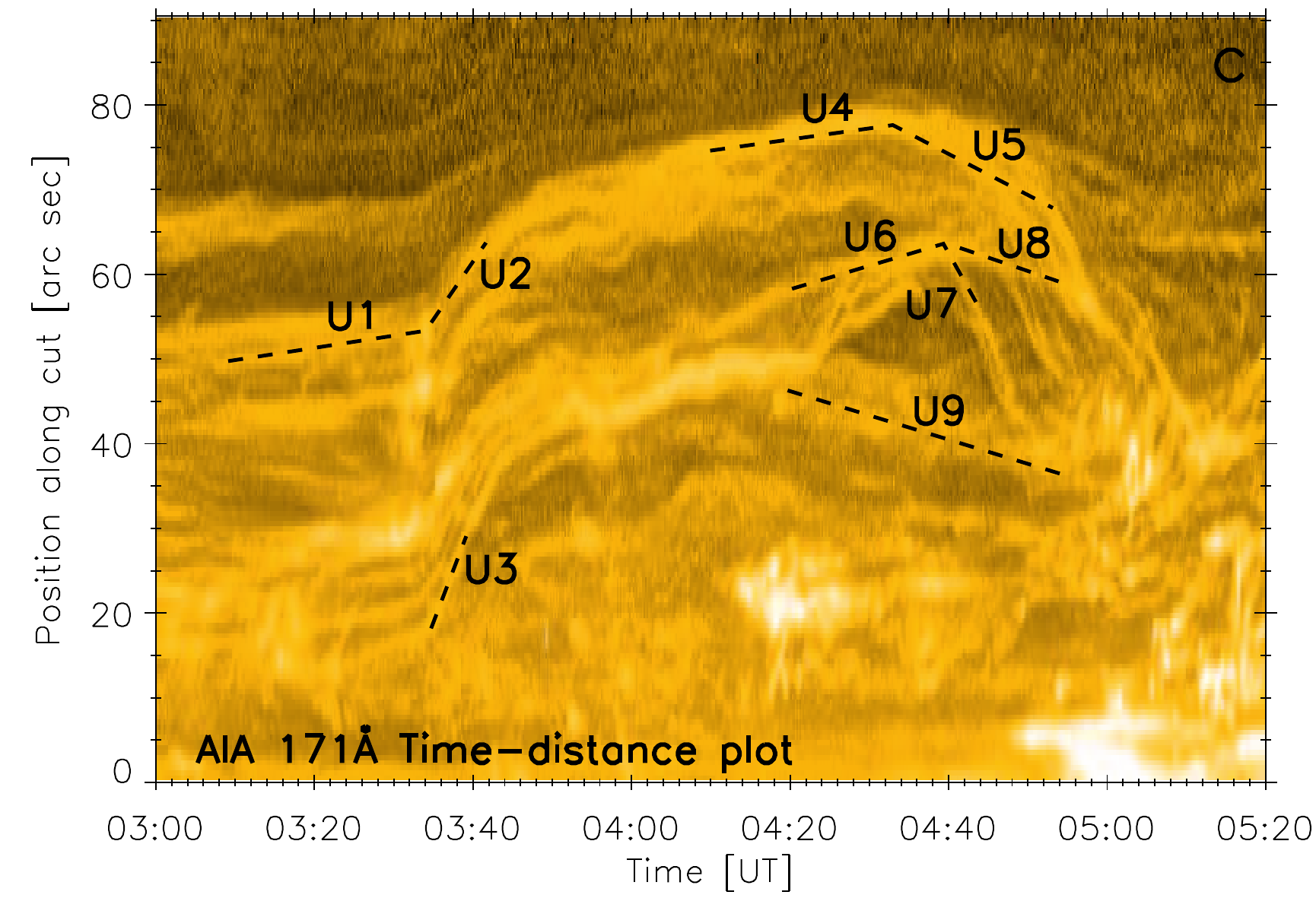}
	\includegraphics[width=8.60cm,clip,viewport=0  0 498 340] {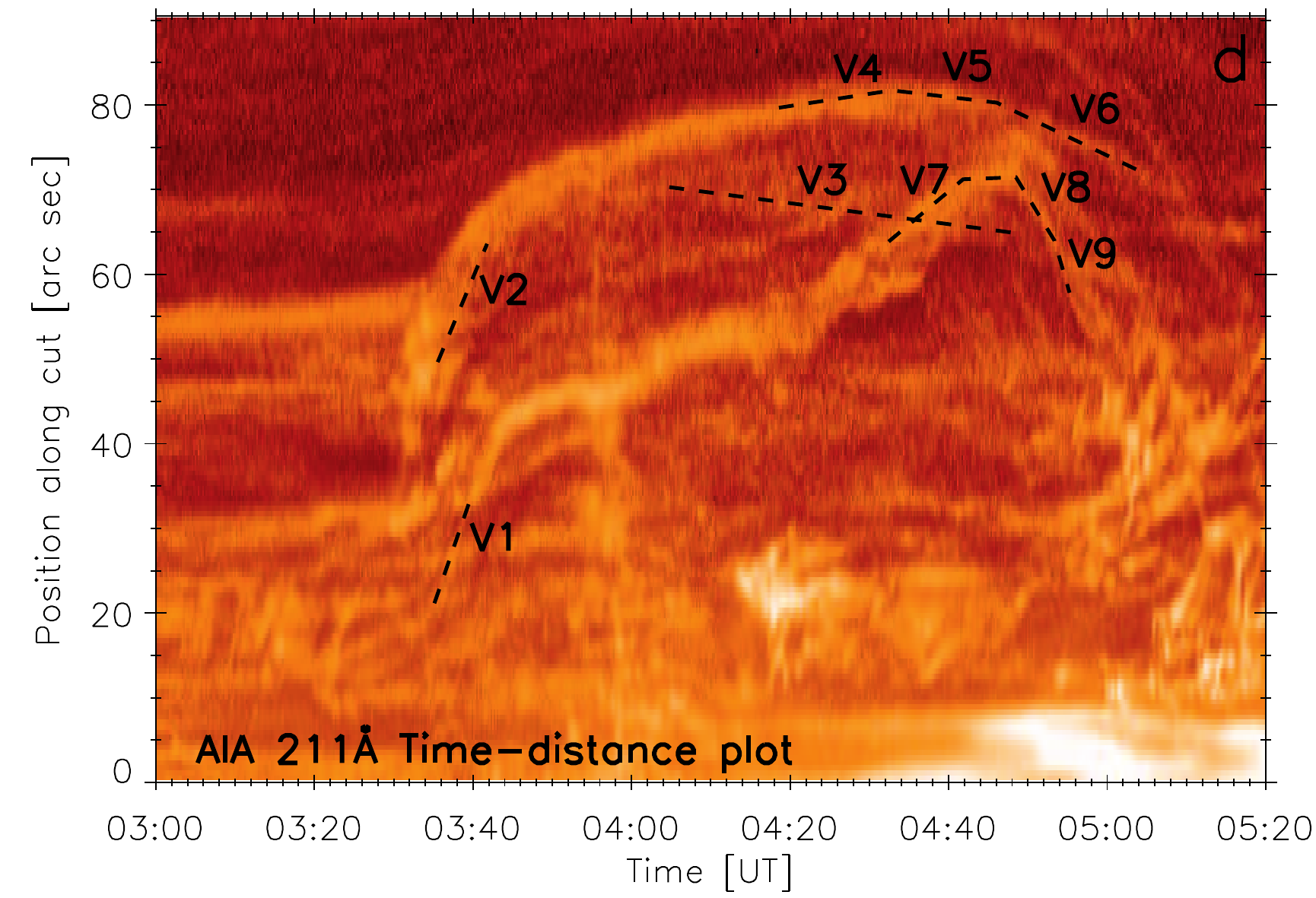}
\caption{Time-distance plots constructed along the cut shown in Figure \ref{Fig:X_Vortex_loops}. Panels (a)--(b) and (c)--(d) are scaled using the inverse and normalized MGN intensity, respectively. Velocities corresponding to individual dashed segments are given in Table \ref{Table:1}.}
\label{Fig:X_Stackplots_loops}
\end{figure*}
%
%

%
\subsection{Vortex motion of coronal loops}
\label{Sect:3.2}

As already mentioned, the AR 11429 contains an arcade of peripheral coronal loops located in its diskward portion. These closed coronal loops are visible in 171\,\AA, 193\,\AA, and 211\,\AA~passbands of AIA. Considering the projection near the eastern limb, these loops are likely highly inclined with respect to the local vertical, similarly to the blue/cyan field lines shown in Figure \ref{Fig:Model}. These loops are studied in the remainder of this Section. We note that some coronal loop dynamics can be discerned at the opposite (limbward) side of the AR at location of about Solar $[X,Y]$\,$\approx$\,$[-800,-350]$ (Figure \ref{Fig:X_Eruption}h), but the projection effects together with the growth of the flare arcade prevent us from studying them.

During the X-class flare, the peripheral coronal loops located in the diskward portion of the AR displayed expanding-contracting motions, see Figure \ref{Fig:X_Vortex_loops} and the accompanying animation. To study the motions of the coronal loops, and to enhance fainter loops that could be located at higher altitudes e.g. due to hydrostatics, we processed the AIA 171\,\AA~and 211\,\AA~data using the Multi-Gaussian Normalization technique \citep[hereafter, MGN;][]{Morgan14}. This technique enhances structures while suppressing large differences in intensity. The MGN convolves the observed image with a set of several Gaussians of increasing widths, then normalizes each convolved image and transforms it via an arctan function. The output image is a weighted average of such processed ones. The weighted mean across several spatial scales also suppresses noise, albeit less than some other, more computationally expensive techniques such as NAFE \citep{Druckmuller13}. The MGN has been successfully applied to AIA, STEREO/EUVI, and Proba-2/SWAP EUV images at 171 or 174\,\AA, as well as SOHO/LASCO-C2 and MLSO/Mk4 coronagraphic data \citep{Morgan14,Byrne14,Hutton15}. Here we find that this technique is also very useful in conjunction with AIA 211\,\AA~data.

The enhancement of the observed structure produced by the MGN technique is shown in Figure \ref{Fig:X_Vortex_loops}. There, the evolution of coronal loops during 03:20 -- 05:00\,UT is shown in AIA 171\,\AA~and 211\,\AA~passbands. First images at 03:20\,UT are the original AIA images, while the other ones are processed by the MGN technique. We chose these two wavelengths since they represent the warm emission arising in \ion{Fe}{9} (171\,\AA) and \ion{Fe}{14} (211\,\AA). We note that the 211\,\AA~bandpass of AIA contains contributions from cooler Fe ions as well as from unidentified lines (see \citet{ODwyer10}, \citet[][Figure 12]{DelZanna11c}, and \citet{DelZanna13}). Nevertheless, the 211\,\AA~images processed by MGN show coronal loops different from those seen in 171\,\AA, suggesting that their emission originates at temperatures higher than the formation of \ion{Fe}{9}. Furthermore, the open fan loops, protruding from the AR to the south, are suppressed in the 211\,\AA~passband. This makes the eastern footpoints of the closed coronal loops visible in 211\,\AA. Finally, these two bands are sufficient to show the vortex character of the expanding/contracting behavior of these coronal loops.

The dynamics of these peripheral coronal loops is elucidated by using the time-distance technique along a cut in the direction of the expanding/contracting motions. The location of this cut is shown on each frame of Figure \ref{Fig:X_Vortex_loops}. The location of the cut was chosen to minimize contamination from the neighbouring flare loop arcade as well as from transient brightenings from the nearby plage. The corresponding time-distance plots along these cuts are shown in Figure \ref{Fig:X_Stackplots_loops} both in negative and positive color scale. This should help to elucidate the rich dynamics shown by these coronal loops. Several slopes, indicating loops changing position, are marked by dashed lines in the positive images. These slopes are labeled U1--U9 in the 171\,\AA~MGN image and V1--V9 in the 211\,\AA~MGN image. The projected velocities $v$ that correspond to the slopes of lines in the time-distance plot are calculated as $v = (s_1 - s_0)/(t_1 - t_0)$, where $s_{0,1}$ and $t_{0,1}$ are the spatial positions of a loop along the cut, and time of the first and last point of the line, respectively. The uncertainty $\sigma(v)$ is then calculated by error propagation from the uncertainties in the spatial and temporal coordinates of the line endpoints
\begin{eqnarray}
	\nonumber \sigma^2(v) =  &&  \frac{\sigma^2(s_0) +\sigma^2(s_1)}{(t_1 -t_0)^2} \\
				 &+& \left(\frac{s_1 -s_0}{t_1 -t_0} \right)^2 
				 \frac{\sigma^2(t_1) +\sigma^2(t_0)}{(t_1 -t_0)^2} \,.
	\label{Eq:sigma_v}
\end{eqnarray}
For this purpose, we adopt a conservative uncertainty in the spatial position $\sigma(s_i)$\,=\,1.5$\arcsec$, while the uncertainty in time, $\sigma(t_i)$, is taken to be 6\,s, half of the AIA temporal resolution. We note that the resulting uncertainty in velocity is strongly dependent on the duration of the observed feature; fast, short-lived features have larger $\sigma(v)$ uncertainties since $t_1-t_0$ is smaller. 

Before 03:26\,UT, the peripheral loops display slow expanding motion. An example is the expansion of one outer loop seen in 171\,\AA~at cut position 50$\arcsec$, labeled as U1 in Figure \ref{Fig:X_Stackplots_loops}c. This loop expands outward with velocity of 1.7\,$\pm1.0$\,\kps. This expansion is evident also from the animation accompanying Figure \ref{Fig:X_Vortex_loops}. At about 03:34\,UT, i.e., during the impulsive phase characterized by fast eruption, the slow expansion switches to a fast one lasting to about 03:42\,UT, with velocities of up to 32.8\,$\pm$5.7\,\kps (V1). During this time, the inner loops always expand faster than the outer ones; compare U2 with U3, and V1 with V2 (Table \ref{Table:1}). 


Following this episodic fast expansion, the expansion of the loops slows down to about 9.5\,$\pm$2.7 (V7) and 1.6\,$\pm$1.1\,\kps\ (U4). This slow expansion lasts for nearly an hour, ending at about 04:40\,UT (see Table \ref{Table:1}), i.e., well into the flare gradual phase (Fig \ref{Fig:X_Context}a). An interesting feature is that some of these slowly expanding loops exhibit quasi-periodic oscillations; especially the outer loops observed in 171\,\AA~(at cut position 70--80$\arcsec$; Figure \ref{Fig:X_Stackplots_loops}a). This shows that loop oscillations do not always follow a contraction \citep{Russell15}, but can also be present during an expansion.

Most importantly, while some loops still expand, others, such as U9 and V3, have already started contracting. The U9 started contracting at 04:19\,UT with velocity of $-3.5$\,$\pm$0.7\,\kps, while the onset of the V3 contraction occurs at about 04:04\,UT, with a contraction velocity of $-1.5$\,$\pm$0.6\,\kps. We emphasize that the V3 starts contracting sooner and is located further along the cut than U9. This is a clear indication that the  switch from expansion to contraction is not a simple process with a single driver either increasing or decreasing the magnetic pressure. 

Instead, the expanding and contracting motions are observed to co-exist together, as predicted by the MHD Model (Section \ref{Sect:2}). This represents evidence of the existence of vortex motions generated during the initial stages of the eruptions, and that are eventually strengthen by the large-scale pressure-drop left behind by the erupting flux rope (ZAD17).

Finally, from 04:48\,UT onwards, i.e., well into the gradual phase, only contracting motions are observed (U5, U7--U8, V6, and V8--V9). These contracting motions are followed by apparent blob-like motions along the loop-like trajectories from about 05:15\,UT and the disappearance of these loops (see animation accompanying Figure \ref{Fig:X_Vortex_loops}), possibly as a result of catastrophic cooling.

%
\begin{figure*}[ht]
	\centering
	\includegraphics[width=8.8cm,clip,viewport= 0  0  498 475]{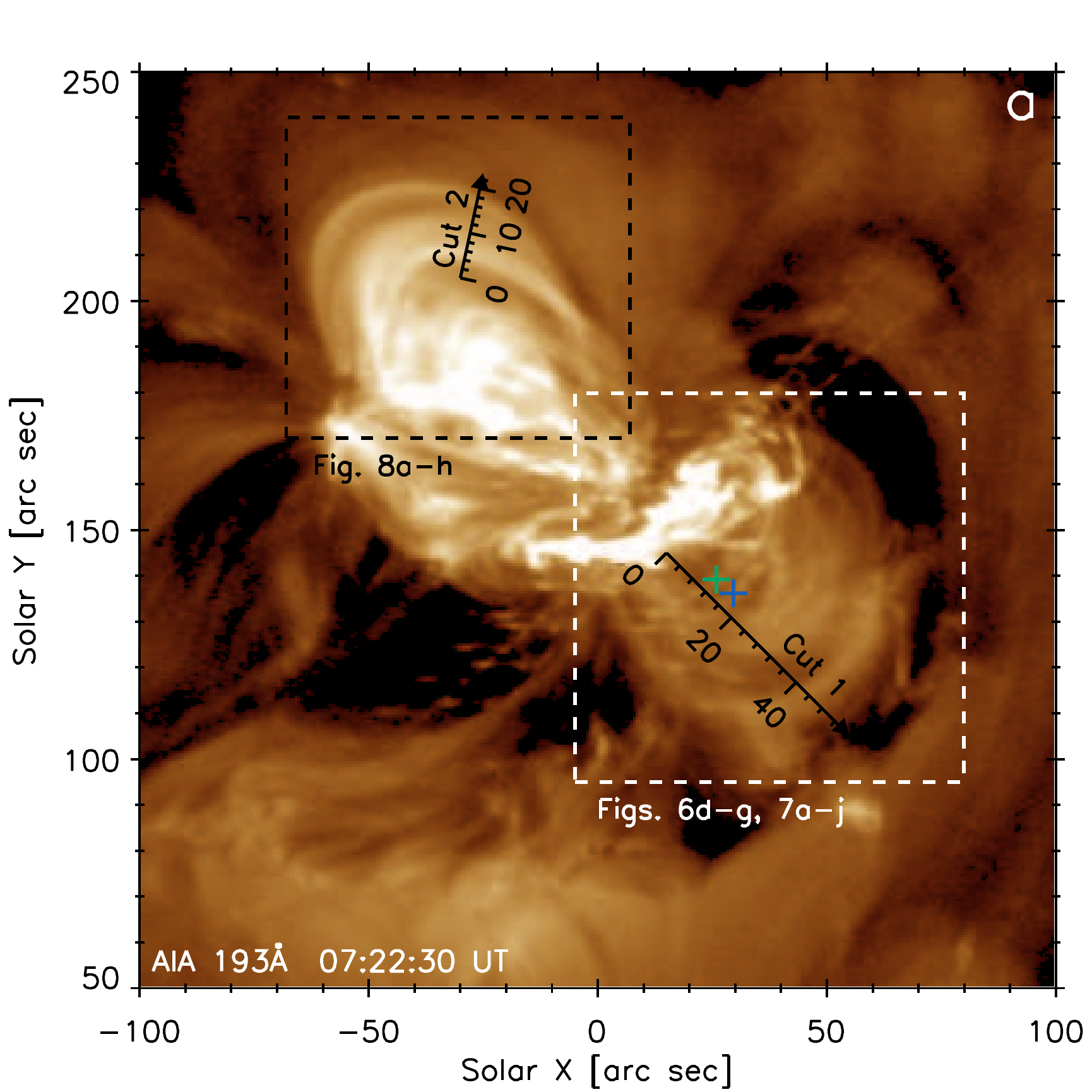}
	\includegraphics[width=8.8cm,clip,viewport= 0  0  498 475]{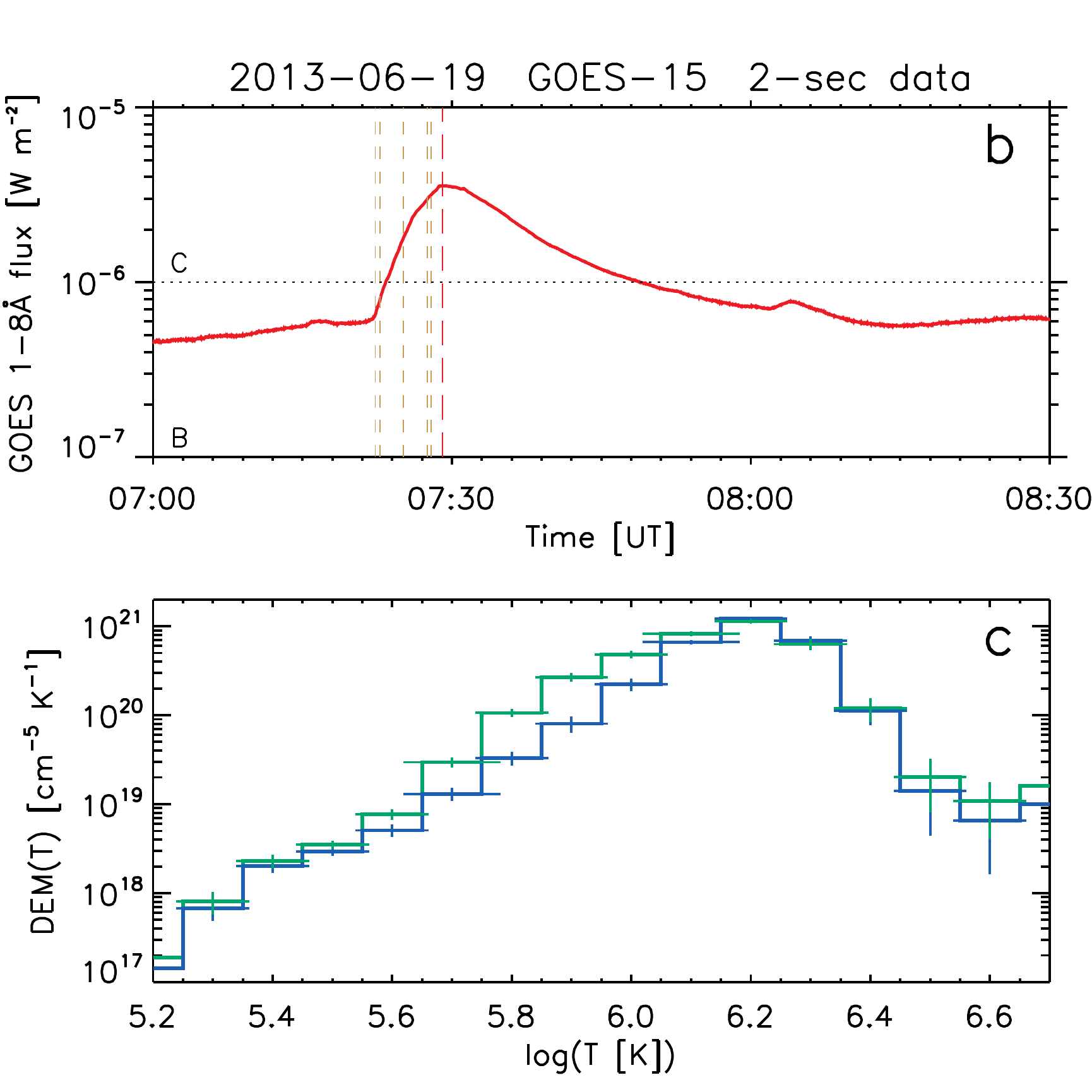}

\caption{(a): AR 11776 observed by AIA 193\,\AA~at 07:22:30\,UT. Cuts used to generate the stackplots along the SW arcade (cut 1, Figure \ref{Fig:C_Vortex_loops_SW}) and along the NE arcade (cut 2, Figure \ref{Fig:C_Vortex_loops_NE}) are shown. (b): GOES X-ray lightcurve of the C-class flare. Brown vertical lines denote the times shown in Figure \ref{Fig:C_Vortex_loops_SW}, panels (a)--(j). (c): Differential emission measures of two loops, obtained from the positions denoted by the green and blue crosses shown in the panel (a). 
\label{Fig:C_Context}}
\end{figure*}
%
%
\begin{figure*}[ht]
	\centering
	\includegraphics[width=4.26cm,clip,viewport= 0 40 244 244]{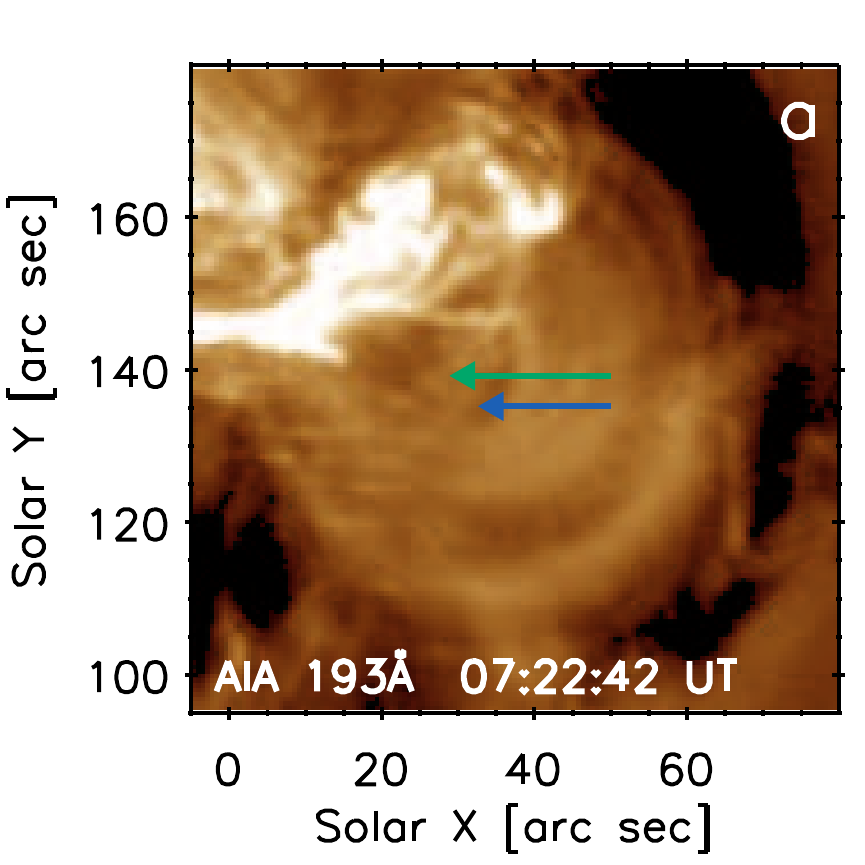}
	\includegraphics[width=3.33cm,clip,viewport=53 40 244 244]{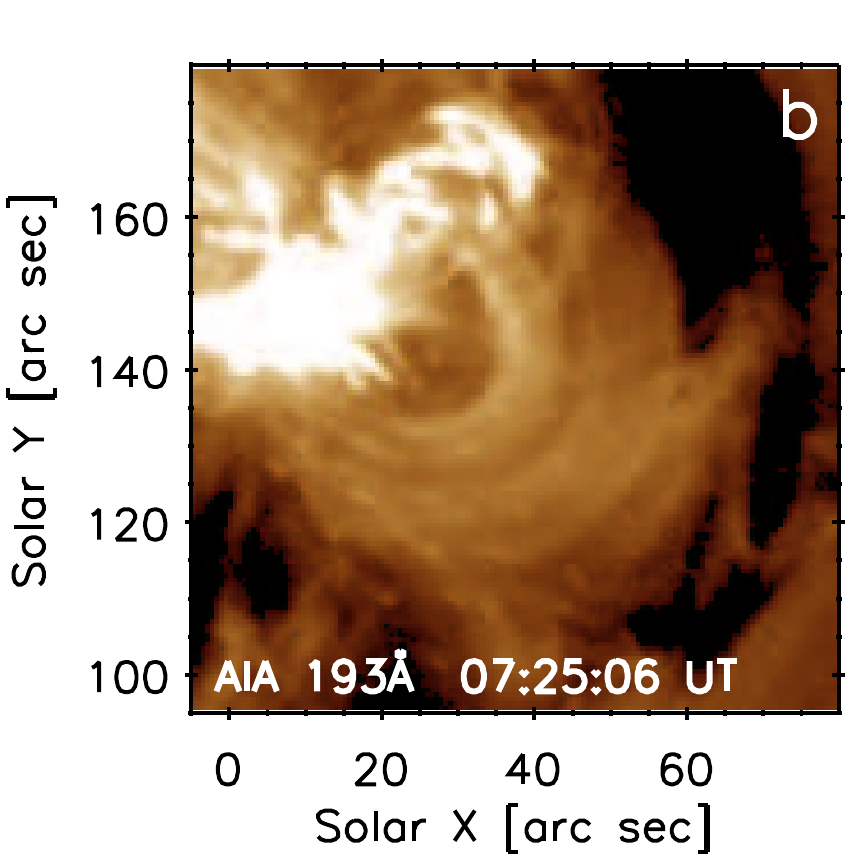}
	\includegraphics[width=3.33cm,clip,viewport=53 40 244 244]{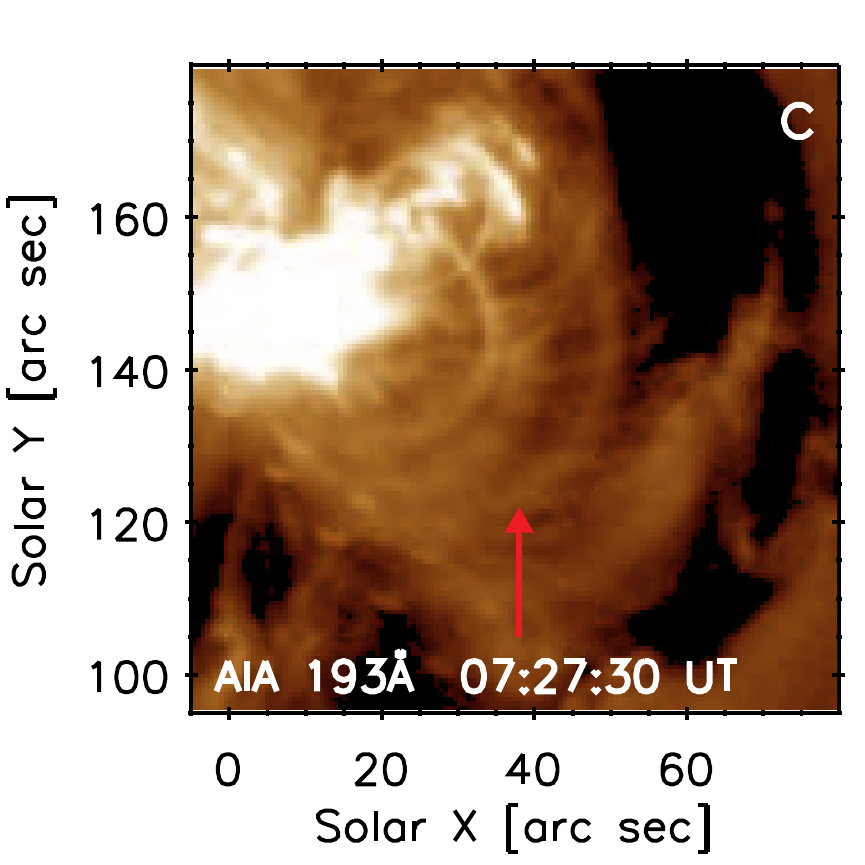}
	\includegraphics[width=3.33cm,clip,viewport=53 40 244 244]{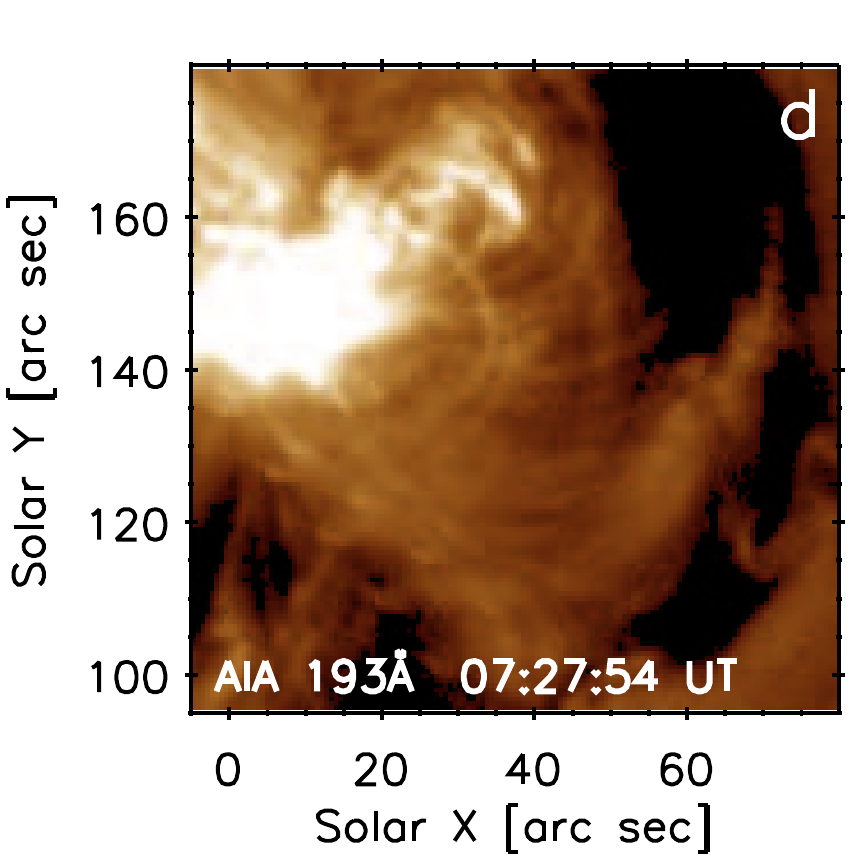}
	\includegraphics[width=3.33cm,clip,viewport=53 40 244 244]{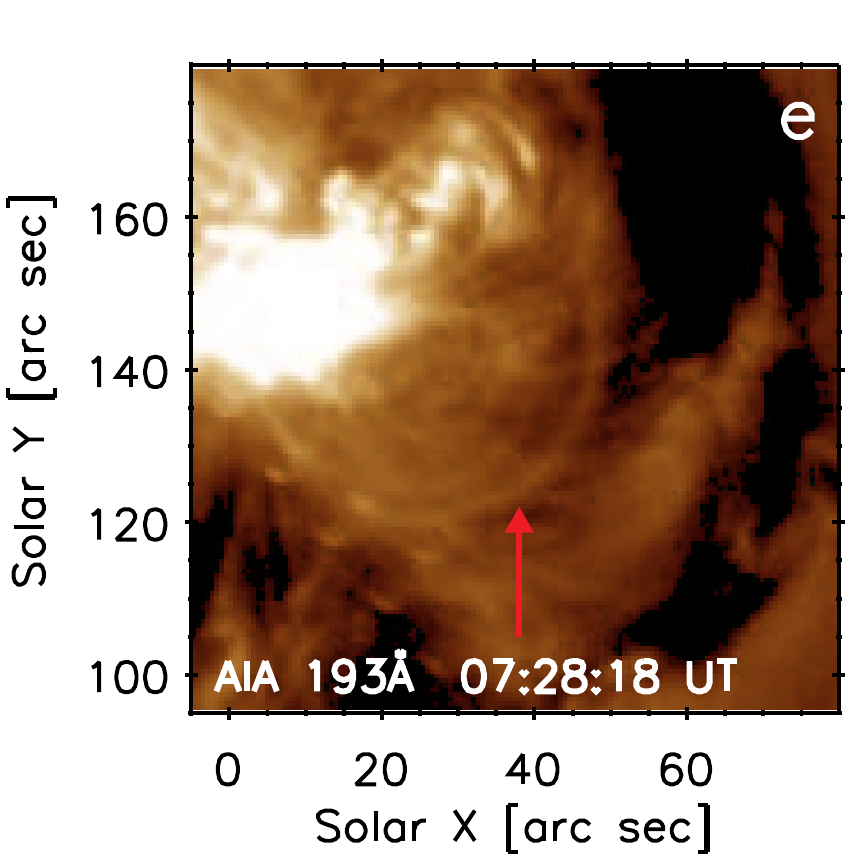}
	\vspace{0.3cm}
	\includegraphics[width=4.26cm,clip,viewport= 0  0 244 235]{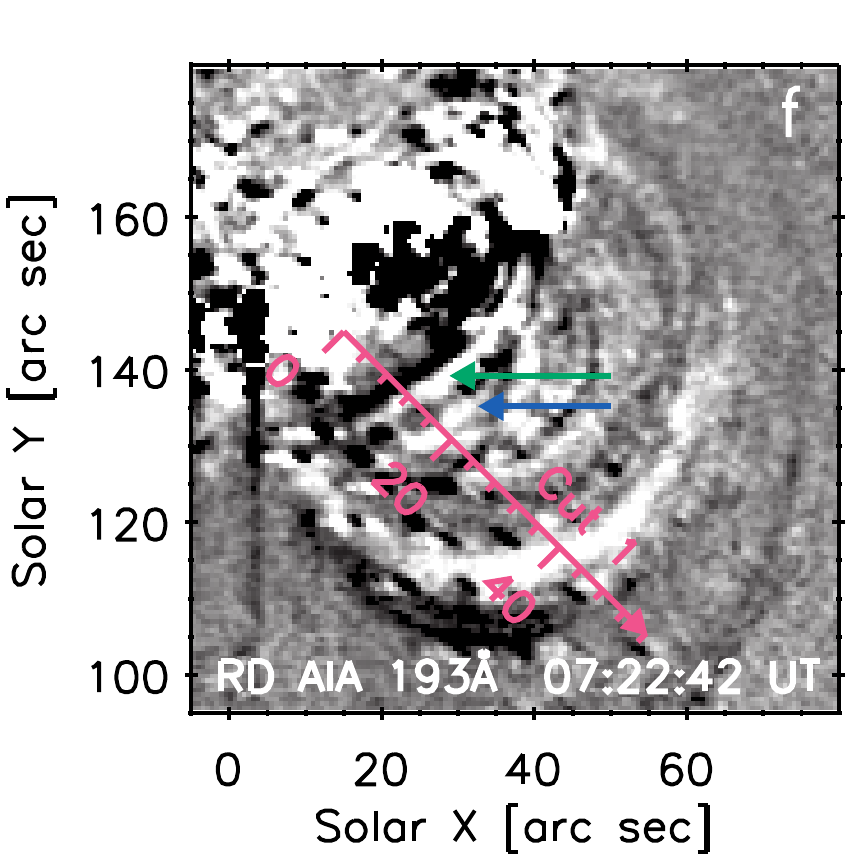}
	\includegraphics[width=3.33cm,clip,viewport=53  0 244 235]{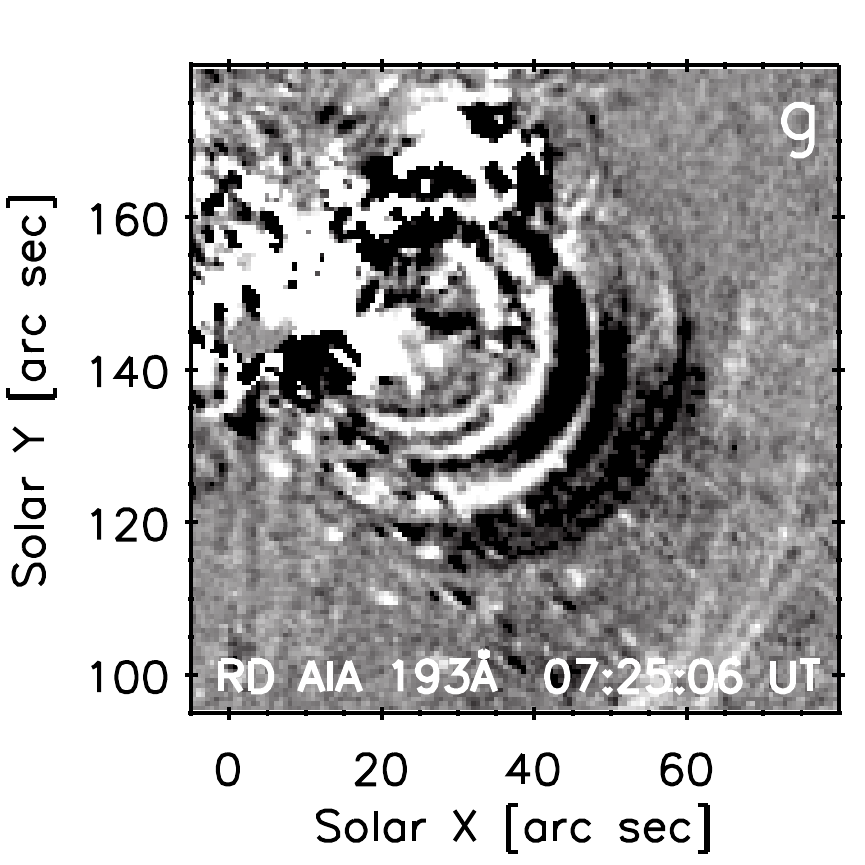}
	\includegraphics[width=3.33cm,clip,viewport=53  0 244 235]{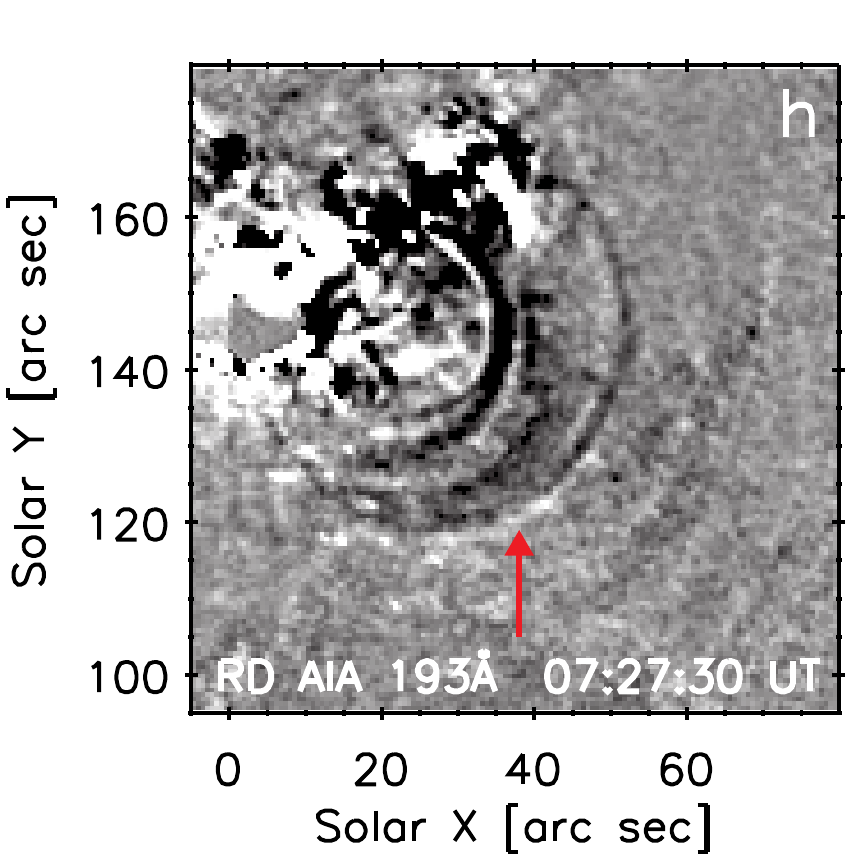}
	\includegraphics[width=3.33cm,clip,viewport=53  0 244 235]{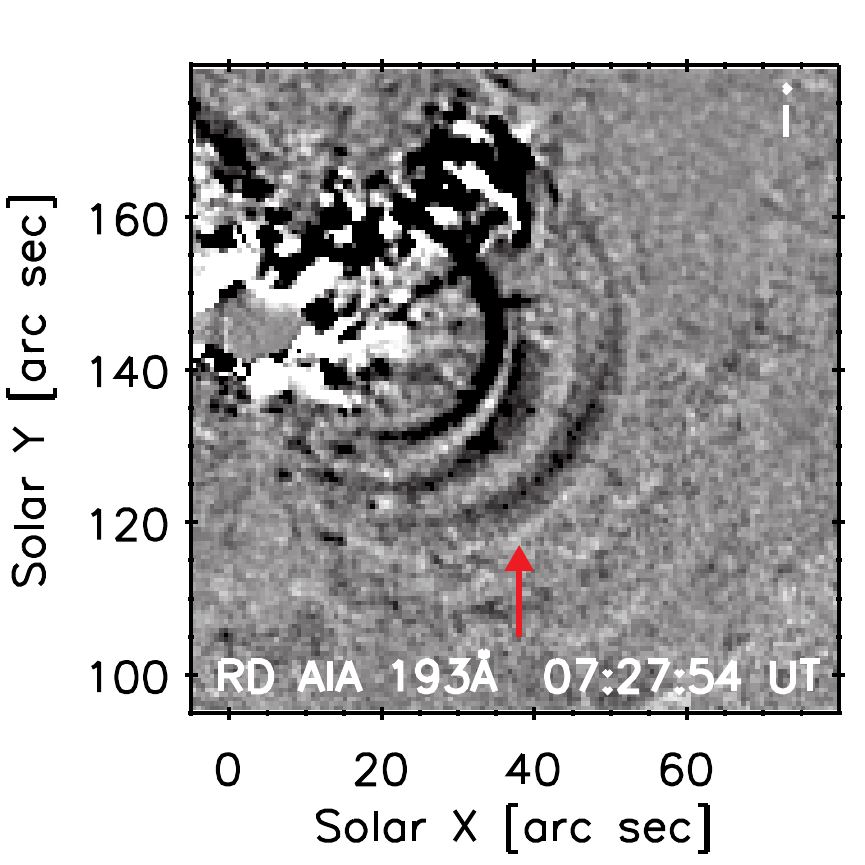}
	\includegraphics[width=3.33cm,clip,viewport=53  0 244 235]{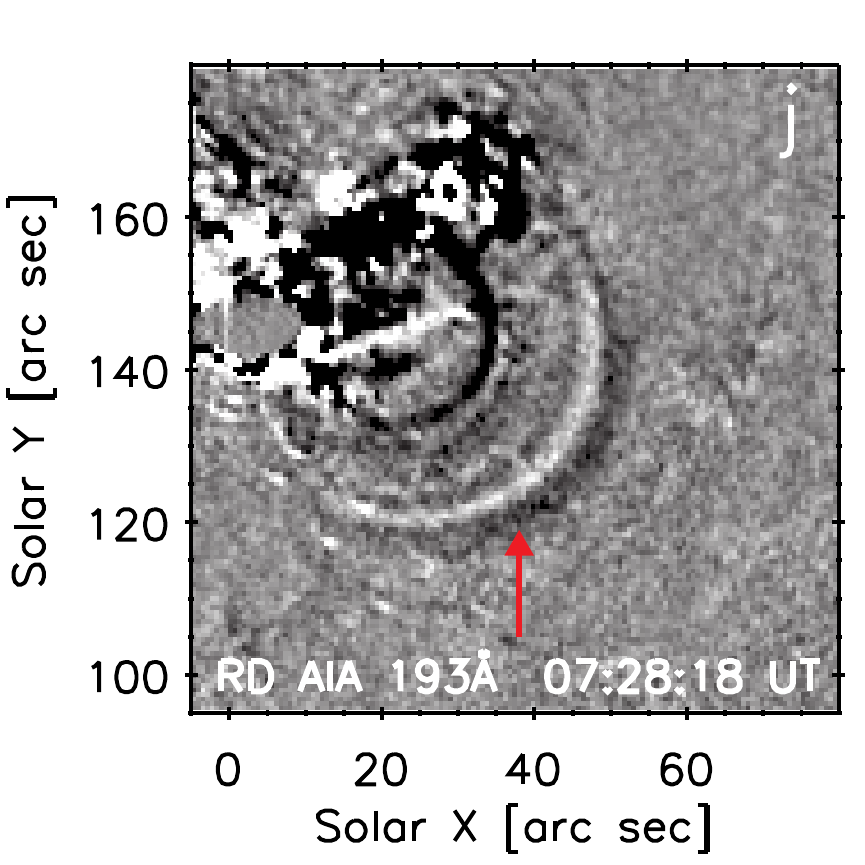}
	\includegraphics[width=9.48cm,clip,viewport= 0 37  490 303]{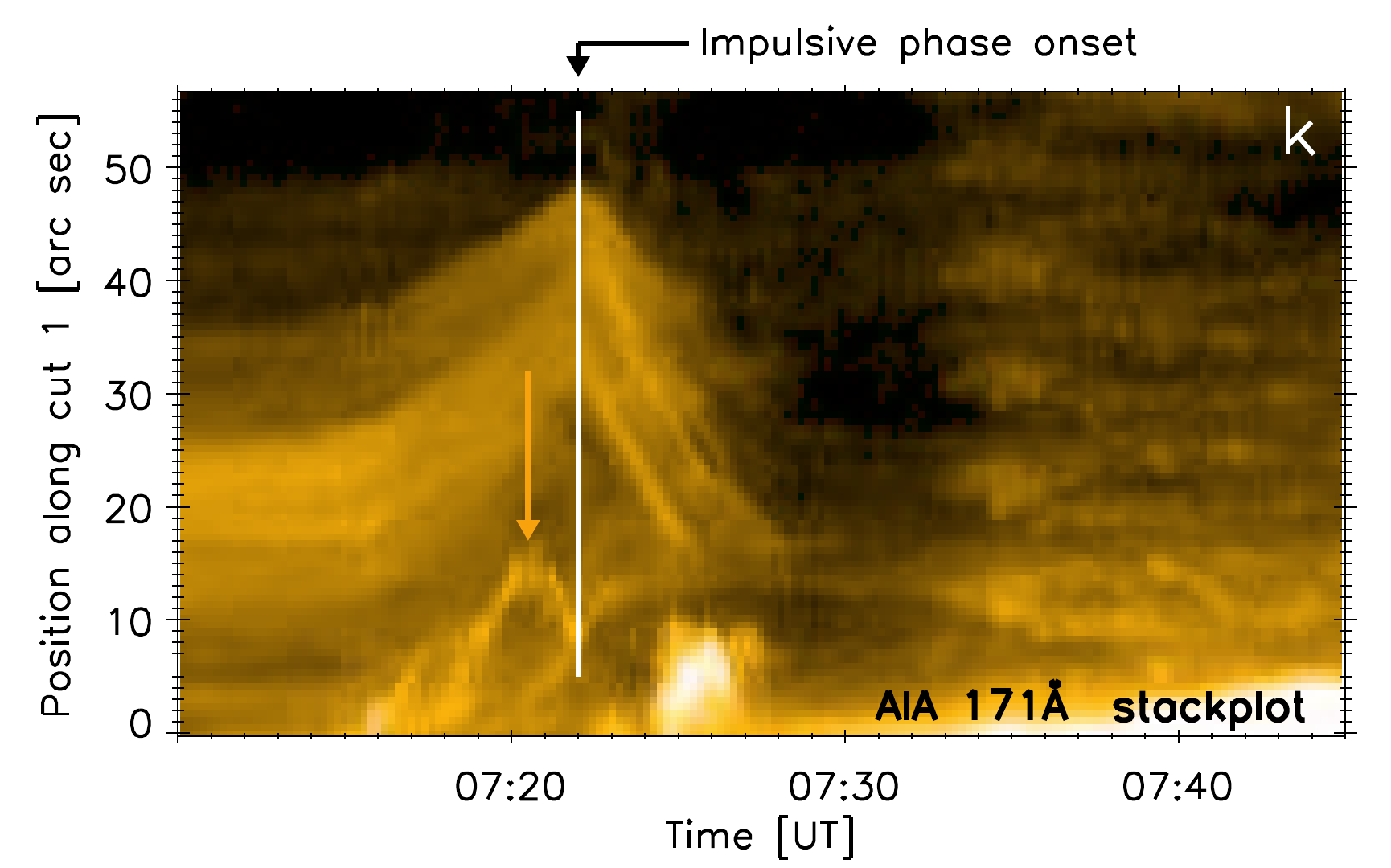}
	\includegraphics[width=8.32cm,clip,viewport=60 37  490 303]{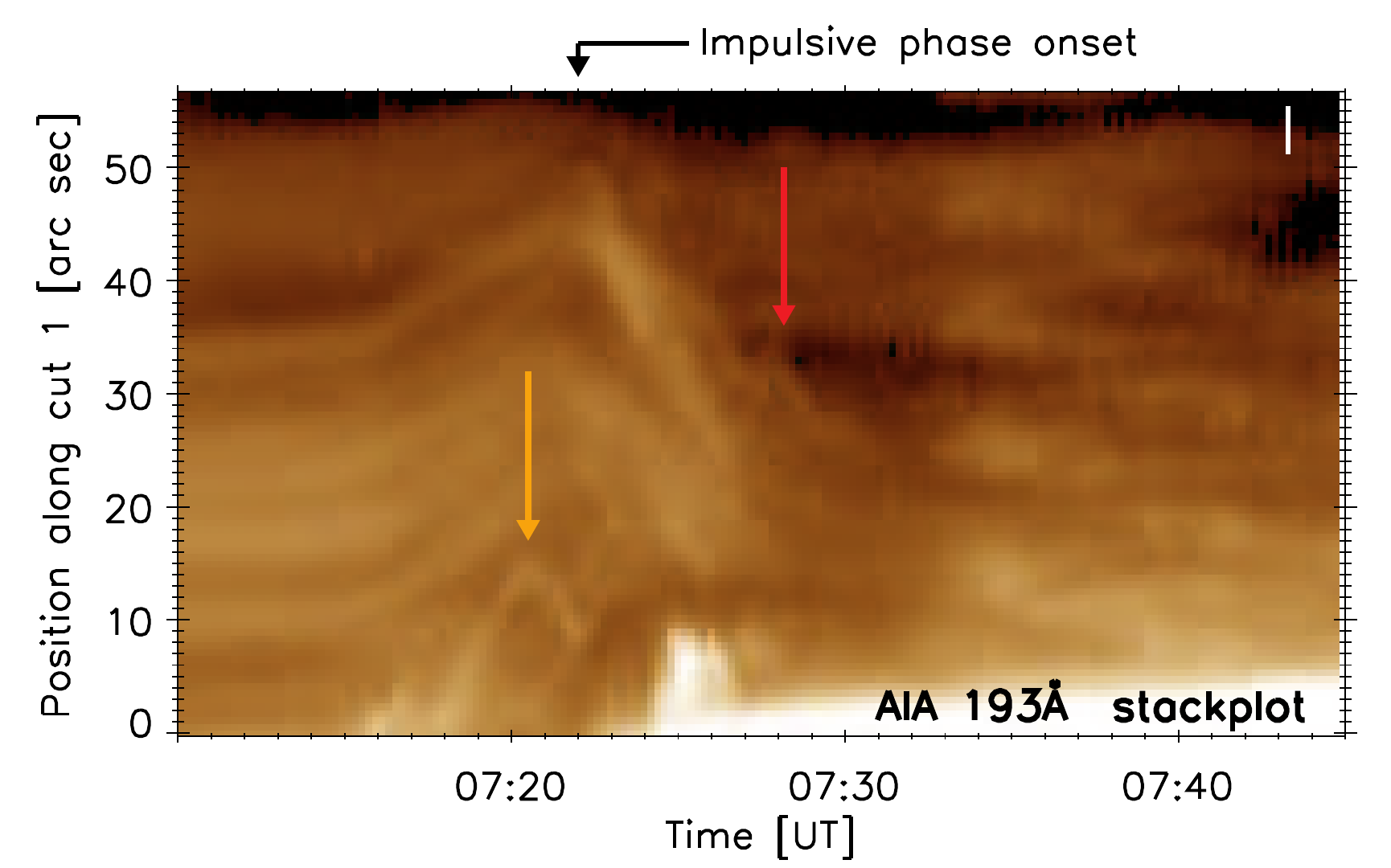}
	\includegraphics[width=9.48cm,clip,viewport= 0  0  490 283]{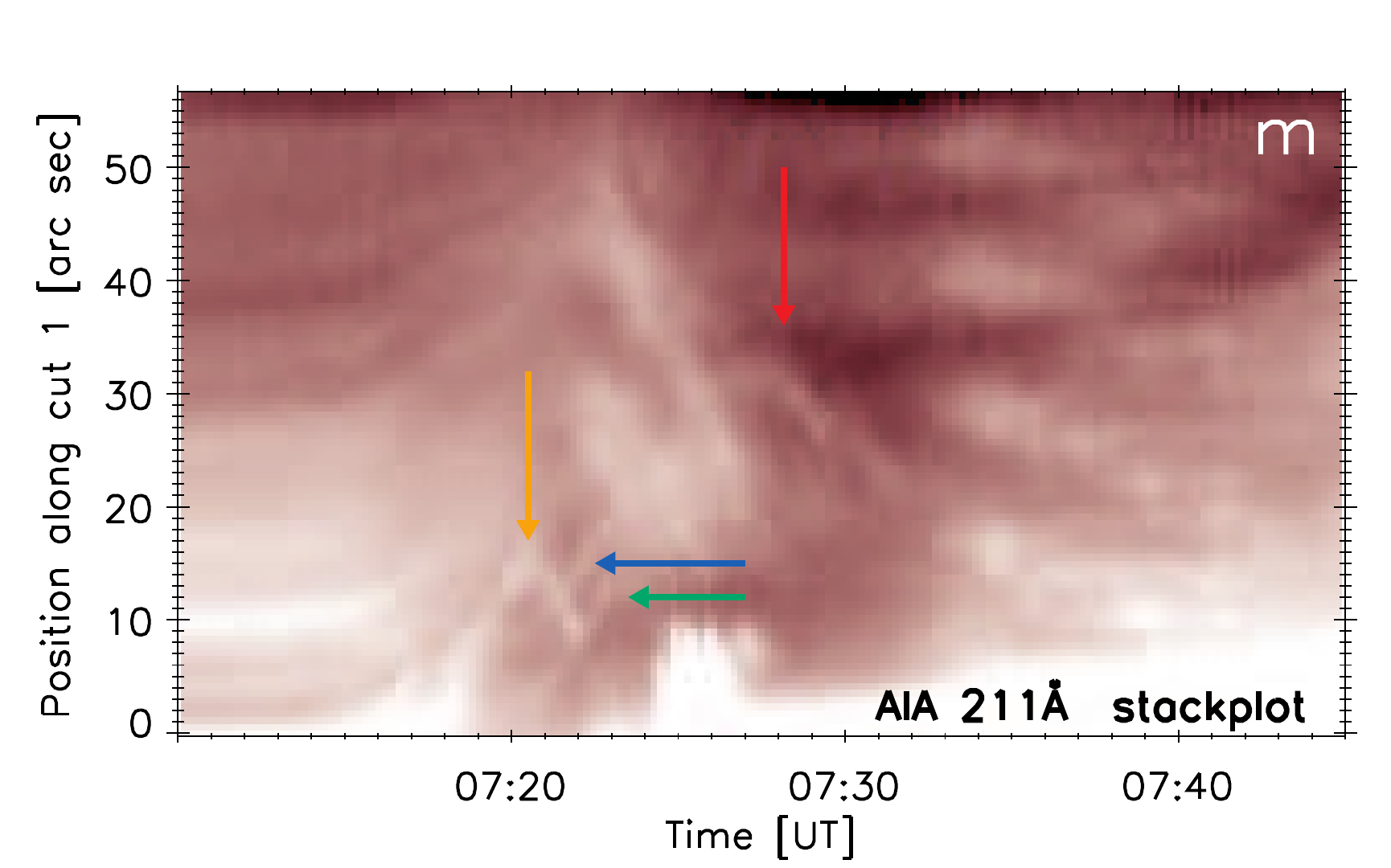}
	\includegraphics[width=8.32cm,clip,viewport=60  0  490 283]{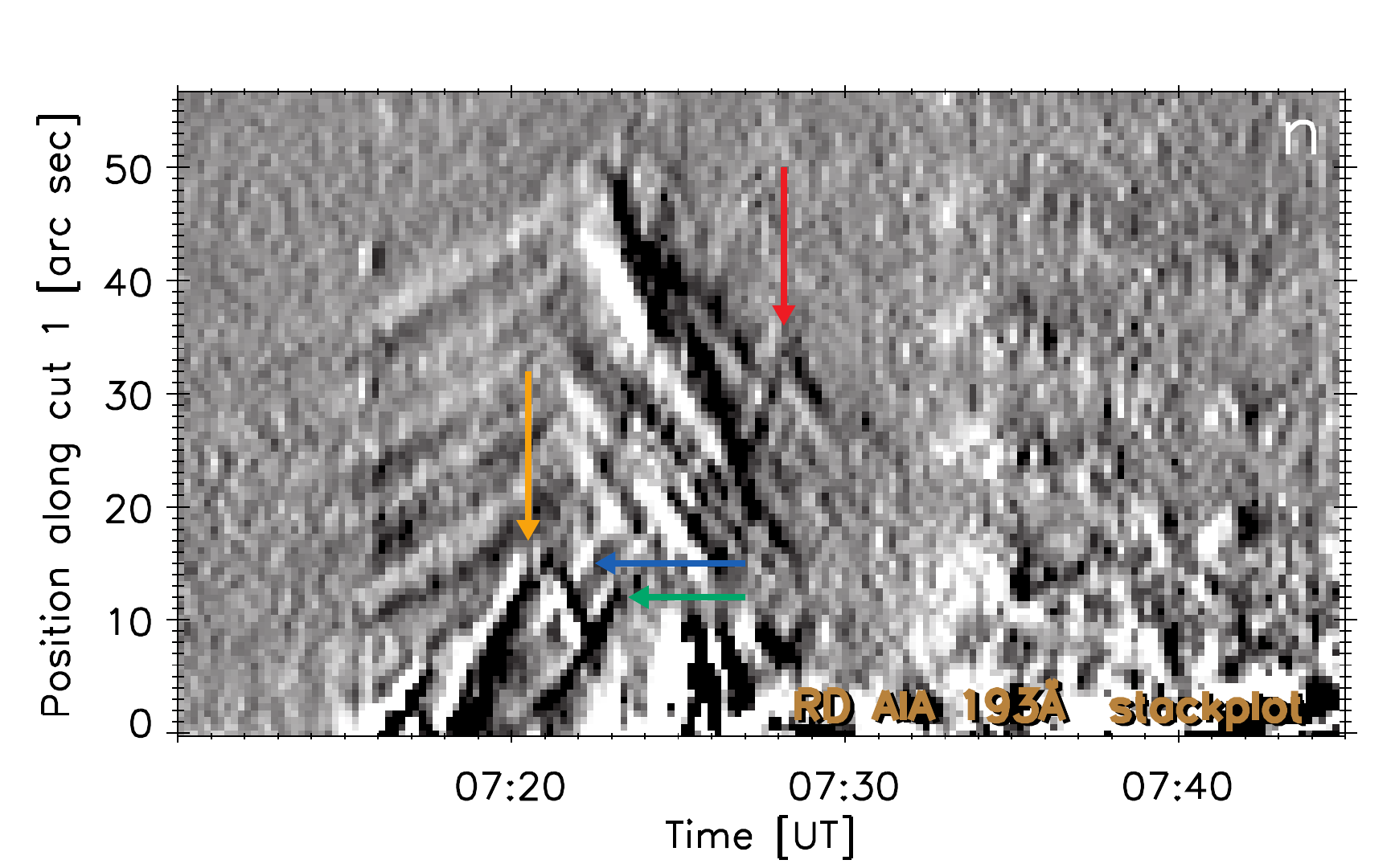}
\caption{(a)--(e) and (f)--(j): Evolution of the SW arcade as observed by AIA 193\,\AA. Intensity images are shown in panels (a)--(e), while the corresponding running difference images are shown in (f)--(j). The location of cut 1 is shown by the pink arrow in panel (f). The field of view shown corresponds to the white box in Figure \ref{Fig:C_Context}a. The red arrow shows a loop switching from expansion to contraction between 07:27:54 and 07:28:18\,UT.
(k)--(n): Time-distance plots constructed for AIA 171\,\AA, 193\,\AA, 211\,\AA, and running difference 193\,\AA~along the cut 1. The orange arrow shows the western part of the erupting filament \citep[discussed in][]{Wang16}. Green and blue arrows show the two expanding loops corresponding to the blue and black crosses in Figure \ref{Fig:C_Context}a. The red arrow points to a loop that switches from expansion to contraction, as shown in panels (h)--(j). The onset of the impulsive phase at 07:22\,UT is denoted by a white vertical line in the panel (k), as well as by a black arrow above the time-distance plots.
}
\label{Fig:C_Vortex_loops_SW}
\end{figure*}
%

%
\section{The 2013 June 19 C-class event}
\label{Sect:4}

We next examine a C3.5-class event that occurred on 2013 June 19 (SOL2013-06-19T07:29) in the active region NOAA 11776. This active region is located near disk center, and contains two arcades of coronal loops (Figure \ref{Fig:C_Context}a). One is located in the south-western part of the AR, at the location of about $[X,Y]$\,=\,$[50\arcsec,130\arcsec]$. On the opposite, north-eastern side, a second, more compact arcade is found.

This C3.5-class event was previously studied by \citet{Wang16}, who reported an apparent implosion of the south-western coronal loop arcade. Here, we re-examine the AIA observations of the event. To do that, the AIA data were processed using the same standard procedure outlined by \citet{Wang16}, including a correction for differential rotation with respect to 07:00 UT on 2013 June 19. A snapshot of the flaring AR 11776 in the AIA 193\,\AA~passband, taken shortly after the onset of the impulsive phase, is shown in Figure \ref{Fig:C_Context}a. The evolution of the X-ray flux as observed by the GOES-15 satellite is shown in Figure \ref{Fig:C_Context}b. The event was eruptive, with a filament eruption exhaustively examined and reported on in detail by \citet{Wang16}. The filament eruption does not need to be re-examined here. We however note that the onset of filament eruption, flare impulsive phase, and strong contraction of the arcade, all began at $\approx$07:22\,UT. Subsequently, the flare soft X-ray flux at 1--8\,\AA~reached its maximum at about 07:29:08\,UT.

%
\begin{figure*}[ht]
	\centering
	\includegraphics[width=5.20cm,clip,viewport= 0 40 245 225]{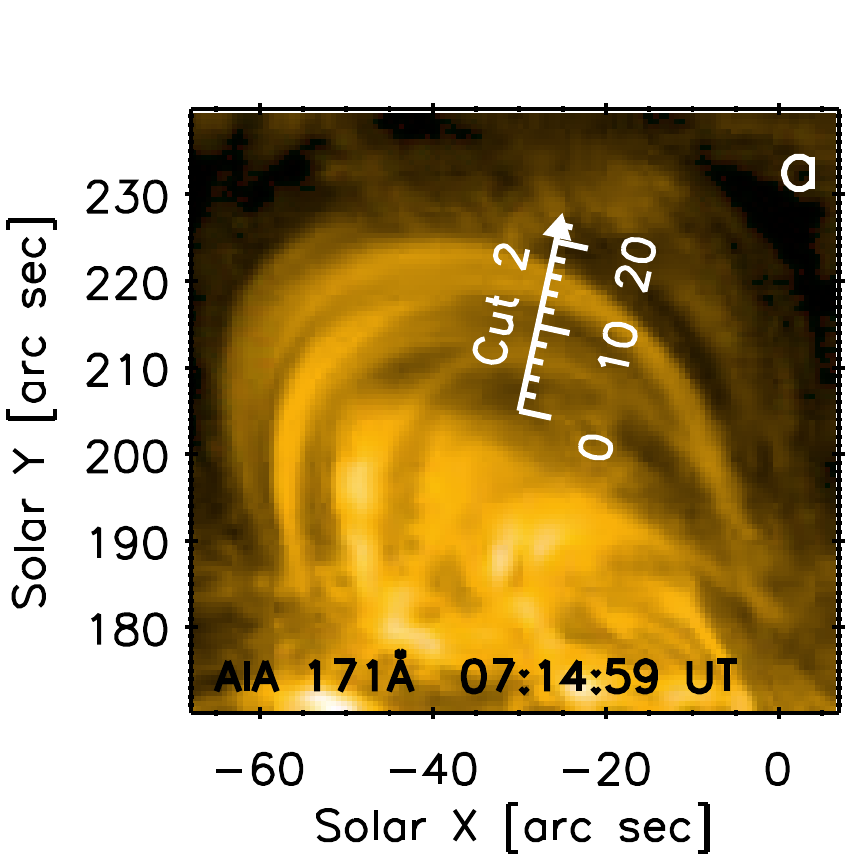}
	\includegraphics[width=4.13cm,clip,viewport=50 40 245 225]{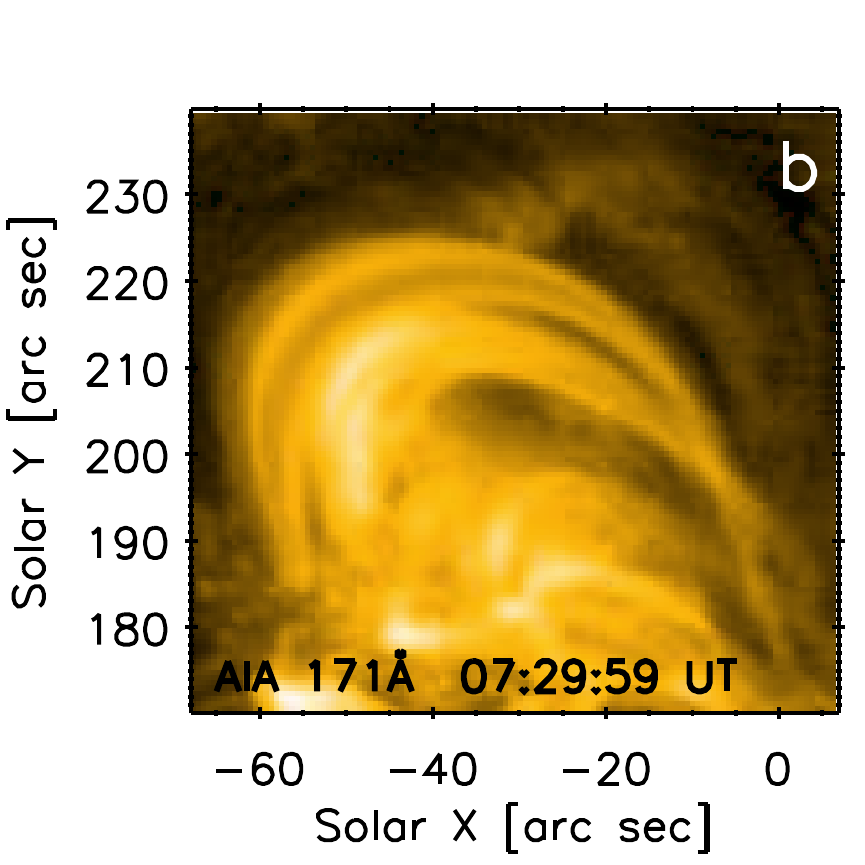}
	\includegraphics[width=4.13cm,clip,viewport=50 40 245 225]{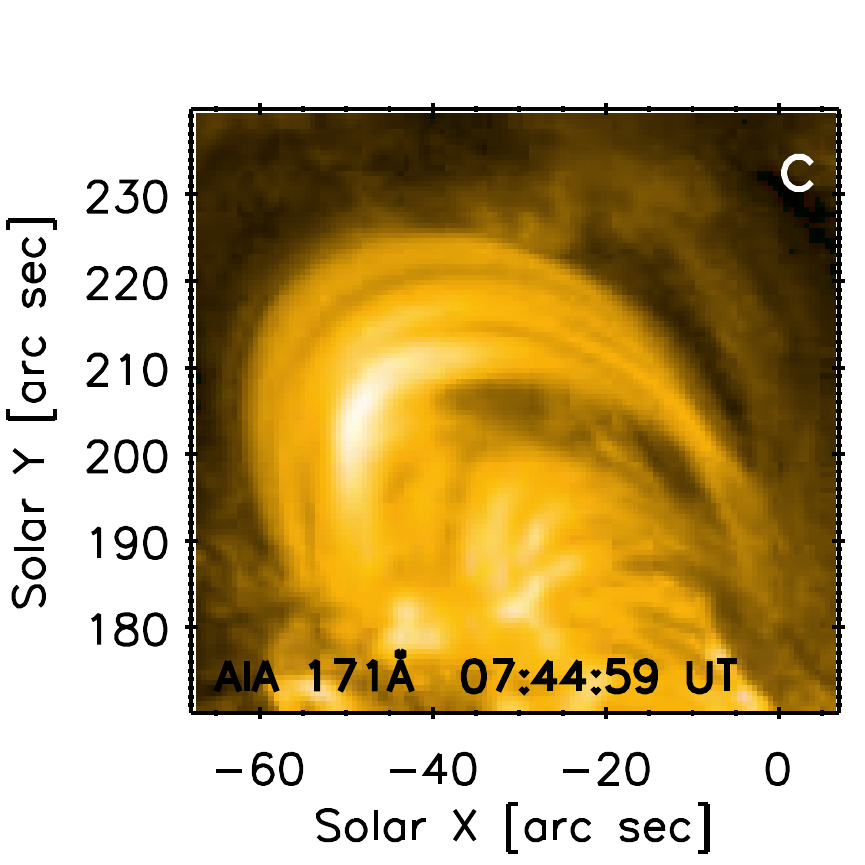}
	\includegraphics[width=4.13cm,clip,viewport=50 40 245 225]{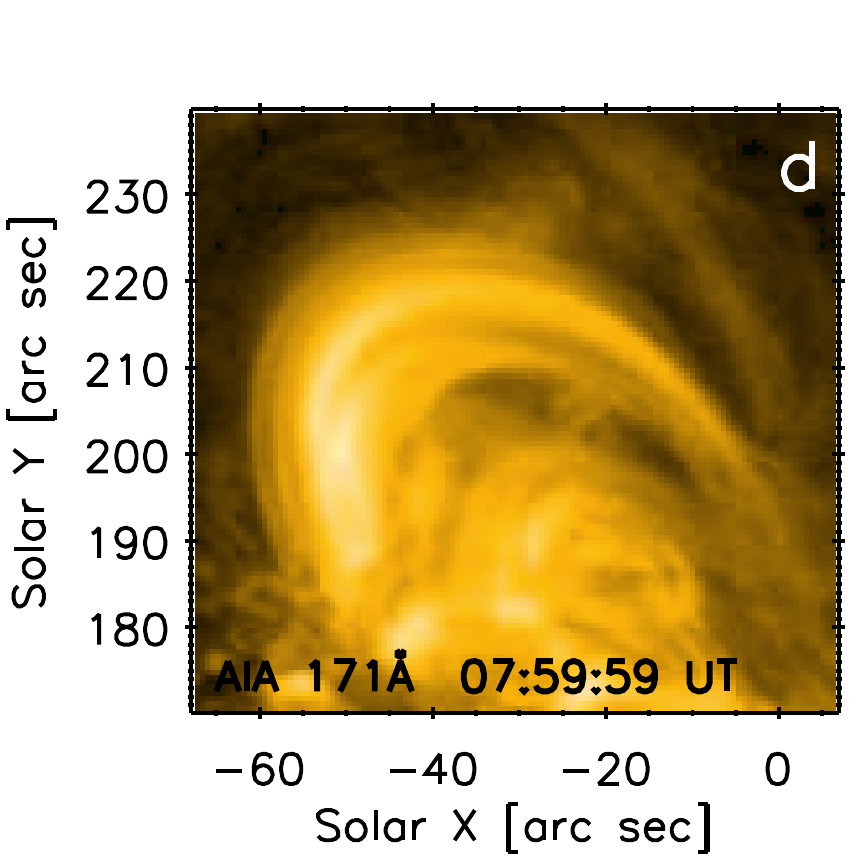}
	\includegraphics[width=5.20cm,clip,viewport= 0  0 245 220]{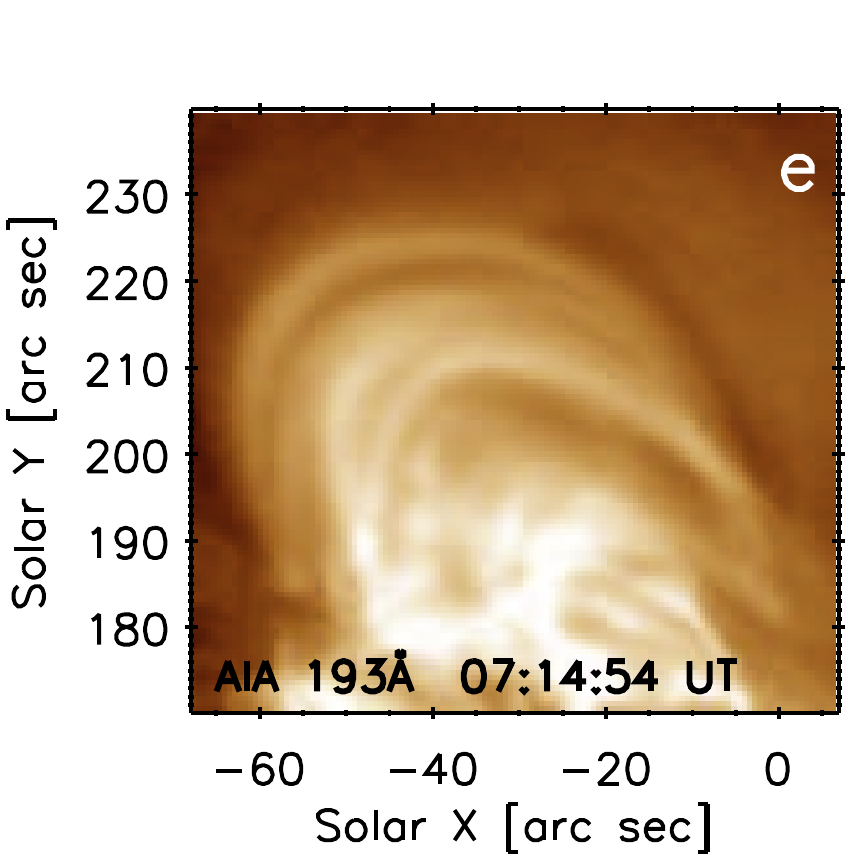}
	\includegraphics[width=4.13cm,clip,viewport=50  0 245 220]{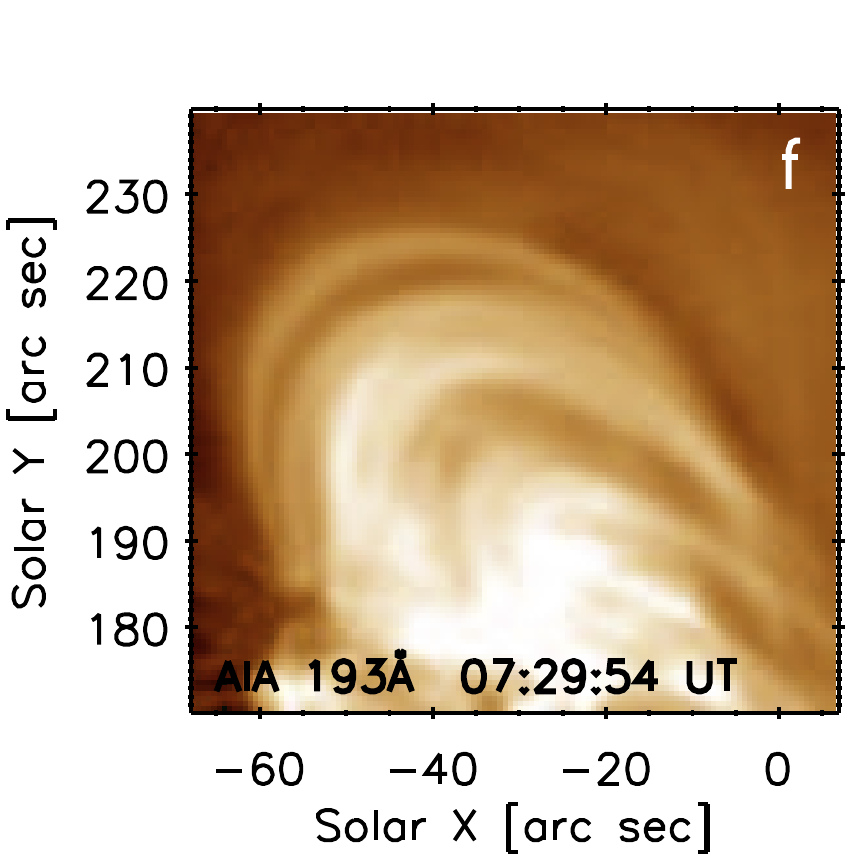}
	\includegraphics[width=4.13cm,clip,viewport=50  0 245 220]{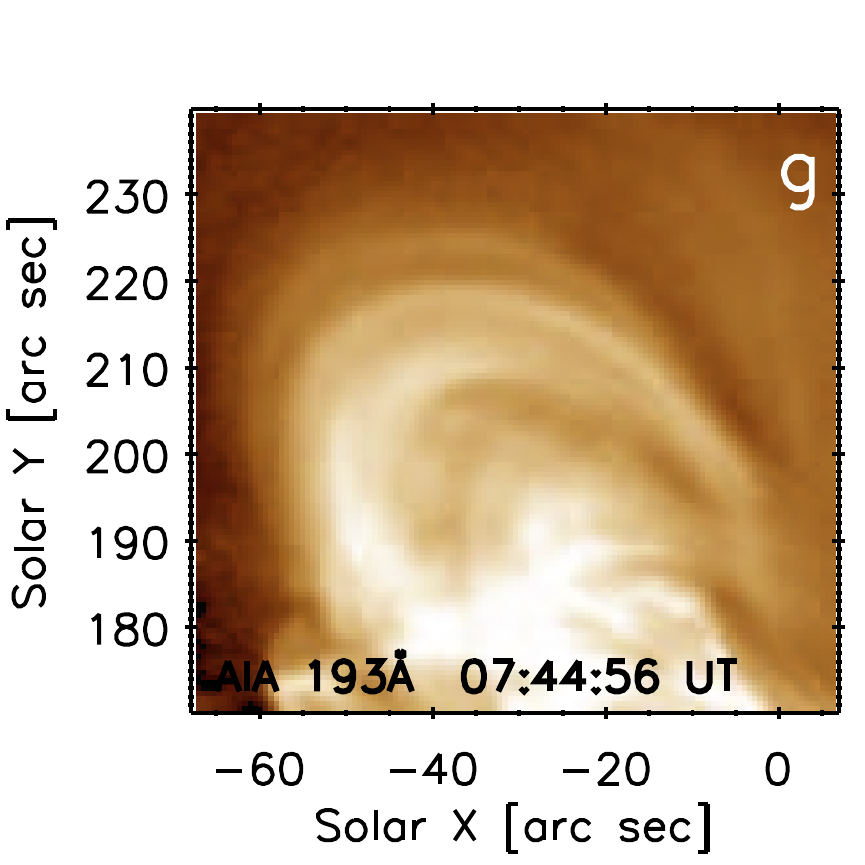}
	\includegraphics[width=4.13cm,clip,viewport=50  0 245 220]{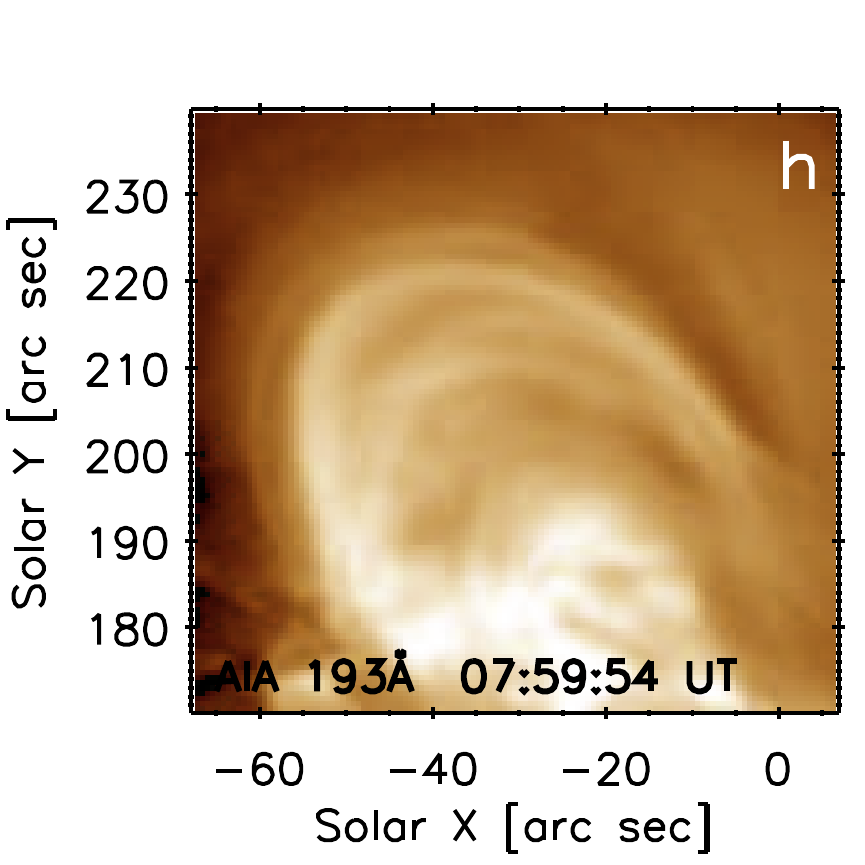}
	\includegraphics[width=9.48cm,clip,viewport= 0  0  490 260]{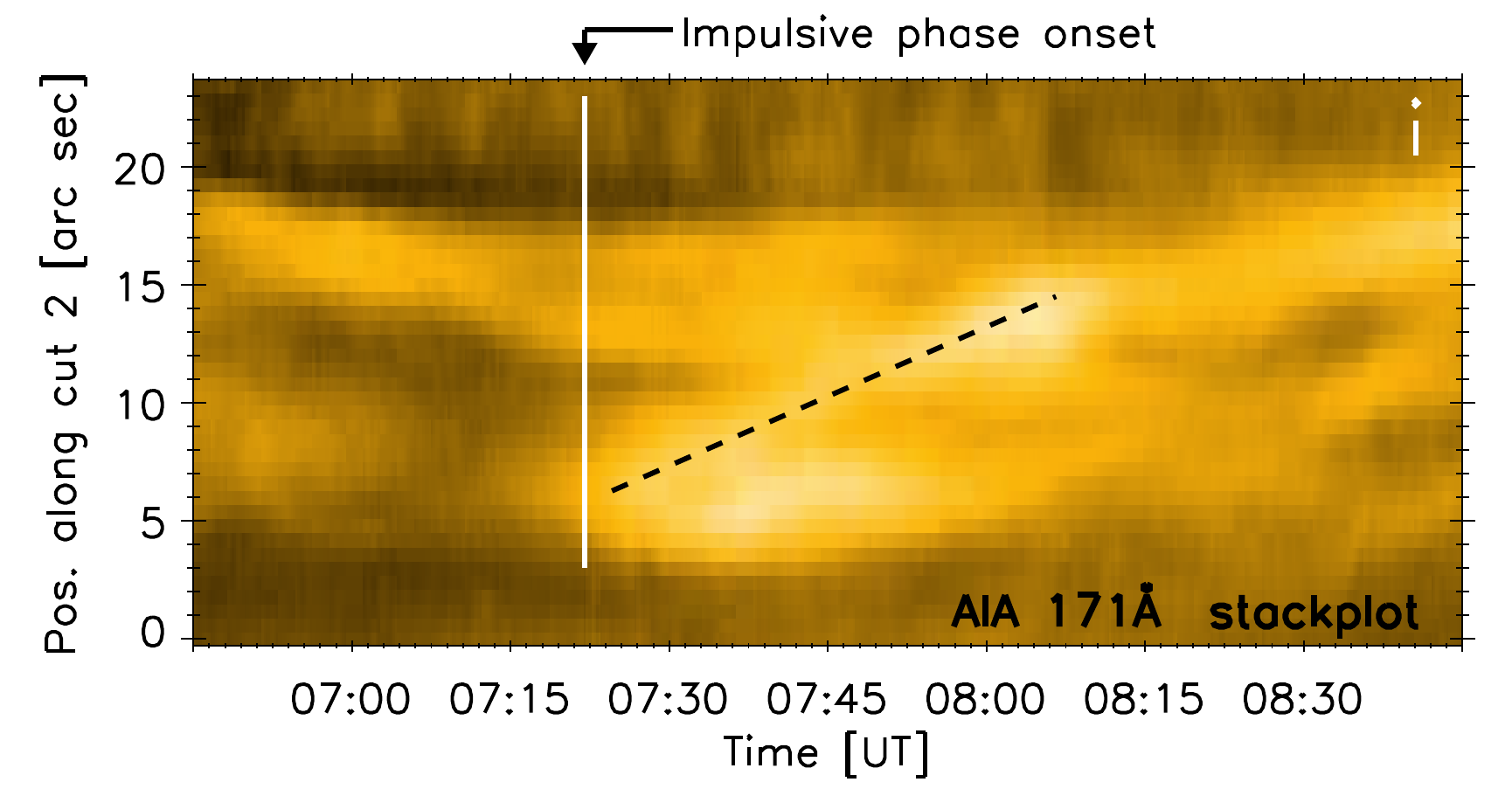}
	\includegraphics[width=8.32cm,clip,viewport=60  0  490 260]{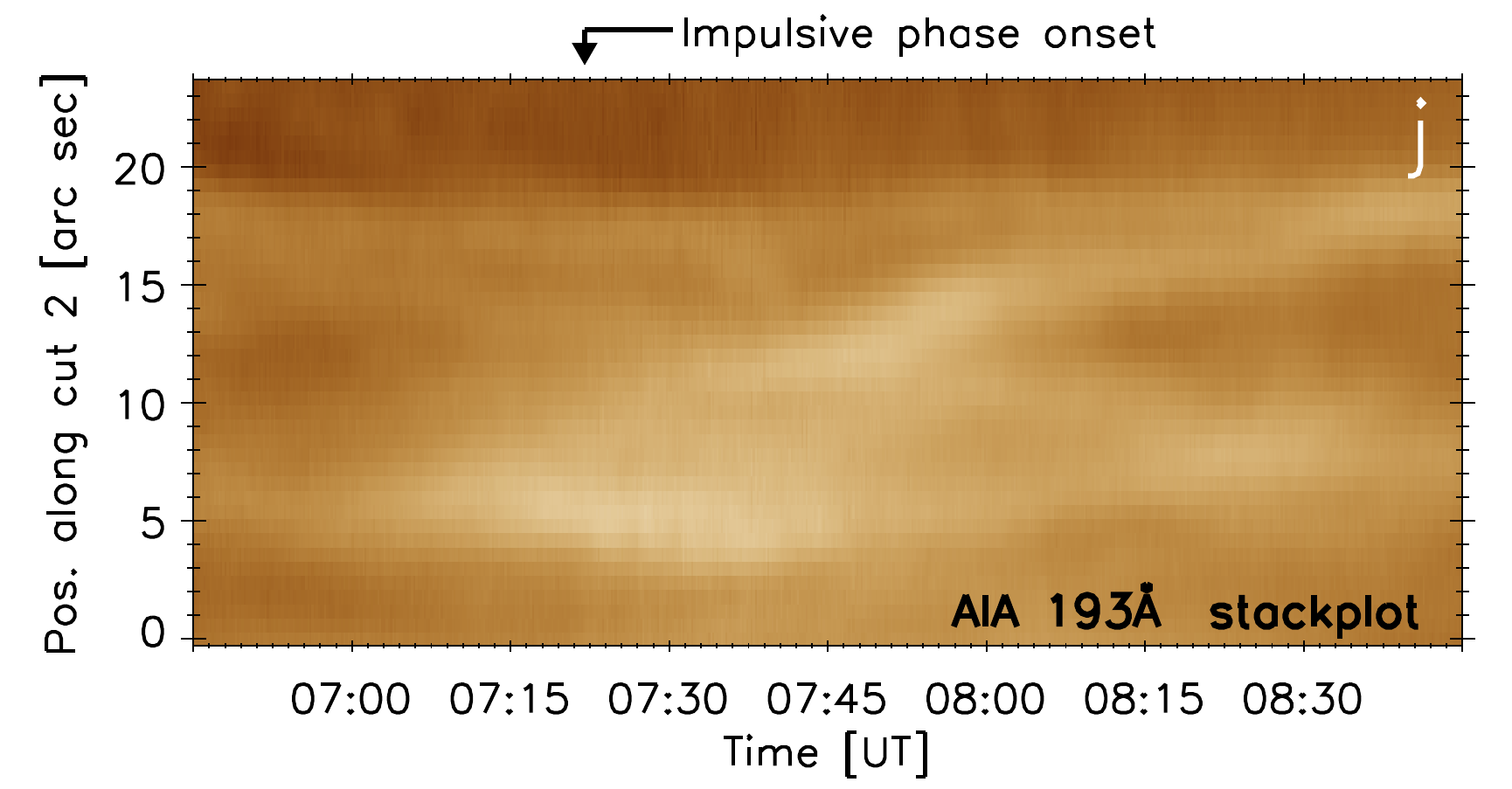}
\caption{(a)--(d) and (e)--(h): Evolution of the NE arcade. The field of view shown corresponds to the black box in Figure \ref{Fig:C_Context}.
(i) and (j): Time-distance plots constructed for AIA 171\,\AA~and 193\,\AA~along the cut 2. The dashed black line denotes an expanding loop in 171\,\AA.
\label{Fig:C_Vortex_loops_NE}}
\end{figure*}
%

\subsection{The South-Western Arcade}
\label{Sect:4.1}

To study the behavior of the apparently imploding arcade, we take a cut across it (labeled as cut 1), avoiding the flare and its diffraction pattern in every AIA filter as much as possible. The cut originates at [$X$, $Y$]\,=\,[15$\arcsec$, 145$\arcsec$], and extends across the arcade towards [55$\arcsec$, 105$\arcsec$] (see Figure \ref{Fig:C_Context}a). The corresponding time-distance plots are shown in Figure \ref{Fig:C_Vortex_loops_SW}, where the onset time of the impulsive phase at 07:22\,UT is marked. We note that this cut is not located across the central portion of the loop arcade. Its location however helps to reveal presence of fainter expanding components, which are not detectable with a central cut, such as the cut 2 of \citet[][see Figure 2 therein]{Wang16}.

The AIA 171\,\AA~time-distance plot along cut 1 in Figure \ref{Fig:C_Vortex_loops_SW} shows loop expansion, lasting until $\approx$07:22\,UT, followed by near-uniform contraction lasting until about 07:27\,UT. This is in complete agreement with the results reported by \citet{Wang16}. The bright feature denoted by the orange arrow corresponds to the western portion of the filament, whose behavior has been reported on by \citet{Wang16}. In the AIA 193\,\AA~and 211\,\AA~time-distance plots however, we detect two expanding loops at a time when the arcade seen in 171\,\AA~is already contracting. These two loops are denoted by blue and green arrows, respectively. 

We performed a differential emission measure (DEM) analysis of the AIA observations at 07:22\,UT using the method of \citet{Hannah12,Hannah13}. The results are shown in Figure \ref{Fig:C_Context}c. The DEMs of the two loops are similar, both peaking at log($T$ [K])\,=\,6.2, which corresponds to formation temperatures of \ion{Fe}{12}, indicating that these are coronal loops. We note that features similar to these two loops are detected also in 304\,\AA~processed images \citep[see Figure 3 of][]{Wang16}. Our DEM results mean that these loops are visible in 304\,\AA~because of the \ion{Si}{11} 303.33\,\AA~emission line \citep[][]{ODwyer10}, which is also formed at log($T$ [K])\,$\approx$\,6.2. 
  
More importantly, at about 07:28\,UT we detect a faint loop that does not follow the expansion/contraction with the same timing behavior as observed in 171\,\AA. This loop is denoted by red arrows in the 193\,\AA~and 211\,\AA~time-distance plots and is not visible in 171\,\AA\ (Figure \ref{Fig:C_Vortex_loops_SW}k--m). The loop expands until $\approx$07:28\,UT, then switches to contraction. We emphasize that this switch from expansion to contraction happens about 6 min later than for the arcade observed in 171\,\AA, and that this time approximately coincides with the end time of the collapse phase seen in 171\,\AA.

To study the behavior of this loop, in Figure \ref{Fig:C_Vortex_loops_SW} we show the AIA 193\,\AA~images (panels a--e) together with the running-difference (hereafter, RD) 193\,\AA~images (panels f--j). The RD helps to enhance the visibility of this loop. The delay time for the RD 193\,\AA~maps was chosen to be 24\,s; i.e., twice the AIA cadence. This is because of the AIA automated exposure control, which shortens the exposure times for every even-numbered frame during flares. Chosing a 24\,s interval for RD ensures that we subtract AIA images having the same exposure time. In the resulting RD 193\,\AA~maps, white corresponds to current position of the loop, and black to the position 24\,s ago. Therefore, contracting loops produce white--black concentric stripes along the direction of cut 1; with black color always on the outside portion of the concentric stripe. At 07:27:30\,UT, the loop denoted by red arrow shows a reversed black--white signature, meaning that it is expanding (Figure \ref{Fig:C_Vortex_loops_SW}h). At 07:27:54\,UT, the loop has moved outward, while decreasing its intensity (Figures \ref{Fig:C_Vortex_loops_SW}h--i), and finally at 07:28:18\,UT, the loop switches to contraction, having a white--black RD 193\,\AA~signature (Figure \ref{Fig:C_Vortex_loops_SW}j). We emphasize that at all these snapshots, the loop is concentric with other, apparently imploding loops. Thus it is unlikely that the expansion of this loop can be attributed to a different magnetic configuration than that of the arcade; apart perhaps from a modest change of inclination, which could be enough for this loop to be caught in a different, upper part of the vortex. Nevertheless, the presence of this loop is in line with the model-predicted coexistence of expanding/contracting loops at the sides of the legs of the erupting flux rope (see Section \ref{Sect:2}).

\subsection{Expanding Loop in the North-Eastern Arcade}
\label{Sect:4.2}

Having found the coexistence of expanding and contracting loops in the SW arcade, we next investigated the arcade on the opposite side of AR 11776. This north-eastern arcade, which has not been investigated by \citet{Wang16}, does not neighbor directly with the erupting filament (their mutual distance is at least several tens of arc sec), and is not reached by the EUV wave or its flanks visible in AIA 211\,\AA, reported by \citet{Wang16}, which generally moves southwards.

To study the behavior of the north-eastern arcade, we produced time-distance plots along the cut 2 (Figure \ref{Fig:C_Context}a and \ref{Fig:C_Vortex_loops_NE}a). We found a loop, visible in 171\,\AA~and 193\,\AA~(Figure \ref{Fig:C_Vortex_loops_NE}a--h), which starts expanding at roughly the same time as the onset of the filament eruption at 07:22\,UT (black dashed line Figure \ref{Fig:C_Vortex_loops_NE}i). The expansion lasts until about 08:05\,UT, i.e., for more than 40 minutes. The loop travels only about 10$\arcsec$ along the direction of cut 2. We note that this travelling loop is a coronal loop and not a flare one, since it is distinct from the flare arcade both in intensity and location \citep[see animations of Figure 4 and 5 of][]{Wang16}. 

Although we found no direct transfer of matter or any other visible causal agent between the filament eruption and the loop expansion, we note that the ZAD17 model predicts that the vortex flows should be located at both sides of the eruption, and gradually moving outwards. Here, we note that in the 4 hours of AIA observations we investigated (06:00--10:00\,UT), we did not find a loop in the NE arcade exhibiting similar behavior. Although there are other, fainter loops that exhibit expanding or contracting motions, these travel less than 5$\arcsec$, are fainter, and are typically overlapped by other loops. Given that the effective AIA resolution is 1.5$\arcsec$ \citep{Lemen12}, this means it is difficult to distinguish these loops clearly.

%
%
\begin{figure*}[ht]
	\centering
	\includegraphics[width=8.80cm,clip]{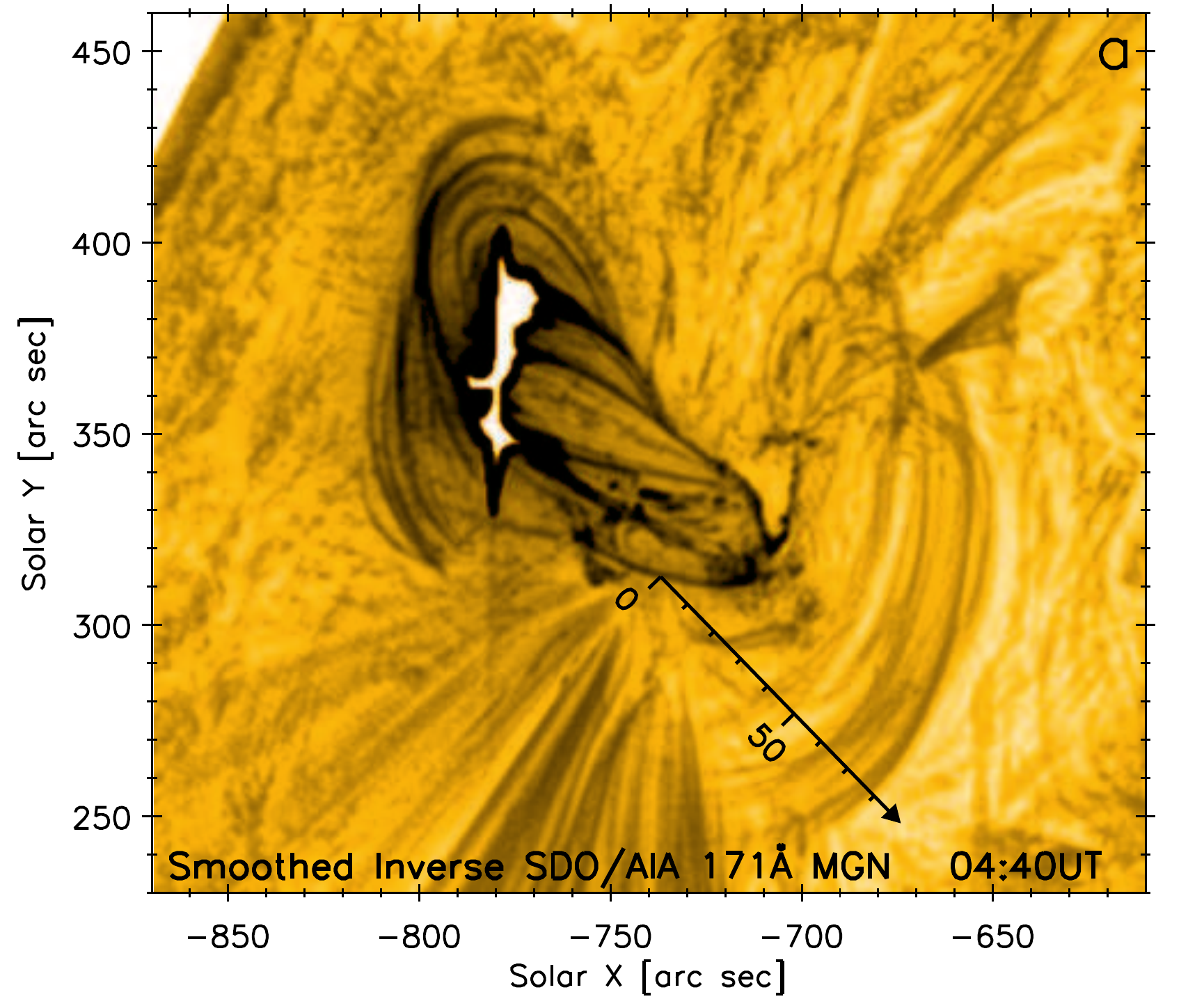}
	\includegraphics[width=8.80cm,clip]{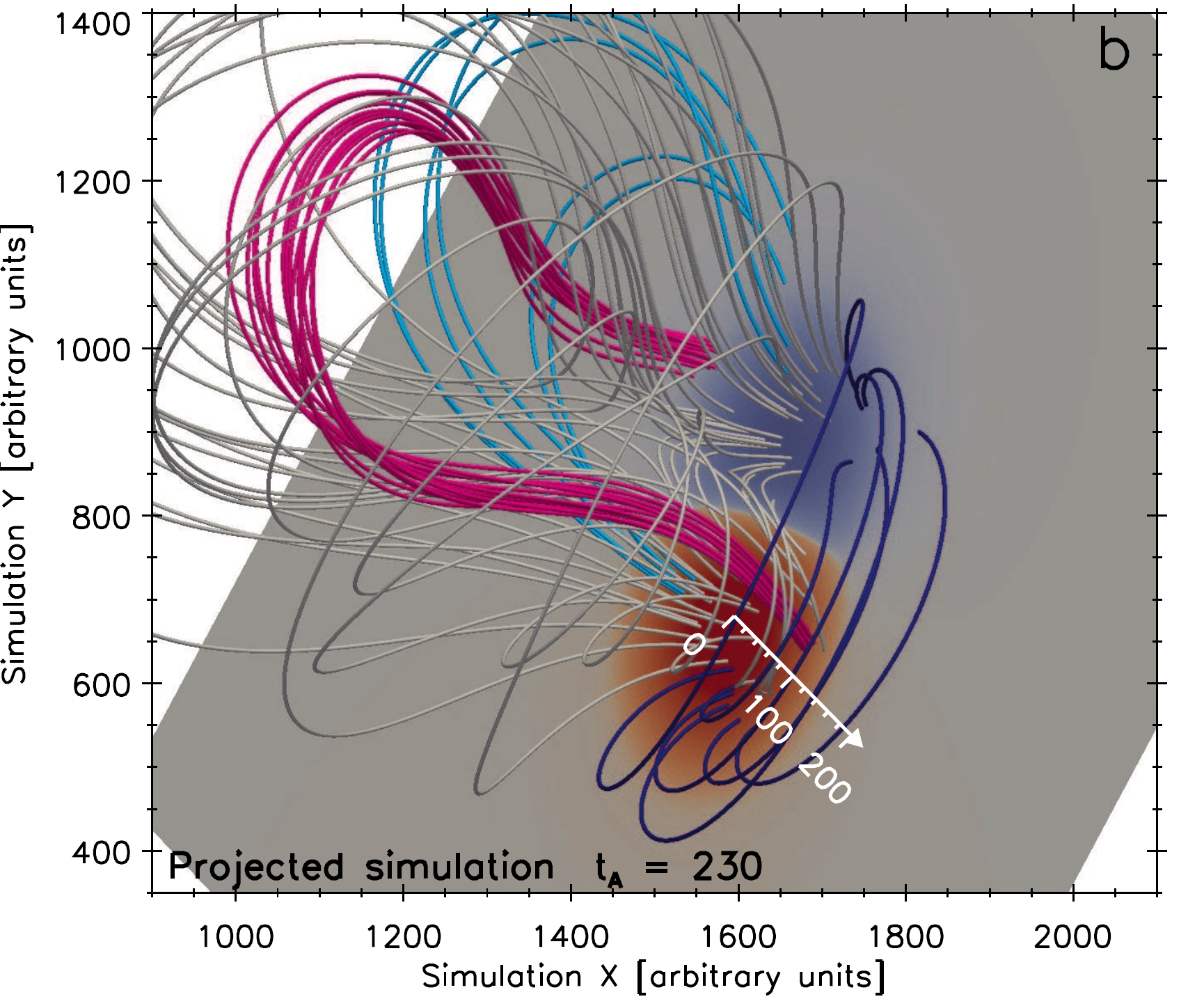}
	\includegraphics[width=8.80cm,clip]{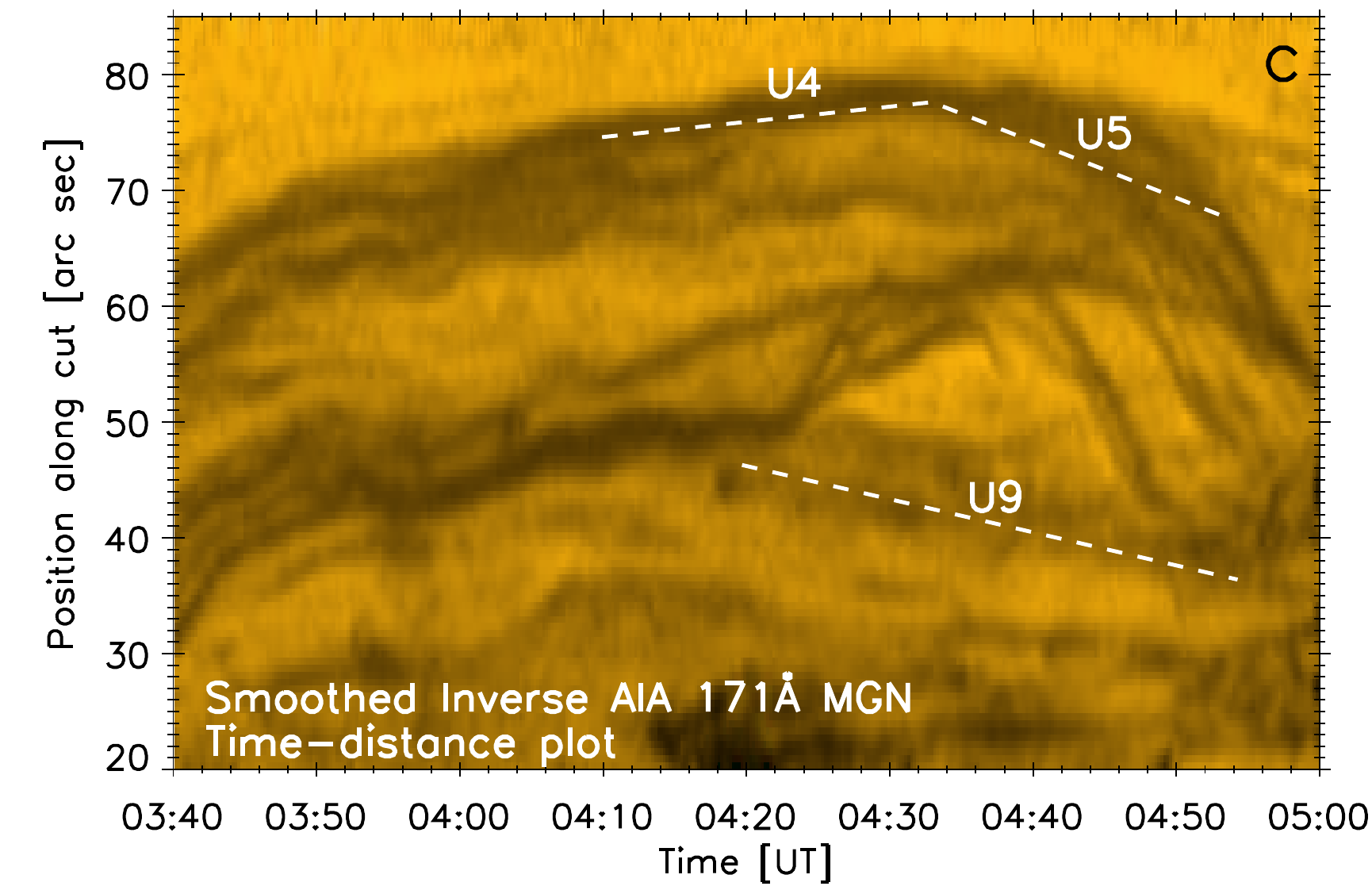}
	\includegraphics[width=8.80cm,clip]{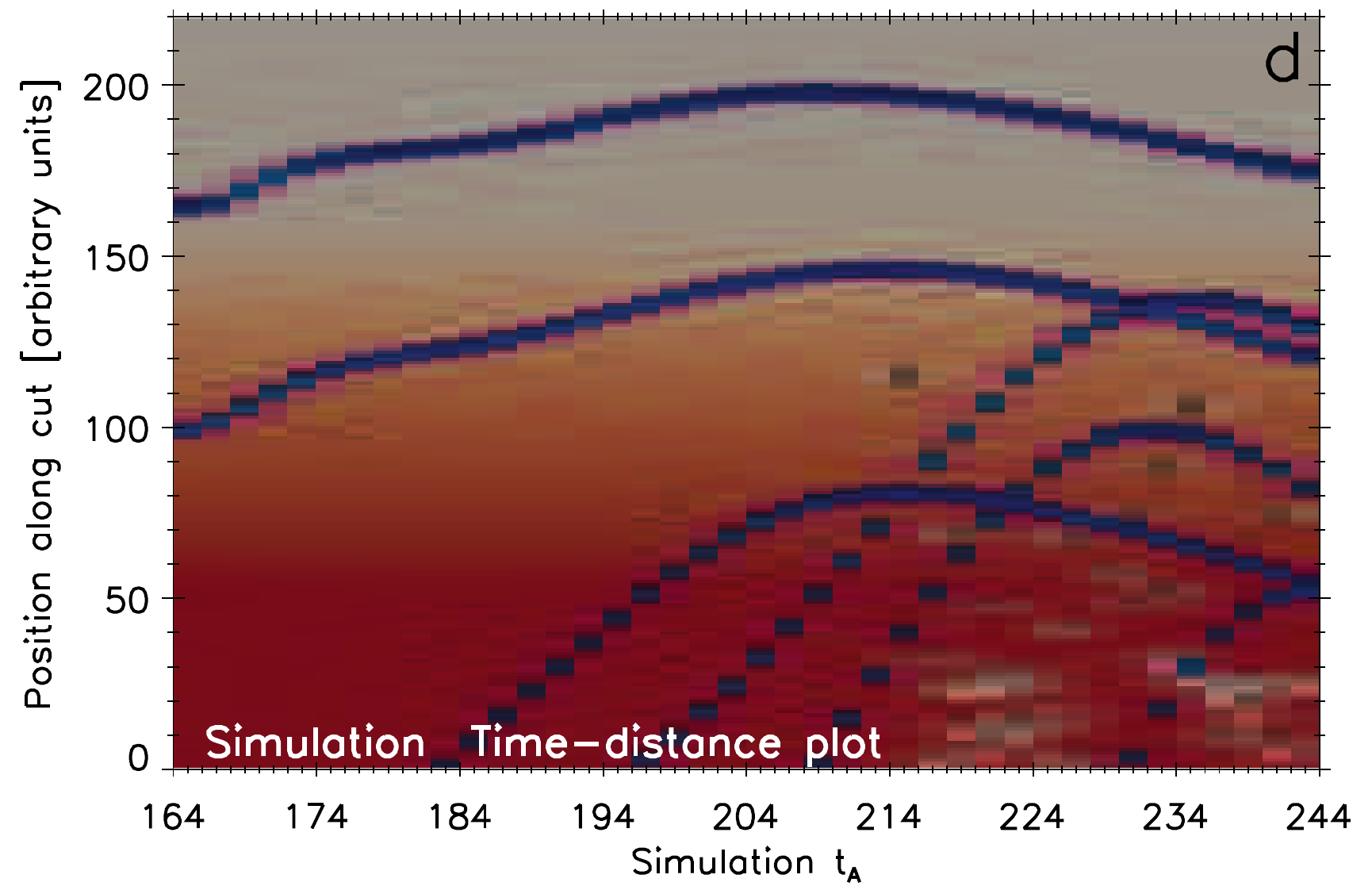}

\caption{Comparison of the observed X-class event (left) and the projected simulation (right). Snapshots of the AIA 171\,\AA~MGN and the simulation are provided in panels (a) and (b), respectively, while the time-distance plots along the respective cuts are given in panels (c) and (d). Note that the time-distance plot in panel (c) has been smoothed by a 3$\times$3 pixel boxcar. The position in the projected simulation is given in arbitrary units, while the time is in $t_\mathrm{A}$.
\label{Fig:Simulation_parallel_proj}}
\end{figure*}
%

\begin{figure*}[ht]
	\centering
	\includegraphics[width=8.80cm,clip]{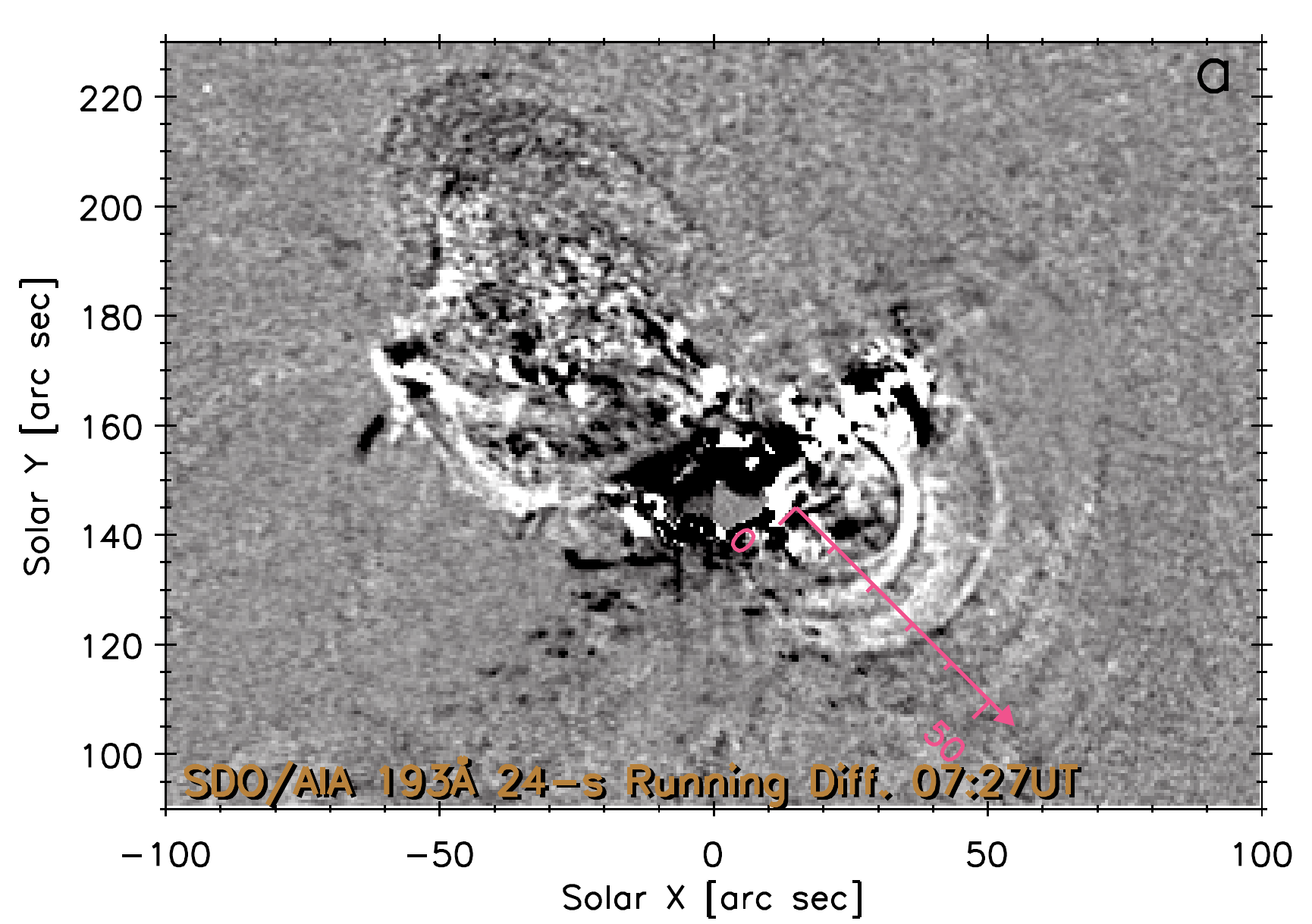}
	\includegraphics[width=8.80cm,clip]{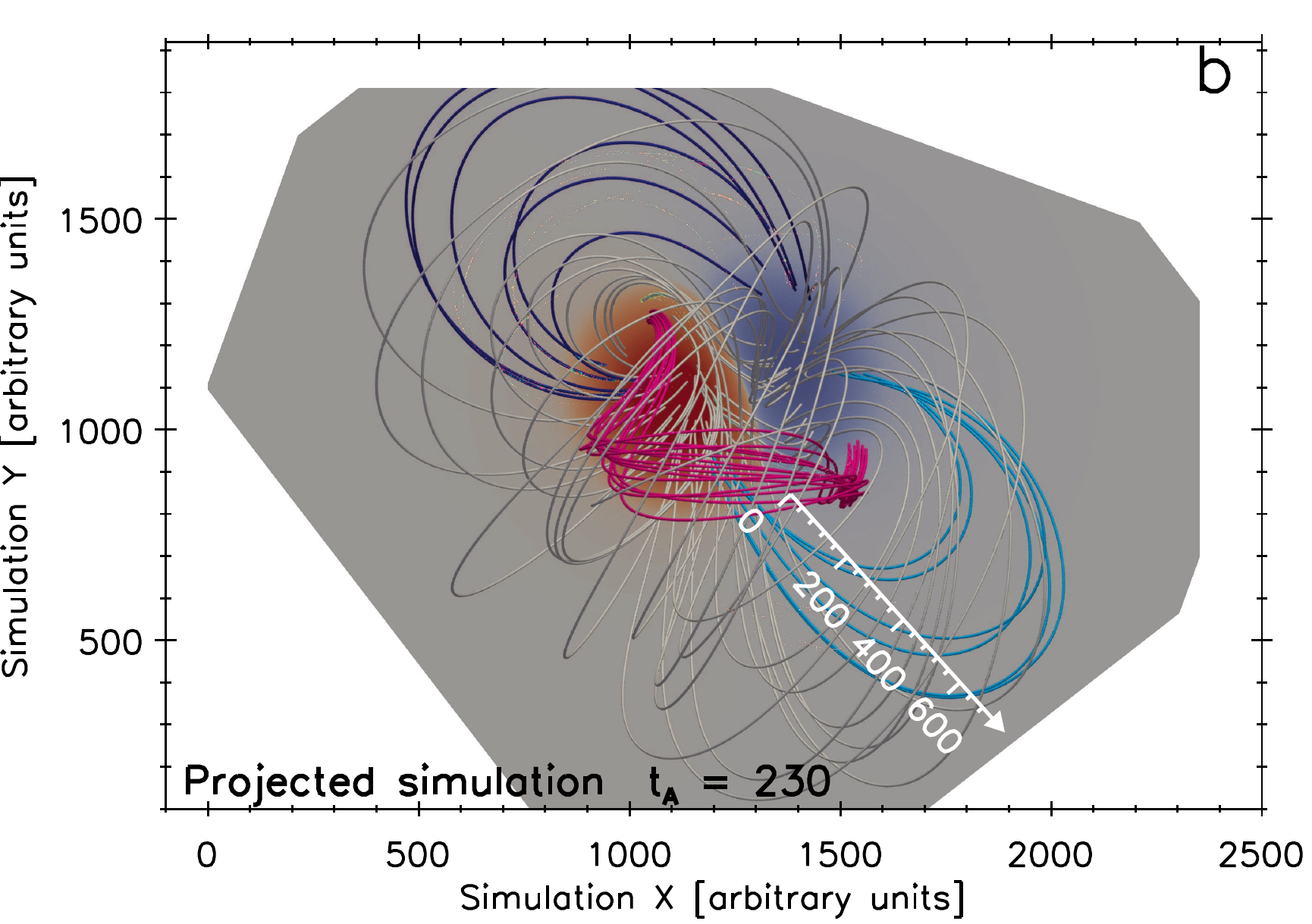}
	\includegraphics[width=8.80cm,clip]{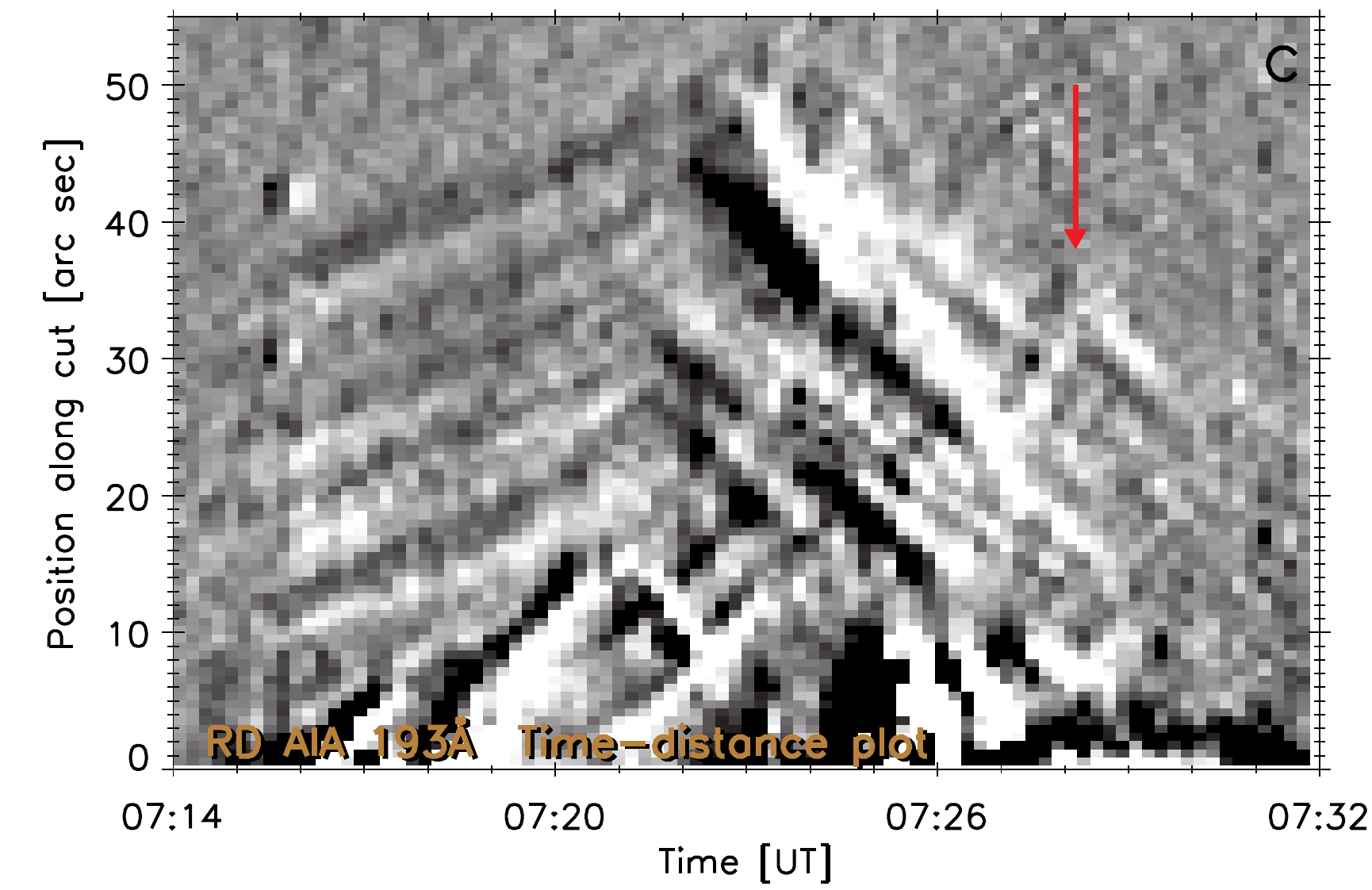}
	\includegraphics[width=8.80cm,clip]{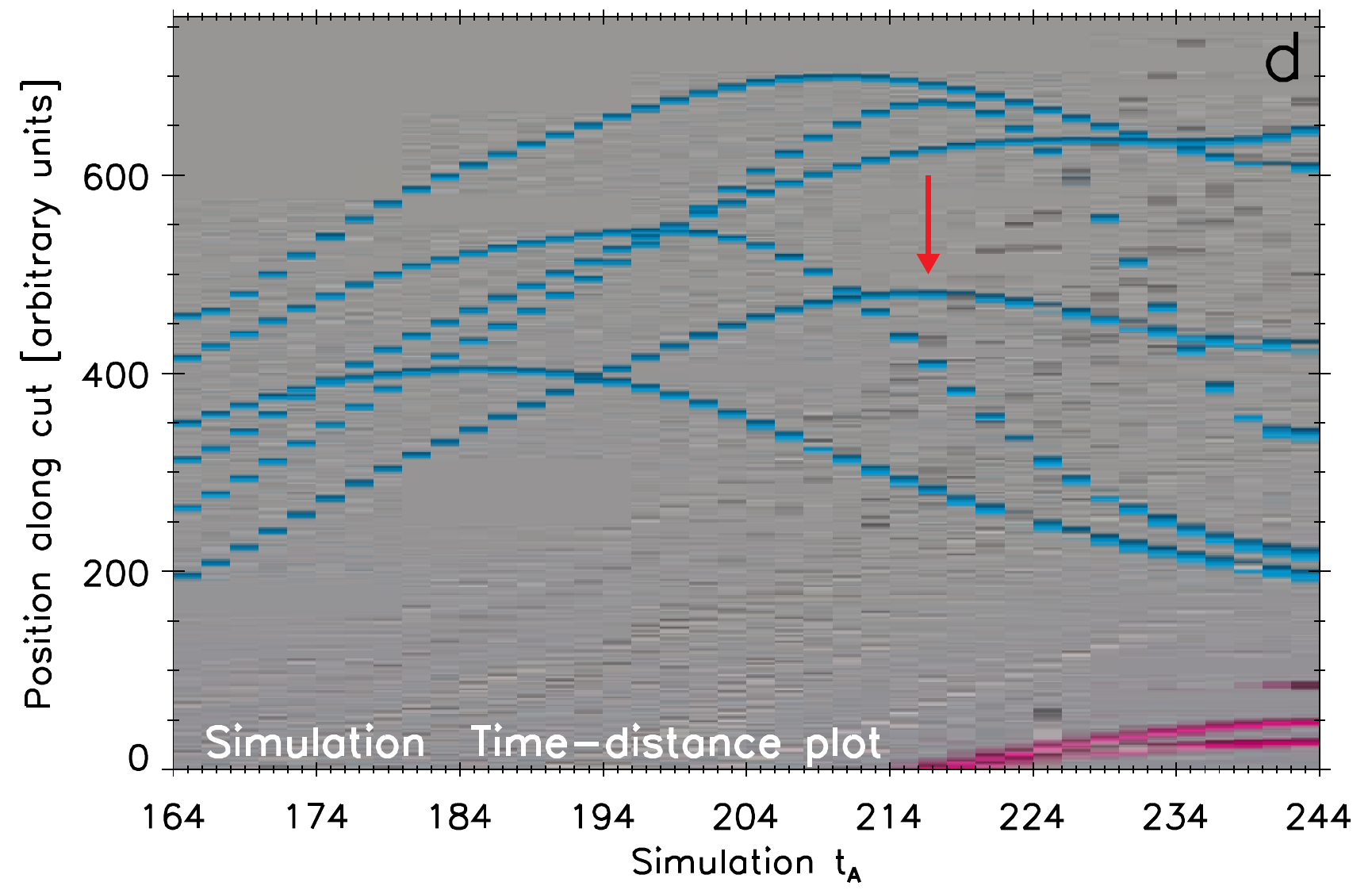}

\caption{Comparison of the observed C-class event (left) and the projected simulation (right). Snapshots of the AIA 193\,\AA~running difference (with 24\,s delay, see Sect. \ref{Sect:4.1}) and the simulation are provided in panels (a) and (b), respectively. The respective time-distance plots along the cuts are given in panels (c) and (d). The position in the projected simulation is given in arbitrary units, while the time is in $t_\mathrm{A}$.
\label{Fig:Simulation_top_view2}}
\end{figure*}
%
\section{Comparison of the Model and Observations}
\label{Sect:5}

Having shown the coexistence of the expanding and contracting motions in two events, we next proceed to compare the model with the observations in more detail. To do that, we oriented the model similarly to observed events. The observer-like views are shown in Figure \ref{Fig:Simulation_parallel_proj} for the X-class event and in Figure \ref{Fig:Simulation_top_view2} for the C-class event. We note again that the MHD model of ZAD17 is generic, i.e., it was \textit{not} designed to match a specific event. For example, there are only two photospheric magnetic polarities in the model, while the observed events both contained multiple magnetic polarities \citep[see Figure \ref{Fig:X_Context}b and Figure 11 of][]{Wang16}; in particular $\delta$-spots or magnetic tongues \citep[see also][]{Poisson16}. These additional complexities in the solar active regions, as well as the presence of other large-scale fields could in principle modify the speed of the vortices and possibly their shape. For these reasons, it cannot be expected that the simulation will capture every aspect of the observed behavior, and we will focus only on comparing the simulation and the observations in general terms.

We also note that the blue and cyan loops shown in Figure \ref{Fig:Model} have been selected for display purposes already in ZAD17. These loops possess a range of inclinations with respect to the local vertical. This is unlike the observed cases, where the range of inclinations could be more restricted as the loops have similar appearance and thus pressure scale-lengths. Nevertheless, we use the observed loops and their size to obtain an approximate spatial scale and thus the field of view for the model. Having done so, we use one set of the modeled loops per event, with the blue field lines approximating the loops observed in the X-class event (Figure \ref{Fig:Simulation_parallel_proj}) and the cyan loops for the C-class event (Figure \ref{Fig:Simulation_top_view2}). We remind the reader that, as stated already in Section \ref{Sect:2.3} and in ZAD17, the behavior of the blue and cyan arcades is qualitatively similar.

Finally, the behavior of the modeled blue and cyan field lines is elucidated by treating the model as synthetic observables, and taking cuts across the projected loop locations. This means that individual pixels in the projected model images are thought to represent an observed situation. The finite width of individual field lines, as well as finite image cadence, can be thought of as consequences of ``real" spatial resolution and temporal cadence. In observations, coronal loops are always several pixels wide \citep[e.g.,][]{Aschwanden05b,Aschwanden08,Aschwanden17,Brooks12,Brooks13,Peter13}. In our model, the temporal cadence of the snapshots is limited and does not allow to produce smooth curves in the simulation time-distance plots. However, these time-distance plots, even if jagged, are sufficient to capture the character of the modeled field line evolution.

In the X-class event, we take a cut across the arcade of blue field lines facing the observer (Figure \ref{Fig:Simulation_parallel_proj}b). The direction of the cut is approximately the same as in the observed X-class flare; however, in the simulation, the cut is shortened near the flare locus to avoid excessive intrusion of the flux rope core (pink) and its envelope (grey) field lines into the simulation time-distance plot. We smoothed the observed AIA 171\,\AA~MGN time-distance plot by a 3$\times$3 boxcar to reduce its resolution for better comparison with the simulation. 

Both the observed and simulated time-distance plots are similar in several ways. First, both show coexistence of expanding and contracting motions, i.e., the scenario predicted by ZAD17. Second, some loops show slowest expansion and relatively flat profile on the time-distance plots. In the observations, these loops are, for example, the loops U4--U5 or the less intense U9 (Figure \ref{Fig:Simulation_parallel_proj}c, see also Figure \ref{Fig:X_Stackplots_loops}a). Similarly, in the simulation, some blue field lines have relatively flat profiles. These loops are located in the outer portion of the cut, starting at positions $\approx$95 and 165 in Figure \ref{Fig:Simulation_parallel_proj}d. Panel b of the same Figure shows that these are the most inclined loops, i.e., those closest to the photosphere. At these altitudes, the vortex flow is relatively slower than higher up, where the other blue loops are located (see e.g. Figure 2 in ZAD17); hence leading to flatter profiles. According to the simulation, this would be due to a boundary-layer resulting from the proximity of the static photosphere, and to the surrounding denser low-corona. In the observations, the U4--U5 loop with a flat profile have likely lower inclination than the most inclined blue field lines, while the loops with largest inclinations and thus closest to the photosphere (such as U9) can not be visible at some times (see animation accompanying Figure \ref{Fig:X_Vortex_loops}). In both the simulation and the X-class event, this  behavior is at least in part a projection effect, as the loops expand and evolve mostly along the LOS direction.

Third, the inner (along the simulation time-distance plot) ``late'' loops show faster velocities than the outer ones (Figure \ref{Fig:Simulation_parallel_proj}d). In the simulation the large range of initial inclinations implies that some blue field lines enter from position~0 only at later times. These field lines have the lowest initial inclinations with the respect to the local vertical. Finally, the simulation does not contain the fast contraction phase that occurs after 04:40\,UT in the X-class event; i.e., more than an hour after the onset of the fast eruption (Figure \ref{Fig:X_Eruption}). As mentioned in ZAD17 and in Section \ref{Sect:2.1}, the simulation is halted at $t$\,=\,244\,$t_\mathrm{A}$, when the flux rope is still present and erupting. Thus, the simulation cannot capture this fast collapse phase.

For the C-class event, we selected the cyan field lines. This is for two reasons. First, the range of their inclinations at the onset of the flux rope eruption is lower than in the blue arcade (see Figure 1 in ZAD17). Second, the eruption is asymmetric and moves in the direction of the cyan field lines (see Figure \ref{Fig:Model5}), while in the C-class event, the filament erupts towards the SW arcade \citep[see][Figures 3--4]{Wang16}. To achieve better comparison with the observations, the model is mirrored along its X-axis. The observed C-class event and the oriented simulation are shown in Figure \ref{Fig:Simulation_top_view2} (a) and (b). We again take a cut in the approximate direction through the arcade, and compare it with the observational results in panels (c) and (d). Both time-distance plots show coexistence of expanding and contracting motions. We note that several gray field lines, indicating the envelope of the erupting flux rope, invade the location of the cut as they expand following the erupting flux rope core shown in pink. 

The behavior of the observed loop arcade is however different from the modeled cyan field lines. In the simulation, the cyan loops of ZAD17 are not concentric as is the case in the observed event. Furthermore, the observed AIA 171, 193, and 211\,\AA~loops change their behavior from expanding to contracting almost at the same time ($\approx$07:22\,UT, Section \ref{Sect:4.1}), although the outer 193 and 211\,\AA~loops seem to exhibit a slow delay of less than 1 minute (see Figure \ref{Fig:C_Vortex_loops_SW}) compared to the inner loops. 

In the simulation time-distance plot, some loops follow a behavior such that the longer the loop, the later it switches from expansion to contraction, with a linear-like delay. This is similar to the previous reports of observed contraction events seen near disk center \citep[e.g.,][]{Sun12,Gosain12,Simoes13a,Petrie16}. Furthermore, the expansion for these loops is slower than the subsequent contraction, similarly as in the our observation \citep[see also Figure 6b of][]{Wang16} and other observations mentioned. This behavior is likely caused by strengthening and reaccelerating of the vortex in time, as explained in Section 3.2.3 of ZAD17.

Loops not following this pattern exist however, similarly as reported in Section \ref{Sect:4.1}. In particular, the innermost cyan field line (starting from the position of 200 at 164 $t_\mathrm{A}$ switches to contraction only after 214 $t_\mathrm{A}$, a behavior similar to the loop denoted by red arrow (Figure \ref{Fig:Simulation_top_view2}c), although the rate of contraction in the simulation is lower than in the observations.

Summarizing, despite using a generic eruption model, we have shown that the modeled behavior in the simulation time-distance plots is similar to the observations discussed. This provides further support for the vortex interpretation of the observed coexistence of the expanding and contracting loops.

%
\section{Summary and Discussion}
\label{Sect:6}

We have shown, for the first time, that co-temporally expanding and contracting loops exist in eruptive flares, both X-class and C-class. Such behavior was recently predicted by MHD modeling to be generally present during solar eruptive events \citep{Zuccarello17}. In the model, regions of enhanced vorticity develop around the legs of the erupting flux rope. The rotational motions within the vortices are such that outward flows from the eruption site are located at higher altitudes, and reversing flows occur in the lower part of the vortex. These vortices are a hydromagnetic analogue of a well-known hydrodynamic process creating vortex rings (toroidal vortices), such as smoke rings. In solar eruptions, such vortices naturally translate to coexistence of expanding and contracting motions of coronal loops, as loops with different inclinations may be caught in different part of the vortex. We note that the predominance of reported contraction \citep[e.g.,][]{Gosain12,Simoes13a,Wang16} may be simply due to the return flows at the lower part of the vortex affecting highly inclined loops, which are more readily visible because of the hydrostatic effects changing the apparent density scale heights with inclination.

Our analysis have shown that the coexistence of expanding and contracting loops in both flares occurs after the onset of eruption and impulsive phase for both events. This is in accordance with the general prediction of the model, with the observed behavior being similar to the simulated one (see animation of the model Figure \ref{Fig:Model}, as well as Figures \ref{Fig:Simulation_parallel_proj} and \ref{Fig:Simulation_top_view2}). In the long-duration X-class event of 2012 March 05, the observed coexistence of expanding and contracting coronal loops within the same arcade occurs for more than 35 minutes, even if the arcade exhibits a general expansion followed mostly by contraction. In the C-class event, a single expanding 193\,\AA~loop was found to be concentric with the overally imploding arcade in 171\,\AA, with the expansion of the loop lasting nearly until the end of the overall contraction phase as seen in 171\,\AA.

Previously, contracting loop behavior during flares have been regarded as a validation of the implosion conjecture of \cite{Hudson00}, later rephrased in terms of coronal loop arcade collapse as a consequence of the decrease of magnetic pressure, due to the energy release that occurs during the flare \citep{Russell15} or eruption \citet{Wang16}. 

In this context, the expanding 193\,\AA~loop in the C-class flare is of special importance, as it is observed to be both expanding long after the filament has erupted, while being concentric with the apparently imploding arcade. Similarly to the MHD model, it is unlikely that the expansion of this loop can be attributed to a different magnetic configuration than that of the arcade, apart perhaps from a modest change of inclination.


Finally, we note that the prediction of the vortex flows is a general one, and thus more observational evidence of vortex flows, especially at different projections, should be forthcoming.

\acknowledgments
The authors thank the anonymous referee for comments that helped improve the manuscript. J.D. acknowledges support from the Grant Agency of the Czech Republic, Grant No. 17-16447S, and institutional support RVO:67985815 from the Czech Academy of Sciences. J.D. and F.P.Z. also acknowledge support from the Royal Society via its Newton Alumni Programme. F.P.Z. is a Fonds Wetenschappelijk Onderzoek (FWO) research fellow. 
This work was granted access to the HPC resources of MesoPSL financed by the R\'{e}gion Ile de France and the project
Equip@Meso (reference ANR-10-EQPX-29-01) of the programme Investissements d’ Avenir supervised by the Agence Nationale pour la Recherche. AIA and HMI data are courtesy of NASA/SDO and the AIA and HMI science teams. The SOHO/LASCO data used here are produced by a consortium of the Naval Research Laboratory (USA), Max-Planck-Institut f \"{u}r Sonnensystemforschung (Germany), Laboratoire d’Astrophysique Marseille (France), and the University of Birmingham (UK). SOHO is a project of international cooperation between ESA and NASA.
\facilities{AIA (SDO), HMI (SDO), LASCO (SOHO), GOES.}






\bibliographystyle{aasjournal}
\bibliography{Vortex_Obs_Paper1}



\end{document}